\documentclass[reqno,10pt]{article}
\usepackage{amsmath,amsfonts,amsgen,amstext,amsbsy,amsopn,amsthm, amssymb}
\usepackage{multirow, verbatim, cancel, mathabx}
\usepackage{bbold}
\usepackage{enumitem}
\usepackage[colorlinks=true,hyperindex=true]{hyperref}

\numberwithin{equation}{section}

\newcounter{smallarabics}

\newcounter{smallroman}
\newenvironment{romanenumerate}
{\begin{list}{{\normalfont\textrm{(\roman{smallroman})}}}
  {\usecounter{smallroman}\setlength{\itemindent}{0cm}
   \setlength{\leftmargin}{5ex}\setlength{\labelwidth}{4ex}
   \setlength{\topsep}{0.75\parsep}\setlength{\partopsep}{0ex}
   \setlength{\itemsep}{0ex}}}
{\end{list}}

\newcommand{\ben}{\begin{romanenumerate}}  
\newcommand{\een}{\end{romanenumerate}}  
\newcommand{\fhL}{\mathfrak{h}_\Lambda}

\newcommand{\doublearrow}{\leftrightarrow}

\newtheorem{theoreme}{Theorem }[section]
\newtheorem{theorem}[theoreme]{Theorem}
\newtheorem{proposition}[theoreme]{Proposition}
\newtheorem{lemma}[theoreme]{Lemma}

\newtheorem{corollary}[theoreme]{Corollary}
\newtheorem{remark}{Remark}[section]

\newcommand\nn\nonumber
\renewcommand\leq\varleq
\renewcommand\geq\vargeq
\newcommand{\sign}{\mathrm{sign}}
\renewcommand{\star}{*}
\newcommand{\1}{\mathbb 1}
\newcommand{\RR}{\mathcal{R}}

\renewcommand{\proof}{\noindent \emph{Proof. }}
 
 \newcommand{\R}{\mathbb{R}}
 \newcommand{\N}{\mathbb{N}}
\newcommand{\Z}{\mathbb{Z}} \newcommand{\C}{\mathbb{C}}
 
\newcommand{\E}{\mathcal{E}} \renewcommand{\H}{\mathcal{H}}

\newcommand{\norm}[1]{\left\| #1 \right\|}
\newcommand{\K}{\mathcal K}
\renewcommand{\O}{\mathcal O}
\renewcommand{\d}{\mathrm{d}}
\renewcommand{\i}{\mathrm{i}}
\newcommand{\sx}{\sigma^{(1)}}
\newcommand{\sy}{\sigma^{(2)}}
\newcommand{\sz}{\sigma^{(3)}}
\newcommand{\su}{\sigma^{(+)}}
\newcommand{\sd}{\sigma^{(-)}}
\newcommand{\UJW} {U_{\mathrm{JW}}}

\def\fh{\mathfrak{h}}
\newcommand{\ES}{\mathrm{ES}}
\renewcommand{\epsilon}{\varepsilon}
\newcommand{\eps}{\varepsilon}
\renewcommand{\P}{\mathbb{P}}
\renewcommand{\sp}{\mathrm{sp}}
\newcommand{\dto}{\downarrow}
\newcommand{\ac}{\mathrm{ac}}
\newcommand{\T}{\mathcal{T}}
\newcommand{\GC}{\mathrm{GC}}
\renewcommand{\iff}{\Leftrightarrow}
\newcommand{\A}{\mathcal{A}}
\newcommand{\remarkk}{\noindent \emph{Remark. }}
\newcommand{\FCS}{\mathrm{FCS}}
\newcommand{\JW}{\mathrm{JW}}
\newcommand{\loc}{\mathrm{loc}}
\newcommand{\rest}{\upharpoonright}
\newcommand{\chl}{\chi_l}
\newcommand{\chr}{\chi_r}
\newcommand{\chlr}{\chi_{l/r}}
\newcommand{\chrl}{\chi_{r/l}}
\newcommand{\cha}{\chi_a}
\newcommand{\chb}{\chi_b}

\DeclareMathOperator{\tr}{tr} 
\DeclareMathOperator{\im}{Im} \DeclareMathOperator{\re}{Re}
\DeclareMathOperator{\spec}{sp}
\DeclareMathOperator{\ran}{Ran}
\DeclareMathOperator*{\slim}{s-lim}
\DeclareMathOperator{\supp}{supp}

\begin{document}

\begin{titlepage}

\begin{center}
{\Large Entropic fluctuations of XY quantum spin chains}

\vspace{10 pt}

Benjamin Landon

Department of Mathematics and Statistics

McGill University, Montreal

\vspace{10 pt}
August, 2013

\vspace{10 pt}

A thesis submitted to McGill University in partial fulfillment \\
of the requirements of the degree of M.Sc. in Mathematics

\vspace{10 pt}

\copyright Benjamin Landon, 2013

\end{center}

\end{titlepage}

\title{Entropic fluctuations of XY quantum spin chains}
\author{Benjamin Landon\\ Department of Mathematics and Statistics \\ McGill University, Montreal}
%\date{May 2013}

%\maketitle

\begin{abstract} We consider an XY quantum spin chain that consists of a left, center and right part initially at thermal equilibrium at  temperatures $T_l$, $T_c$, and $T_r$, respectively.  The left and right systems are infinitely extended thermal reservoirs and the central system is a small quantum system linking these two reservoirs. If there is a temperature differential, then heat and entropy will flow from one part of the chain to the other.  We consider the Evans-Searles and Gallavotti-Cohen functionals which describe the fluctuations of this flux with respect to the initial state of the system and the non-equilibrium steady state reached by the system in the large time limit. We also define the full counting statistics for the XY chain and consider the associated entropic functional, as well a natural class of functionals that interpolate between the full counting statistics functional and the direct quantization of the variational characterization of the Evans-Searles functional which appears in classical non-equilibrium statistical mechanics. The Jordan-Wigner transformation associates a free Fermi gas and Jacobi matrix to our XY chain. Using this representation we are able to compute the entropic functionals in the large time limit in terms of the scattering data of the underlying Jacobi matrix. We show that the Gallavotti-Cohen and Evans-Searles functionals are identical in this limit. Furthermore, we show that all of these entropic functionals are equal in the large time limit if and only if the underlying Jacobi matrix is reflectionless.

\vspace{10 pt}

\end{abstract}

\tableofcontents

\subsection*{Acknowledgements}

First and foremost, I would like to thank my parents, Rocky and Petra, for their continuing and boundless support throughout my education.  I would like to thank my supervisors Dr. Vojkan Jak\v{s}i\'c and Dr. Robert Seiringer whose guidance and direction has proven invaluable to my education. I am greatly indebted to them for the academic opportunities they have provided me.  I would also like to mention Dr. Claude-Alain Pillet and thank him for his help in my learning of quantum statistical mechanics and spectral theory, as well as mention the mathematical physics group at McGill who have made my time a positive learning experience. I would like to thank the Department of Mathematics and Statistics at McGill University for the opportunity to study here, and NSERC for financial support. I also owe a lot to the teachers I have had throughout my education. To name a few I would like to thank Dan Kimmerer, James Young, John Kitney, Cindy Eccles, Geoff Stewart, Todd Kartye, Brenda Scarlett, Dan Lalonde, Sean Allen, Brian Nesbitt, Dr. Gantumur Tsogtgerel and the many others who have had a profound and positive effect on my education.  In addition, I would like to thank the administrative staff at the Department for their help with all of the logistics that come with being a student; they have made my time here infinitely easier and very enjoyable. Yariv Barsheshat provided me with valuable assistance with regards to the French translation of the abstract.  Finally, I thank Leah Weiner for her constant support throughout the completion of my degree. By necessity, a list of acknowledgements omits many to whom thanks are due; there are many other friends, colleagues and instructors who have helped me but are not listed here and I give them my thanks as well.

%%\begin{comment}
\section*{Preface}

This document is a thesis submitted in partial requirements of the degree of M. Sc. in Mathematics, sought by the author, Benjamin Landon.  The author was supervised by Dr. Vojkan Jak\v{s}i\'c and Dr. Robert Seiringer.  The thesis has been written by the author. Dr. Jak\v{s}i\'c and Dr. Claude-Alain Pillet provided guidance where necessary as to the content of the thesis and also suggested methods by which to complete the proofs.
%%\end{comment}

\section{Introduction}\label{sec:int}

Entropic fluctuations in statistical mechanics concerns the study of the relative probability that entropy will increase or decrease over a time interval in a physical system.  We provide a brief historical review of the subject; the interested reader may consult \cite{RM} for a more exhaustive treatment and additional references. The study of fluctuations in statistical mechanics began in 1905 with the seminal paper of Einstein \cite{E1} on Brownian motion in which the first fluctuation-dissipation relation was given. In his 1910 paper \cite{E2}, Einstein gave a link between the entropy formula of Boltzmann and the probability of fluctuations out of an equilibrium state. Subsequent developments were made by Ornstein and Nyquist; Ornstein obtained a fluctuation-dissipation relation for a random force acting on a Brownian particle \cite{Or} and Nyquist computed spectral densities and correlation functions of thermal noise in linear electrical circuits in terms of their impedance \cite{Ny}. The classic result of Onsager is the converse \cite{O1, O2}; he obtained a formula for the transport coefficients or impedances in terms of thermal fluctuations.  Progress in the theory of transport coefficients and fluctuation-dissipation relations was continued in the works of Green \cite{G1, G2} and Kubo \cite{Kub} (e.g., the derivation of the well-known Green-Kubo formulas).

These early results concerned primarily the close-to-equilibrium regime; e.g., the transport coefficients refer to the first order response of a physical system in equilibrium to thermodynamic or mechanical forces pushing it out of equilibrium. While the equilibrium theory is considered satisfactory and complete, the same is not true of the non-equilibrium theory \cite{RM}.

Modern fluctuation theorems were first suggested numerically in the paper \cite{ECM}. There, the authors were interested in violations of the second law of thermodynamics; in the course of studying a deterministic particle system they found the relationship  (roughly speaking) that
\begin{align}
P_t ( \phi ) = e^{ - t \phi } P_t ( - \phi ) \label{eqn:intro1}
\end{align}
where $P_t ( \phi)$ is the probability of measuring a decrease in the entropy of the system of magnitude $\phi$ over a time interval of length $t$. That is, decreases of entropy can occur, but the probability of an entropy decrease occuring is exponentially small in the magnitude of the decrease compared to measuring an increase of the same magnitude. This exponential suppression has an equivalent formulation in terms of the cumulant generating function $E_t ( \alpha)$ for the random variable describing the entropy change over the time interval $[0, t]$. To be more precise, (\ref{eqn:intro1}) holds iff the cumulant generating function obeys $E_t ( \alpha ) = E_t ( 1 - \alpha)$; such a relationship has come to be known in the literature as a \emph{fluctuation relation}.  Such fluctuation relations were established theoretically for the first time by Evans and Searles \cite{ES} and Gallavotti and Cohen \cite{GC1, GC2}. One of the novel features of such modern fluctuation theorems is that they hold for systems far from equilibrium, and in the linear regime near equilibrium reduce to Green-Kubo formulas and Onsager relations, an observation made by Gallavotti in 1996 \cite{Ga, JPR}. They are therefore a generalization or a far-from-equilibrium version of the Green-Kubo formulas and Onsager relations which have played an important role in the development of non-equilibrium statistical mechanics.

While the majority of work in entropic fluctuations concerns classical mechanics, comparatively little is known in the quantum case and establishing quantum analogues of the existing results is an emerging and active area of research. The modern fluctuations theorems represent some of the few exact and general results in non-equilibrium statistical mechanics and it is therefore desireable to obtain a generalization to the quantum regime.  As it stands, both the classical and quantum theory admit an axiomatization \cite{JPR, JOP}; starting with a general classical or quantum dynamical system, the basic objects of the theory such as the entropy production observable and the finite time entropic functionals can be defined at a great level of generality. The axioms concern the existence and regularity of the entropic functionals in the large time limit. In the quantum case, it is generally extremely difficult to verify these axioms in physically interesting models \cite{JOPP}. 

The mathematical tools required to deal with the quantum case are in general very demanding. Virtually all aspects of Tomita-Takesaki modular theory play a role in the description of entropic fluctuations in quantum non-equilibrium statistical mechanics. For example, the Araki-Masuda non-commutative $L^p$ spaces take the place of the familiar $L^p$ spaces of measure theory appearing in the classical case, and the Connes cocycle and relative modular operators take the place of Radon-Nikodyn derivatives \cite{JOPP}. As the quantum case is somewhat overshadowed by these technical requirements, it makes sense to consider a specific, exactly solvable model, such as the XY spin chain, and compute the quantities of interest in this specific case. Due to the exact solvability, the complete description of the entropic fluctuations of the XY quantum spin chain will require no more technical tools than the basics of trace-class scattering theory, a subject accessible to beginning graduate students or even advanced undergraduates. The XY spin chain has seen a great deal of study in quantum statistical mechanics - we refer the reader to \cite{JLP} for a sampling of the existing literature on the subject. For example, the first proofs of the existence of a non-equilibrium steady state were given in the context of XY chains \cite{AH, AP}.

As previously mentioned, the goal of this thesis is to test the emerging theory of entropic fluctuations in the context of XY quantum spin chains. In particular, we compute the large time limit entropic functionals of the XY chain.  These functionals describe the fluctuations of entropy production in the XY spin chain.  The results of this paper were already announced in \cite{JLP}, and some proof sketches appeared there. The model considered here is more general, and we complete the calculations that were omitted in \cite{JLP}. Our principle results concern the existence and the properties of the entropic functionals and are the content of Theorem \ref{thm:main}; we summarize the results in the sequel.

\subsection{Outline of paper}

This paper is organized as follows.  Section \ref{sec:confined} deals with the finite volume XY chain and studies some of its properties.   We initially define an XY chain confined to a finite interval in $\Z$. We then define the finite volume open XY chain which consists of three pieces, each of which is by itself an XY chain confined to a finite interval. We refer to these pieces as the left, center, and right parts of the chain. The parts are coupled at their end points; initially the left/center/right part of the chain is at thermal equilibrium at inverse temperature $\beta_{l/c/r}$. If there is a temperature differential, then energy and entropy will flow from one part of the chain to the other.  We study the properties of this flux and associated entropy production, and introduce the finite time and finite volume entropic functionals. 

The entropic functionals introduced are the following. The Evans-Searles functional is the direct quantization of the corresponding functional in classical non-equilibrium statistical mechanics. In the classical case, the functional describes fluctuations of the entropy production of the system with respect to its initial state.  However,  the direct quantization results in a functional that fails to have the desired $\alpha \doublearrow 1 - \alpha$ symmetry characterizing the aforementioned fluctuation relation of non-equilibrium statistical mechanics (see Section \ref{sec:essym}). 

At this point we introduce an entropic functional associated to the full counting statistics of the XY chain, which is a probability measure associated to a repeated quantum measurement protocol of the entropy of the XY chain (see Section \ref{sec:fcs} for more details). We introduce a functional which is the direct quantization of the variational characterization of the Evans-Searles functional in classical non-equilibrium statistical mechanics (see Proposition \ref{prop:finiteentropic}(vii)), and a class of functionals which interpolate between the variational functional and the full counting statistics functional. We mention that the full counting statistics, variational and interpolating functionals all have the $\alpha \doublearrow 1 - \alpha$ symmetry.

In the latter half of Section \ref{sec:confined} we introduce the Jordan-Wigner transformation. The Jordan-Wigner transformation maps the confined XY chain to a free Fermi gas, and associates to the Hamiltonian of the XY chain a Jacobi matrix.  In the remainder of the paper, we work exclusively in the Fermi gas representation of the XY chain. The Jordan-Wigner representation allows us to derive simple formulas for the entropic functionals which prove useful in the remainder of the paper.

In Section \ref{sec:therm} we take the thermodynamic limit in which the volume of the chain tends to infinity. In our set-up, the center part of the chain is fixed, while the right and left parts of the chain become semi-infinite chains (that is, they have one fixed end, to which the center part is coupled, and one end at infinity). We call the resulting object the extended XY chain. We show that the initial state and dynamics of the finite volume chains converge to a state and dynamics of the extended XY chain. We also show that the finite volume entropic functionals have infinite volume limits and compute them in closed form.

Section \ref{sec:time} concerns the large time limit, and contains our main results, Theorem \ref{thm:main}. In the previous section we fixed the time $t$ and took the thermodynamic limit, resulting in, for each fixed $t$, an extended XY chain. In this section, we will take the limit $t \to \infty$ of the extended XY chain. We begin by introducing the necessary facts from spectral and scattering theory that we will require to prove our resuts. In the case of the XY chain, the large time limit is easy to control via trace-class scattering theory. We prove that in the large time limit, the state of the XY chain converges to a non-equilibrium steady state (NESS) and we compute the steady state heat fluxes in terms of the scattering data of the underlying Jacobi matrix of the XY chain. Except in trivial cases, there is a non-zero steady state entropy production as long as the left and right parts of the chain are initially at different temperatures. At this point we introduce the Gallavotti-Cohen entropic functional which measures fluctuations with respect to the NESS.

Subsequently, we show that the entropic functionals have large time limits and compute them in closed form in terms of the scattering data of the Jacobi matrix associated to the XY chain. We prove that, except in trivial cases, all the entropic functionals are identical in the large time limit iff the Jacobi matrix associated to the XY chain is reflectionless (the definition of which is given in Section \ref{sec:refl}). If the Jacobi matrix is not reflectionless, then the full counting statistics, variational and interpolating functionals are all different, and are different from the Gallavotti-Cohen and Evans-Searles functionals. Regardless of whether or not the underlying Jacobi matrix is reflectionless, the Evans-Searles and Gallavotti-Cohen functionals are equal in the large time limit. In particular, the Evans-Searles functional recovers the $\alpha \doublearrow 1 - \alpha$ symmetry if and only if the Jacobi matrix is reflectionless.

\section{Finite volume XY chain} \label{sec:confined}
\subsection{The XY chain confined to an interval} 
A finite dimensional quantum system is a pair $( \H , \O_\H )$ where $\H$ is a finite dimensional complex Hilbert space and $\O_\H$ is the algebra of matrices acting on $\H$. $\O_\H$ is called the \emph{algebra of observables}. Throughout this paper the inner product on a Hilbert space $\H$ will be denoted $\langle \cdot , \cdot \rangle_\H$ and will be taken to be linear w.r.t. the second variable. When the meaning is clear we will drop the subscript and write instead $\langle \cdot , \cdot \rangle$. The self-adjoint elements of $\O_\H$ are those $A$ s.t. $A = A^*$ and the positive elements of $\O_\H$ are those $A$ s.t  $\langle \psi , A \psi \rangle \geq 0$ for every $\psi \in \H$. The \emph{spectrum} of $A \in \O_\H$ is denoted $\spec (A)$ and is the set of complex numbers $z$ s.t. $A - z$ is not invertible.

If $\lambda$ is an eigenvalue of $A$, we denote the projection onto the corresponding eigenspace by $\1_\lambda (A)$. We set $|A| = \sqrt{A^* A}$ and for $p \in ]0, \infty [$ set
\begin{align}
\norm{A}_p = ( \tr |A|^p )^{1/p}. %\notag
\end{align}
Furthermore, we define $\norm{A}_\infty$ to be the largest eigenvalue of $|A|$. It follows that $\lim_{ p \to \infty} \norm{A}_p = \norm{A}_\infty$.

We now consider an XY spin chain confined to a finite interval in $\Z$. To each $x \in \mathbb Z$ we associate the Hilbert space $\mathcal{H}_x = \mathbb C ^2 $. The algebra of matrices acting on $\H_x$ is denoted $\O_x$ and is equal to $M_2 (\C )$, that is, $2 \times 2$ matrices with complex entries. A basis for $\O_x$ given by the Pauli matrices 
\begin{align} %\label{eqn:pauli_basis}
\sigma_x^{(1)}=\left( \begin{matrix}0&1\\ 1&0\end{matrix}\right)\,,\quad
\sigma_x^{(2)}=\left(\begin{matrix}0&-\i\\  \i&0\end{matrix}\right)\,,\quad
\sigma_x^{(3)}=\left(\begin{matrix}1&0\\ 0&-1\end{matrix}\right)\,,
\end{align}
together with the identity matrix $\mathbb 1 _x$. The Pauli matrices satisfy the relations
\begin{align}
\sigma_x ^{j)} \sigma_x ^ {(k)}= \delta_{jk} \mathbb 1_x + i \varepsilon ^{jkl} \sigma_x ^{(l )} \label{eqn:pauli_rel},
\end{align}
with $\delta_{jk}$ the Kronecker delta and $\varepsilon^{jkl}$ the Levi-Civita symbol.

 Let $\Lambda = [N, M]$ be a finite interval in $\mathbb Z$. The \emph{XY chain confined to $\Lambda$} is the quantum system described by the Hilbert space and algebra of observables
\begin{align}
\mathcal{H}_{\Lambda} = \bigotimes_{x \in \Lambda } \mathcal{H}_{x} , \quad \mathcal{O}_{\Lambda} = \bigotimes_{x \in \Lambda} \mathcal{O}_x. %\notag
\end{align}
For simplicty of notation, we identify the operator $A_x \in \mathcal{O}_x$ at a single site with the operator $ ( \otimes_{y \in \Lambda \backslash \{x \} } \mathbb 1 _y ) \otimes A_x$ which is an element of  $\mathcal{O}_{\Lambda}$, the algebra of observables of the confined XY chain. Similarly, if $\Lambda' \subseteq \Lambda$, we identify an element $A \in \O_{\Lambda'}$ with the element $A \otimes\1_{\Lambda \backslash \Lambda'} \in \O_\Lambda$. The Hamiltonian of the XY chain confined to $\Lambda$ is given by
\begin{align}
H_{\Lambda} = \frac{1}{2} \sum_{x \in [N, M [ } J_x ( \sigma_x ^{(1)} \sigma_{x + 1} ^{(1)} +  \sigma_x ^{(2)} \sigma_{x + 1} ^{(2)} ) + \frac{1}{2} \sum_{x \in \Lambda} \lambda_x \sigma_x^{(3)}. \label{eqn:ham_def}
\end{align}
Here, $\{ J_x \}_{x \in \mathbb Z}$ and $\{ \lambda_x \} _{x \in \mathbb Z} $ are assumed to be bounded sequences of real numbers. $J_x$ is the nearest neighbour coupling at each site, and $\lambda_x$ is the strength of a magnetic field in the direction $(3)$ at the site $x$.

\subsection{Finite volume open XY chain} \label{sec:openxy}
In the previous section we described an XY chain confined to an interval in $\Z$. The open XY chain is an XY chain confined to an interval where the initial state of the system consists of three distinct pieces, each at thermal equilibrium at a different temperature. This will be made precise in this section.

 Let $\Lambda = [- M, M]$ be a finite interval in $\mathbb Z$, and consider an XY chain confined to $\Lambda$ as defined in Section \ref{sec:confined}. Let $N \in \Z_+$ (here, $\Z_+ = \N \cup \{ 0 \}$), with $N < M$. We denote $\Lambda_L = [-M, -N -1]$, $\Lambda_C = [-N, N]$ and $\Lambda_R = [N+1, M]$. $\Lambda_{L/C/R}$ is the left/center/right part of the chain. The Hamiltonian of the XY chain confined to $\Lambda$ can be written as
\begin{align}
H_{\Lambda} = H_L + H_C + H_R + V_L + V_R
\end{align}
where $H_{L/C/R}$ is the Hamiltonian of an XY chain confined to $\Lambda_{L/C/R}$, as defined by (\ref{eqn:ham_def}), and
\begin{align}
V_L = \frac{1}{2} J_{-N - 1} ( \sigma_{-N - 1} ^{(1)} \sigma_{-N} ^{(1)} + \sigma_{-N-1}^{(2)} \sigma_{-N} ^{(2)}), \quad V_R =  \frac{1}{2} J_{N} ( \sigma_{N} ^{(1)} \sigma_{N+1} ^{(1)} + \sigma_{N}^{(2)} \sigma_{N+1} ^{(2)} ),
\end{align}
is the coupling energy of the left and right chains to the center. We define $H_0 = H_L + H_C + H_R$ and $V = V_L + V_R$. Furthermore, until the end of the section, $\Lambda = [-M, M]$ is fixed and we will omit the subscript $\Lambda$ when the context is clear (i.e., we write $H = H_\Lambda, \mathcal{H} = \mathcal{H}_\Lambda$, etc.)

A \emph{density matrix} $\rho$ on the algebra of observables of a quantum system is a positive matrix with $\tr ( \rho) =1$. The \emph{state} associated to the density matrix $\rho$ is the linear functional on the algebra of observables given by $\rho (A) = \tr ( \rho A)$, for $A$ an observable. Since every positive linear functional $\nu$ acting on the algebra of observables with $\nu (\1) =1$ is given by $ \nu (A) = \tr (\nu A)$, with $\nu$ a density matrix, we will abuse the notation slightly and identity the state with the density matrix.

If $\A$ is an algebra of observables of a quantum system, then a $\star $-automorphism of $\A$ is a linear bijection $\vartheta : \A \to \A$ so that $\vartheta(AB) = \vartheta(A) \vartheta(B)$ and $\vartheta(A^*) = \vartheta(A)^*$. It follows that $\spec ( \vartheta (A) ) = \spec(A)$ and that $\norm{ \vartheta (A) } = \norm{A}$. The set of $\star$-automorphisms of $\A$ is denoted $\mathrm{Aut} (\A)$.

A \emph{dynamics} on the algebra of observables $\A$ of a quantum system is a continuous one-parameter subgroup of $\star$-automorphisms of $\A$. That is, it is a map $\R \ni t \to \tau^t \in \mathrm{Aut} (\A )$ so that $\tau^{ t + s} = \tau^t \circ \tau^s$ and $\lim_{t \to 0} \norm{ \tau^t (A) - A } = 0$ for every $A$ in $\A$. If $B$ is a self-adjoint element of $\A$, then $\tau^t (A) := e^{ \i t B} A e^{ - \i t B}$ is a dynamics on $\A$.

The Hamiltonian of the XY chain induces the dynamics on the algebra of observables, defined for $A \in \O$ as
\begin{align} \label{eqn:hamdyn}
\tau^t (A) = e^{\i tH} A e^{-\i tH}.
\end{align}
In the Heisenberg picture observables evolve forward in time and in the Schr\"odinger picture states evolve backwards in time and so we define,
\begin{align}
A_t = \tau ^t (A), \qquad \rho_t = \tau^{-t} ( \rho).
\end{align}
With this convention, $\rho_t (A) = \rho (A_t)$.

The initial state of the open XY chain is
\begin{align}
\omega = \frac{e^{- \beta_L H_L + -\beta_C H_C -\beta_R H_R}}{\tr (e^{- \beta_L H_L + -\beta_C H_C -\beta_R H_R})}.
\end{align}
With this initial state, the left, center and right systems are initially at thermal equilbrium at inverse temperatures $\beta _L, \beta_C$ and $\beta_R$, respectively. In fact, in the absence of the coupling between the chains, the state $\omega$ is a steady state. That is, if the dynamics in (\ref{eqn:hamdyn}) were induced by $H_0$ instead of $H$, then $\omega$ would be invariant under the dynamics as $[ H_0, \omega] = 0$ (here, $[A,B] = AB - BA$ denotes the commutator of observables $A$ and $B$). Note also that $\omega$ is a \emph{faithful} state. That is, $\ker \omega = \{ 0 \} $. We shall call the quantum dynamical system $(\O_\H, \tau, \omega)$ defined on the Hilbert space $\H$ the \emph{open XY chain confined to $\Lambda$}.

From now until the end of the paper we take,
\begin{align}
\beta_C = 0.
\end{align}
This convention is for notational simplicity and does not affect our main results.

We also note that the XY chain is \emph{time reversal invariant} (TRI). That is, there exists an anti-linear involutive $\star$-automorphism $\Theta$ of $\mathcal{O}$ s.t. $\tau^t \circ \Theta = \Theta \circ \tau^{-t}$ and $\Theta ( \omega ) = \Theta$. $\Theta$ is called a \emph{time reversal} and will be described in Section \ref{sec:jw}.

For future reference, we define the \emph{relative entropy} of a state $\rho$ with respect to a state $\nu$ to be
\begin{align}
S ( \rho \vert \nu ) = \begin{cases} \tr ( \rho ( \log \nu - \log \rho ) ) & \mbox{if ker } \nu \subseteq \mbox{ker } \rho , \\ -\infty & \mbox{otherwise.} \end{cases}
\end{align}
With the convention that $0 \times \infty = 0$, the relative entropy is well-defined and as a consequence of Klein's inequality (see Theorem 2.1 in \cite{JOPP}), $S (\rho \vert \nu ) \leq \tr ( \nu - \rho ) = 0$ with equality iff $\rho = \nu$.

\subsection{Observables of the open XY chain}
\subsubsection{Entropy production and heat fluxes}
In this section we define some of the basic observables of the open XY chain confined to an interval.  Much of the discussion here coincides with the abstract framework put forth in \cite{JOPP} in the case of general finite dimensional quantum systems, and our definitions are the same as those in \cite{JLP}. As the Hamiltonian $H_{L/R}$ is associated with the energy in the left/right part of the chain, the \emph{heat fluxes} out of the left and right parts of the chain at $t = 0$ are
\begin{align}
\Phi_{L/R} = -  \frac{\mathrm{d}}{\mathrm{dt}} \tau^t ( H_{L/R} ) \bigg\vert_{t = 0} = -\i [H, H_{L/R} ] = \i [H_{L/R}, V_{L/R} ].
\end{align}
It is easy to compute these fluxes using the relations (\ref{eqn:pauli_rel}):
\begin{align}
\Phi_{R} &= \frac{i}{4} J_N J_{N+1} \left( \sigma_N^{(1)} [\sigma_{N+1} ^{(2)} , \sigma_{N+1} ^{(1)} ] \sigma_{N+2}^{(2)} + \sigma_N^{(2)} [\sigma_{N+1} ^{(1)} , \sigma_{N+1} ^{(2)} ] \sigma_{N+2}^{(1)} \right) \nonumber \\
&+ \frac{i}{4} J_N \lambda_{N+1}  \left( \sigma_{N}^{(1)} [ \sigma_{N+1} ^{(3)} , \sigma_{N+1} ^{(1)} ] + \sigma_{N}^{(2)}[ \sigma_{N+1} ^{(3)} , \sigma_{N+1} ^{(2)} ]  \right) \nonumber \\
&= \frac{1}{2} J_N J_{N+1} \sigma_{N+1} ^{(3)} \left( \sigma_N ^{(1)} \sigma_{N+2} ^{(2)} - \sigma_N ^{(2)} \sigma_{N+2} ^{(1)}  \right) + \frac{1}{2} J_N \lambda_{N+1}  \left( \sigma_{N}^{(2)}\sigma_{N+1}^{(1)} -\sigma_{N}^{(1)} \sigma_{N+1}^{(2)} \right).
\end{align}
A similar computation shows that
\begin{align}
\Phi_L = \frac{1}{2} J_{-N-1} J_{-N-2} \sigma_{-N-1} ^{(3)} \left( \sigma_{-N-2} ^{(2)} \sigma_{-N} ^{(1)} - \sigma_{-N-2} ^{(1)} \sigma_{-N} ^{(2)}  \right) + \frac{1}{2} J_{-N-1} \lambda_{-N-1}  \left( \sigma_{-N}^{(2)}\sigma_{-N-1}^{(1)} -\sigma_{-N}^{(1)} \sigma_{-N-1}^{(2)} \right).
\end{align}
The \emph{entropy production observable} (as defined by thermodynamics) is
\begin{align}
\sigma =  -\beta_L \Phi_L - \beta_R \Phi_R = - \i [H, \log \omega ]
\end{align}
Note that the heat fluxes and the entropy production observable change sign under the time reversal $\Theta$:
\begin{align}
\Theta ( \Phi_{L/R} ) = - \Phi_{L/R} , \quad \Theta ( \sigma ) = - \sigma \label{eqn:tri_fluxes}
\end{align}
This is a consequence of the time reversal invariance of $\omega$. The \emph{mean entropy production rate} over the time interval $[0, t]$ is
\begin{align}
\Sigma ^t = \frac{1}{t} \int_0 ^t \sigma_s \mathrm{d}s.
\end{align}
If $ t < 0$, we define $\Sigma^t$ by the above equation as well.
We summarize a few of the basic properties of these observables in the proposition below. These results as well as further details can be found in \cite{JOPP} (see also \cite{JLP}).
\begin{proposition}\label{prop:basic} The folowing hold:
%% \begin{enumerate}[label=(\roman{*}), ref=(\roman{*}),font=\normalfont]
\ben
\item $\log \omega_t = \log \omega + t \tau^{-t} ( \Sigma^t ),$ 
\item $S(\omega_t \vert \omega ) = -t \omega ( \Sigma^t ),$ 
\item $\tau^t ( \Sigma^{-t} ) = \Sigma^t$, 
\item $\Sigma^t = - \tau^t ( \Theta (\Sigma ^t ) ),$  
\item The spectrum of $\Sigma^t$ is symmetric with respect to $0$. In fact, $\dim \mathbb1 _\phi ( \sigma^t ) = \dim \mathbb 1 _{-\phi} ( \sigma ^t )$ for all $\phi \in \spec (\Sigma ^t )$.
%%\end{enumerate}
\een
\end{proposition}
\proof
(i) We differentiate and integrate to obtain,
\begin{align}
\log \omega_t - \log \omega = \int_0 ^t \frac{\mathrm{d}}{\mathrm{ds}} \log \omega_s \mathrm{d}s = \int_0 ^t  \tau^{-s} (- \i [ H, \log \omega] ) =t \tau^{-t} ( \Sigma^t ).
\end{align}
(ii) Using (i), we have
\begin{align}
S(\omega_t \vert \omega ) = \omega_t ( \log \omega - \log \omega_t ) = - t \omega ( \tau^t ( \tau^{-t} (\Sigma ^t ) ) ) = - t \omega (\Sigma^t ).
\end{align}
(iii) Using (i), we have $\tau^t ( \Sigma ^{-t} ) = (\log \omega - \log \omega_{-t})/t$. On the other hand, (i) implies that $\Sigma^t = \tau^t (\log \omega_t - \log \omega ) / t = (\log \omega - \log \omega_{-t} )/t .$ \newline
(iv) By (\ref{eqn:tri_fluxes}), $\Theta ( \Sigma^t ) = - \Sigma^{-t}$. By (iii), $\Sigma^t = \tau^t (\Sigma^{-t} ) = \tau^t (\Theta ( \Theta (\Sigma^{-t} ) ) ) = - \tau^t (\Theta (\Sigma^t ) ).$ \newline
(v) By Exercise 3.1 in \cite{JOPP} (see also \cite{L}), $\Theta (A) = U_{\Theta} A U_{\Theta} ^{-1}$ for an anti-unitary $U_{\Theta}$. Therefore (iv) implies that for every eigenvector $v$ with eigenvalue $\phi$ of $\Sigma^t$, $e^{\i tH} U_{\Theta} v$ is an eigenvector of $\Sigma^t$ with eigenvalue $-\phi$, and the claim follows.
\qed

Proposition~\ref{prop:basic} has a few consequences. Firstly, part (i) allows one to interpret the entropy production observable as the \emph{quantum phase space contraction rate}. Part (ii) together with the fact that the relative entropy of two states is always nonpositive implies that the average entropy production over the interval $[0,t]$ is nonnegative. By this fact,
\begin{align}
0 \leq t \omega ( \Sigma ^t ) = - \beta _L \int_0 ^t \omega ( \Phi_{L, s}  ) \mathrm{d} s - \beta_R \int_0 ^t \omega ( \Phi_{R,s} ) \mathrm{d}s = - \beta_L \Delta_{L,t} -\beta_R \Delta_{R,t}
\end{align}
where $\Delta_{L/R, t} = \int_0 ^t  \omega ( \Phi_{L/R,s} ) \mathrm{d}s$ is the average heat flow out of the left/right part of the chain during the time interval $[0,t]$. This inequality is the finite time second law of thermodynamics; on average, heat flows from the hotter to the colder part of the chain. 

Additionally, parts (ii) and (v) of Proposition \ref{prop:basic} imply that
\begin{align}
\omega ( \Sigma^t ) = \sum_{\phi \in \spec{( \Sigma^t )} }  \phi p_{\phi} ^t =  \sum_{\substack{\phi \in \spec{( \Sigma^t )} \\ \phi > 0 } }   \phi ( p_{\phi} ^t  - p _{-\phi} ^t ) \geq 0,
\end{align}
where $p^t _\phi =\omega ( \mathbb 1 _\phi ( \Sigma ^t ) )$, the probability of measuring the mean entropy production rate to be $\phi$ over the time interval $[0, t]$.

\subsubsection{Evans-Searles Symmetry} \label{sec:essym}

As stated in Section \ref{sec:int}, the first numerical evidence for a fluctuation relation was obtained in \cite{ECM} in the context of deterministic \emph{classical} $N$ particle systems. There, they studied the non-equilibrium steady state of a fluid under an external sheer and conjectured that
\begin{align}
\frac{P_t (-A) }{P_t (A)} = e^{-t A} \label{eqn:esold}
\end{align}
where $P_t (A)$ is the probability of measuring $A$ for the average dissipation of power over the time interval $[0, t]$. Evans and Searles \cite{ES} were the first to prove that such a relation holds, and we call this relation of this sort the \emph{Evan-Searles fluctuation relation}. 

The direct quantization of the relation (\ref{eqn:esold}) to our current setting is
\begin{align}
 p_{- \phi}^t= e^{ - t \phi } p_\phi^t.
\end{align}
This equality implies that the Evans-Searles functional
\begin{align}
\ES_t (\alpha ) = \log \omega ( e^{- \alpha t \Sigma^t } ) = \log \sum_{\phi \in \spec (\Sigma^t )} e^{-\alpha t \phi } p_{\phi} ^t ,
\end{align}
has the symmetry
\begin{align}
\ES_t (\alpha ) = \ES_t (1 - \alpha) .\label{eqn:naive}
\end{align}
We remark that we sometimes use the notation $e_{\ES, t} (\alpha) = \ES_t (\alpha)$. It is easy to see that (\ref{eqn:naive}) fails in general unless the system is in a steady state. More precisely, we prove the following:
\begin{proposition} \label{prop:esfails}
The functional $\ES_t ( \alpha) $ has the symmetry (\ref{eqn:naive}) if $[H, \omega ]= 0.$ If $[H, \omega ] \neq 0$, then the fluctuation relation (\ref{eqn:naive}) can hold only if $t$ is in a countable set described below.
\end{proposition}
\proof  If $[H, \omega] = 0$ then $\ES_t (\alpha) \equiv 0$ and so (\ref{eqn:naive}) holds. Conversely, suppose the fluctuation relation holds for some $t$. Then $0 =\ES_t ( 0 ) = \ES_t (1) = \log \omega (e^{- t \Sigma^t } )$ and therefore $\omega (e^{- t \Sigma^t } )= 1$. Furthermore,  
\begin{align}
\omega (e^{- t \Sigma^t } ) = \tr ( e^{\log \omega } e^{ \log \omega_{-t} - \log \omega } ).
\end{align}
By the Golden-Thompson inequality (see Corollary 2.3 of \cite{JOPP}),
\begin{align}
\tr ( e^{\log \omega } e^{ \log \omega_{-t} - \log \omega } ) \geq \tr (e^{\log \omega_t } ) = 1,
\end{align}
and equality holds only if $\omega$ and $\omega_{-t}$ commute. Let $\omega = \sum_k \lambda_k \vert e_k \rangle \langle e_k \vert$, with $\{e_k \}_k$ orthonormal. Consider the functions
\begin{align}
f_{jk}(t) = \langle e_j , e^{ \i t H} w e^{ - \i t H} e_k \rangle .
\end{align}
As analytic functions, they either vanish identically or have an isolated, countable set of zeros. Furthermore, $\omega_{-t}$ and $\omega$ commute iff $f_{jk}(t) = 0$ for every $(j, k) \in \A := \{ (j, k) : \lambda_j \neq \lambda_k \}$. If there is at least one $(j, k) \in \A$ s.t. $f_{jk}(t)$ is not identically $0$, then the set
\begin{align}
\{ t: f_{jk} (t) = 0 \mbox{  } \forall (j, k) \in \A \}
\end{align}
is countable, and $t$ must be an element of it if the fluctuation relation holds. If all of these functions vanish identically then we have that $\omega_{-t}$ and $\omega$ must commute for every $t$. It follows from the formula
\begin{align}
\1_E (A) = \frac{1}{2 \pi \i} \oint_{ | z - E | = \eps} \frac{1}{z - A } \d z , 
\end{align}
which holds for self-adjoint $A$ and $\eps$ small enough that the spectral projections of $\omega$ and $\omega_{-t}$ commute. From the Lemma preceding Theorem XII.8 in \cite{RS4} the projection-valued functions
\begin{align}
Q_{jk} (t) = \1_{\lambda_j} ( \omega ) \1_{\lambda_k} ( \omega_{-t} ) = \1_{\lambda_j} ( \omega) e^{ \i t H} \1_{\lambda_k } (\omega) e^{ - \i t H}
\end{align}
have constant kernel (note that it is projection-valued due to the fact that $\1_{\lambda_j} ( \omega)$ and $\1_{\lambda_k} ( \omega_{-t} )$ commute). We conclude that $Q_{jk}(t) = Q_{jk} (0) = 0$ if $\lambda_j \neq \lambda_k$ and that
\begin{align}
\1_\lambda (\omega) &= \1_\lambda ( \omega) e^{\i t H} \left( \sum_{\lambda' \in \spec (\omega ) } \1_{\lambda'} ( \omega ) \right) e^{ - \i t H} = \1_\lambda ( \omega ) e^{ \i t H} \1_\lambda ( \omega ) e^{ - \i t H} \notag \\
& = \left( \sum_{ \lambda' \in \spec ( \omega ) } \1_{\lambda'} ( \omega ) \right) e^{ \i t H} \1_\lambda ( \omega ) e^{ - \i t H} = \1_{\lambda} ( \omega_{-t} ),
\end{align}
from which we conclude $\omega = \omega_{-t}$ for every $t$. Differentiating this at $t=0$ yields $[H, \omega] = 0$. 

Therefore, if $[H, \omega] \neq 0$, then the functions $f_{jk} (t)$ cannot vanish identically for every $(j, k) \in \A$, and so $t$ must be in the countable set described above if the fluctuation relation holds. \qed

\begin{remark} The proof of the above proposition also yields that $\ES_t (1 ) > 0$ unless either $[H , \omega] = 0$ or $t$ is in the countable subset described above.
\end{remark}
Proposition \ref{prop:esfails} tells us that the Evans-Searles functional $\ES_t (\alpha)$ will fail to have the Evans-Searles symmetry (\ref{eqn:naive}) except in trivial cases. In the sequel we will discuss a natural choice of entropic functionals which satisfy the Evans-Searles symmetry.

\subsubsection{Entropic functionals derived from full counting statistics} \label{sec:fcs}

In this section we introduce the concept of the full counting statistics associated to a repeated quantum measurement protocol of the entropy flow of our system. The full counting statistics will lead us to define an entropic functional that satisfies the Evans-Searles symmetry.

We define the \emph{entropy observable} as
\begin{align}
S = - \log \omega = \beta_L H_L + \beta_R H_R +Z,
\end{align}
with $Z = \tr (e^{- \beta_L H_L - \beta_R H_R} )$. Recall that we have set $\beta_C = 0.$ We have $S_t = - \log \omega_{-t}$, and  Proposition \ref{prop:basic} implies that $t \Sigma^t = S_t - S$.

According to the postulates of quantum mechanics, a measurement of the entropy of our system at $t=0$ will yield the eigenvalue $s$ of the observable $S$ with probability $\omega( \mathbb 1 _s (S))$. After this measurement, the system is in the reduced state
\begin{align}
\frac{\omega \mathbb 1_s (S) }{\omega  ( \mathbb 1_s (S) ) }.
\end{align}
At a later time $t$, the system has evolved and is in the state
\begin{align}
e^{-itH}  \frac{\omega \mathbb 1_s (S) }{\omega \mathbb 1_s (S) } e^{itH}.
\end{align}
A measurement of the entropy at time $t$ will yield the eigenvalue $s'$ of $S$ with probability
\begin{align}
\tr \left( e^{-\i t H}  \frac{\omega \mathbb 1_s (S) }{\omega \mathbb 1_s (S) } e^{\i tH} \mathbb 1_{s'} (S) \right) .
\end{align}
The joint probability of these two measurements is
\begin{align}
\tr ( e^{-itH} \omega \mathbb 1_s (S) e^{itH} \mathbb 1_{s'} (S) ).
\end{align}
The mean rate of entropy change is $ \phi = (s' - s) /t$, and we have derived its probability distribution, which is
\begin{align}
\mathbb P_t (\phi ) = \sum_{\substack{s, s' \\ s' - s = t \phi } } \tr ( e^{-\i tH} \omega \mathbb 1_s (S) e^{\i tH} \mathbb 1_{s'} (S) ).
\end{align}
The above discrete probability measure is called the \emph{full counting statistics} for the entropy change over the time interval $[0, t]$ operationally defined by the specified measurement protocol. The entropic functional associated to the full counting statistics is, for $\alpha \in \R$,
\begin{align}
\FCS_t ( \alpha ) = \log \sum_{\phi} e^{- \alpha t \phi } \mathbb P _t (\phi ).
\end{align}
 It is easy to compute,
\begin{align}
\FCS_t (\alpha ) &= \log \sum_{\phi} \sum_{\substack{s, s' \\ s' - s = t \phi } } e^{-\alpha (s' - s )}\tr ( e^{-\i tH} \omega \mathbb 1_s (S) e^{\i tH} \mathbb 1_{s'} (S) ) \nonumber \\
&= \log \tr\left( \sum_{s, s'} e^{-\alpha (s' - s )} e^{-\i tH} \omega \mathbb 1_s (S) e^{\i tH} \mathbb 1_{s'} (S) \right) = \log \tr \left( \omega_t ^{1 - \alpha } \omega ^\alpha \right).
\end{align}
With this identity we can verify that the Evans-Searles symmetry holds:
\begin{align}
\FCS_t (\alpha ) = \log \tr ( \omega_t ^{ 1 - \alpha}\omega^\alpha ) = \log \tr (\omega^{1 - \alpha} \omega_{-t} ^\alpha ) = \log \tr (\omega_{-t} ^\alpha \omega ^{ 1 - \alpha} ) = \log \tr ( \omega_t ^{\alpha} \omega ^{ 1- \alpha } ) = \FCS_t ( 1 - \alpha).
\end{align}
The second to last equality follows from an application of the time reversal. This identity implies that
\begin{align}
\FCS ( \alpha ) - \FCS (1 - \alpha) = \sum_{\phi} e^{ - \alpha t \phi}\left[ \mathbb P _t (\phi ) - e^{t \phi } \mathbb P_t ( - \phi ) \right] = 0. \label{eqn:degeneratesum}
\end{align}
Since this holds for any $\alpha \in \R$, we conclude that
\begin{align}
\mathbb P_t (- \phi ) = e^{-t \phi } \mathbb P_t (\phi). \label{eqn:esnew}
\end{align}
Compare this equality with (\ref{eqn:esold}). This equality says that the probability of measuring a decrease in entropy is exponentially suppressed compared to the probability of measuring an increase.

\subsubsection{Entropic pressure functionals}
The definition of $\FCS_t (\alpha)$ can be generalized. For $p > 0$ and $ \alpha \in \mathbb R$ we define the \emph{entropic pressure functionals} $e_{p, t} ( \alpha )$ by
\begin{align}
e_{p,t} ( \alpha ) = \begin{cases} \log \tr \left( \left[ \omega ^{\frac{1-\alpha}{p} } {\omega_{t}}^{\frac{2 \alpha}{p}  } \omega^{\frac{1-\alpha}{p}} \right]^{\frac{p}{2}} \right) & \mbox{if } 0 < p < \infty , \\ \log \tr \left( e^{ (1- \alpha) \log \omega + \alpha \log \omega_{t} } \right) & \mbox{if } p = \infty. \end{cases} 
\end{align}

In particular, $e_{2, t} (\alpha) = \FCS_t (\alpha)$. In the following proposition we summarize the basic properties of these entropic functionals.

\begin{proposition}\label{prop:finiteentropic} The entropic functionals satisfy
\begin{enumerate}[label=(\roman{*}), ref=(\roman{*}),font=\normalfont]
\item $e_{p,t} (0) = e_{p,t} (1) = 0$. 
\item The functions $\mathbb R \ni \alpha \to e_{p,t} (\alpha)$ are real-analytic and convex.
\item $ e_{p,t} (\alpha ) = e_{p,t} (1 - \alpha)$ and $e_{p, t} ( \alpha) = e_{p, -t } ( \alpha)$.
\item The function $]0, \infty ] \in p \to e_{p,t} (\alpha)$ is continuous and strictly decreasing for $\alpha \neq 0, 1$, and $\lim_{p \to \infty} e_{p,t} (\alpha) = e_{\infty, t} (\alpha)$.
\item $e ' _{p,t} (0) = - t \omega (\Sigma ^t ), e'_{p,t} (1) = t \omega ( \Sigma^t )$. In particular, these derivatives do not depend on $p$. 
\item $e_{2, t} (\alpha) = e_{\FCS , t} (\alpha)$ and
\begin{align}
e''_{2, t } (0) = \ES_t '' (0) = \int_0 ^t \int_0 ^t \omega((\sigma_s - \omega (\sigma_s)) (\sigma_u - \omega (\sigma_u ) ) ) \mathrm{d}s\mathrm{d}u . \label{eqn:finitederiv}
\end{align}
\item $e_{\infty ,t} (\alpha) = \max_\rho S (\rho, \omega ) - \alpha t \rho (\Sigma ^t ),$ where the $\mathrm{max}$ is taken over all states $\rho > 0$.
\end{enumerate}
\end{proposition}

\remarkk We remark that (vii) characterizes $e_{ \infty, t} ( \alpha)$ as the direct quantization of the variational characterization of the Evans-Searles functional in classical non-equilibrium statistical mechanics.

\vspace{3 pt}

\proof 
(i) These identities follow directly from the definition of $e_{p,t} (\alpha)$. \newline
(ii) Real analyticity follows from the strict positivity of $\omega$ and the analytic functional calculus (i.e., the function $f(x) = x^\alpha$ is real analytic on $(0, \infty)$ and so $f(A)$ is analytic for any strictly positive matrix). Corollary 2.4 of \cite{JOPP} implies convexity. \newline
(iii) The second equality is immediate from an application of the time reversal. Unitary invariance of trace norms and the identity $\norm{AB}_p = \norm{BA}_p$ yield
\begin{align}
\norm{\omega_t^{\frac{\alpha}{p}} \omega^{\frac{1 - \alpha}{p} }}_p &= \norm{e^{- \i tH} \omega^{\frac{\alpha}{p} }e^{\i tH} \omega^{\frac{1 - \alpha}{p} }}_p \nonumber \\
&= \norm{\omega^{\frac{\alpha}{p} }e^{\i tH} \omega^{\frac{1 - \alpha}{p} }e^{ - \i tH}}_p \nonumber \\
&= \norm{\omega^{\frac{\alpha}{p} } {\omega_{-t}}^{\frac{1 - \alpha}{p} }}_p = \norm{{\omega_{-t}}^{\frac{1- \alpha}{p} } \omega^{\frac{\alpha}{p} }}_p .
\end{align}
The first equality then follows from an application of the time reversal. \newline
(iv). Continuity is clear: for nonnegative matrices, $p \to A^{1/p}$ is a continuous matrix valued function, and the trace is a continuous function from matrices to $\mathbb \C$. Therefore $p \to e_{pt} $ is continuous as it is a composition of continuous maps. The remainder of the claim follows from Corollary 2.3 of \cite{JOPP} and the fact that we can write $e_{p,t} (\alpha) = \log \norm{ e^{\frac{2 \alpha}{p} S_t} e^{\frac{1 - \alpha}{p} S} }_p ^p$. \newline

We will prove (v) once we have introduced the Jordan-Wigner transformation (see the remarks following the proof of Lemma \ref{lem:entropyformulae}) illustrating the power of the Fermi gas representation. For (vi), direct computation yields
\begin{align}
e''_{2, t} (0) &= \omega \left(  \left( t\Sigma^t \right) ^2 \right) -  \left(\omega \left( t \Sigma^t \right) \right)^2 \nonumber \\
&= \int_0^t \int_0^t \omega ( \sigma_s \sigma_u ) - \omega ( \sigma_s ) \omega ( \sigma_u ) \d s \d u \notag \\
&= \int_0 ^t \int_0 ^t \omega((\sigma_s - \omega (\sigma_s)) (\sigma_u - \omega (\sigma_u ) ) ) \mathrm{d}s\mathrm{d}u .
\end{align}
We refer the reader to \cite{JOPP} for the proof of (vii). This property, like the others in this proposition are algebraic in nature and hold for general finite dimensional quantum systems.
\qed

We close this section with the remark that the variational characterization of $e_{\infty, t} (\alpha)$ given in (vii) is the quantization of the variational characterization of the finite time Evans-Searles functional in classical non-equilibrium statistical mechanics.

\subsection{The Jordan-Wigner transformation}\label{sec:jw}
The basic tool in the study of the XY chain is the Jordan-Wigner transformation which dates back to \cite{JW}. The Jordan-Wigner transformation maps the XY chain to a free Fermi gas, and associates to the Hamiltonian of the XY chain a Jacobi matrix. We assume the reader is familiar with the concept of second quantization, and recall here some definitions for notational purposes. For a pedagogical introduction we refer the reader to Chapter 6 of \cite{JOPP}. 

Let $\K$ be a finite dimensional Hilbert space. For $n \geq 1$, the $n$-fold antisymmetric tensor product is denoted $\Gamma_n ( \K)$ and is the subspace of the $n$-fold tensor product of $\K$ with itself, denoted $\K^{\otimes n}$  spanned by vectors of the form $\psi_1 \wedge ... \wedge \psi_n, \psi_i \in \K$ where
\begin{align}
\psi_1 \wedge ... \wedge \psi_n = \frac{1}{\sqrt{n!}} \sum_{\pi \in S_n} \sign ( \pi )  \psi_{\pi(1)} \otimes ... \otimes \psi_{\pi (n) },
\end{align}
and $S_n$ is the symmetric group on $n$ letters.  By convention, $\Gamma_0 (\K) = \C$. Note that $\Gamma_n (\K) = \{ 0 \}$ for $n > \dim \K$.  For $A \in \O_\K$ and $n \geq 1$, $\Gamma_n (A)$ and $\d \Gamma_n (A)$ are the elements of $\O_{\Gamma_n ( \K ) }$ defined by
\begin{gather}
\Gamma_n (A) ( \psi_1 \wedge ... \wedge \psi_n ) = A \psi_1 \wedge ... \wedge A \psi_n \notag \\
\d \Gamma_n (A) ( \psi_1 \wedge ... \wedge \psi_n ) = A \psi_1 \wedge ... \wedge \psi_n + ... + \psi_1 \wedge ... \wedge A \psi_n.
\end{gather}
For $n=0$ we set $\Gamma_0 (A)$ to be the identity map on $\Gamma_0 (\K)$ and set $\d \Gamma_0 (A) = 0$. The \emph{Fermionic Fock space over} $\K$ is defined by
\begin{align}
\Gamma (\K) = \bigoplus_{n=0} ^{\dim \K } \Gamma_n ( \K) .
\end{align}
We define for $A \in \O_\K$ the elements $\Gamma(A)$ and $\d \Gamma (A)$ of $\O_{\Gamma(\K) }$
\begin{align}
\Gamma(A) = \bigoplus_{n=0}^{\dim \K} \Gamma_n (A) , \qquad \d \Gamma (A) = \bigoplus_{n=0}^{\dim \K} \d \Gamma_n (A).
\end{align}
We have for $A, B \in \O_\K$ and $\lambda \in \C$,
\begin{gather}
\Gamma (A)^* = \Gamma (A^*), \qquad \d \Gamma (A^*) = \d \Gamma (A) ^* , \notag \\
\Gamma(AB) = \Gamma(A)\Gamma(B) , \qquad \d \Gamma (A + \lambda B) = \d \Gamma(A) + \lambda \d \Gamma (B) , \qquad \Gamma (e^A) = e^{ \d \Gamma (A) } .
\end{gather}
Additionally, one has
\begin{align}
[ \d \Gamma (A), \d \Gamma (B) ] = \d \Gamma \left( [A, B] \right),
\end{align}
and for $A$ invertible $\Gamma(A)^{1} = \Gamma(A^{-1})$ and
\begin{align}
\Gamma(A) \d \Gamma (B) \Gamma (A^{-1}) = \d \Gamma (A B A^{-1} ).
\end{align}
Let $\Omega = 1 \in \Gamma_0 (\K)$. The \emph{annihilation} and \emph{creation operators} for $\psi \in \K$ are elements of $\O_{\Gamma(\K)}$ and are denoted by $a (\psi)$ and $a^* (\psi)$, respectively, and are defined by
\begin{align}
a^* (\psi) \Omega = \psi, \qquad a^* ( \psi) ( \psi_1 \wedge ... \wedge \psi_n ) = \psi \wedge \psi_1 \wedge ... \wedge \psi_n
\end{align}
and
\begin{gather}
a ( \psi) \Omega = 0 , \qquad a ( \psi) \psi_1 = \langle \psi , \psi_1 \rangle_\K \Omega , \notag \\
a(\psi ) ( \psi_1 \wedge ... \wedge \psi_n ) = \sum_{j=1}^n (-1)^{1+j} \langle \psi , \psi_j \rangle_\K \psi_1 \wedge ... \wedge \cancel{\psi_j} \wedge ... \wedge \psi_n.
\end{gather}
The maps $\psi \to a^*(\psi)$ and $\psi \to a ( \psi)$ are respectively linear and antilinear, and $a(\psi)^*  = a^* ( \psi)$. They obey the \emph{canonical anticommutation relations} (CAR)
\begin{align}
\{ a(\psi), a (\phi) \} = \{ a^* (\psi) , a^* (\phi) \} = 0, \quad \{ a (\psi) , a^* (\phi) \} = \langle \psi, \phi\rangle_\K \1 _{\Gamma (\K )}. \label{eqn:car}
\end{align}
for any $\psi$ and $\phi$ in $\K$. Here $\{A, B\} = AB + BA$ denotes the anticommutator of two observables. We will use the notation $a^\#$ to refer to both $a$ and $a^*$ simultaneously.

Given finite dimensional Hilbert spaces $\K$ and $\RR$, a  \emph{representation of the CAR over $\K$ on $\RR$} is a pair of maps from $\K$ to $\O_\RR$ denoted by
\begin{align}
\psi \to b(\psi), \quad \psi \to b^* (\psi) ,
\end{align}
where the first is antilinear and the second is linear, $b(\psi)^* = b^* (\psi)$ and (\ref{eqn:car}) is satisfied with $a^\#$ replaced by $b^\#$. The representation is called \emph{irreducible} if the commutant of the set $\{ b^\# (\psi) , \psi \in \K \}$ is trivial, that is, if
\begin{align}
\{ B \in \O_{\RR } : [B, b^\# (\psi ) ] =0 \mbox{ } \forall \psi \in \K \} = \C \1_\RR .
\end{align}
For finite dimensional $\K$ and $\RR$ we have the following characterization of irreducible representations of the CAR over $\K$ on $\RR$ (see Exercise 6.2 of \cite{JOPP} and also \cite{L}).
\begin{proposition}
Let $\K$ be a finite dimensional Hilbert space and $\psi \to b^\# (\psi)$ be an irreducible representation of the CAR over $\K$ on the finite dimensional Hilbert space $\H$. Then, there exists a unitary operator $U: \Gamma (\K ) \to \RR$ such that $U a^\# (\psi ) U^* = b^\# (\psi)$ for all $\psi \in \K$. Moreover, $U$ is unique up to a phase. \label{prop:car}
\end{proposition}

Consider an XY chain confined to a finite interval $\Lambda \subseteq \Z$, as defined in Section \ref{sec:confined}. The \emph{Jordan-Wigner representation} is an irreducible representation of the CAR over $\fhL = \ell ^2( \Lambda )$ on $\H _\Lambda$. The resulting unitary operator $U_{\JW} : \H_\Lambda \to \Gamma(\fhL)$ guaranteed by Proposition \ref{prop:car} is called the \emph{Jordan-Wigner transformation}. In what follows we construct the Jordan-Wigner representation (see also \cite{JOPP}, although the model there is slightly less general).

Consider the spin raising and lowering operators
\begin{align}
\sigma ^{(+)} = \left( \begin{matrix}0&1\\ 0&0\end{matrix}\right)\,,\qquad \sigma ^{(-)} = \left( \begin{matrix}0&0\\ 1&0\end{matrix}\right)\,.
\end{align}
We have that $\sigma_x^{(\pm )} = ( \sigma_x^{(1)} \pm \i \sigma_x^{(2)} )/2$. Moreover, $\sigma_x^{(-)}$ and $\sigma_x ^{(+)} ={\sigma_x^{(-) *} } $ obey
\begin{align}
\{ \sigma_x ^{(+)}, \sigma_x^{(+)} \} = \{ \sigma_x ^{(-)}, \sigma_x^{(-)} \}  = 0  \quad \{\sigma_x ^{(+)} , \sigma_x ^{(-)} \} = \1 _{\H _\Lambda}. \label{eqn:pmcomm}
\end{align}
If $\Lambda = \{ x \}$ where just a single lattice site, then the maps $ \ell ^2 ( \{ x \} )  \ni \alpha \to \alpha \sigma_x^{(+)} \in \O_{\H_{ \{ x \} } } $ and  $ \ell ^2 ( \{ x \} )  \ni \alpha \to \widebar{\alpha} \sigma_x^{(-)} \in \O_{\H_{ \{ x \} } } $  would indeed define a representation of the CAR over $\ell ^2 ( \{ x \} )$. However if $\Lambda$ contains at least two distinct sites $ x \neq y$, then this does not directly generalize, as the raising and lowering operators at different sites will commute, and not anti-commute. For example, if $\Lambda =[0, 1]$, then the maps $\ell^2 (\Lambda ) \ni (\alpha_0, \alpha_1) \to \alpha_0 \sigma_0^{(+)} + \alpha_1 \sigma_1^{(+)} \in \O_\Lambda $ and $\ell^2 (\Lambda ) \ni (\alpha_0, \alpha_1) \to \widebar{\alpha}_0 \sigma_0^{(-)} + \widebar{\alpha}_1 \sigma_1^{(-)} \in \O_\Lambda$ do not define a representation of the CAR.

Let $\Lambda = [A, B]$. A computation using the fact that $(\sigma_x ^{(3)})^2 = \1 $ yields, for $T_x, S_y \in \O_\Lambda$,
\begin{align}
\{ \sigma_A^{(3)} ... \sigma_{x-1}^{(3)} T_x , \sigma_A^{(3)} ... \sigma_{y-1} ^{(3)} S_y \} = \begin{cases} \{\sigma_x^{(3)} , T_x \} \sigma_{x+1}^{(3)} ... \sigma_{y-1} ^{(3)} S_y  & x<y, \\ \{ T_x , S_y \} & x= y, \\ \{ \sz _y , S_y \} \sz _{y+1} ... \sz _{x-1} T_x  & x > y. \end{cases} \label{eqn:anticommcomp}
\end{align}
Define for $x \in \Lambda \backslash \{A \}$,
\begin{align}
b_x = \sz _A ... \sz _{x-1}  \sd _ x , \qquad b_x^*  = \sz_A ... \sz _{x-1} \su _x
\end{align}
and $b_A = \sd_A$, $b_A^* = \su_A$. We have
\begin{proposition} \label{prop:rep} The maps $ \ell^2 (\Lambda ) \ni \alpha \to b(\alpha) := \sum_x \widebar{\alpha}_x b_x \in \O_\Lambda$ and $\ell^2 (\Lambda ) \ni \alpha \to b^* (\alpha ) := \sum_x \alpha_x b_x \in \O_{\Lambda}$ form an irreducible represenation of the CAR over $\ell^2 (\Lambda )$ on $\H_\Lambda$.
\end{proposition}
\proof That $b^\#$ obey (\ref{eqn:car}) follows immediately from (\ref{eqn:pmcomm}) and (\ref{eqn:anticommcomp}). Clearly they are linear (resp., antilinear) and $b(\alpha)^* = b^* (\alpha)$. To see that the representation is irreducible, we require the following lemma, which is Proposition 6.4 in [JOPP].
\begin{lemma}
A representation $\psi \to b^\# (\psi)$ of the CAR over the finite dimensional Hilbert space $\K$ on $\RR$ is irreducible iff the smallest $\star$-subalgebra of $\O_\RR$ containing the set $\{ b^\# (\psi)  \vert \psi \in \K \}$ is $\O_\RR$.
\end{lemma}
Now, define
\begin{align}
V_x =\begin{cases} \1  & \mbox{if }x = A, \\ \prod_{y \in [A, x[ } ( 2 b_x ^* b_x - 1 ) & \mbox{else}. \end{cases}.
\end{align}
We claim that the following holds:
\begin{align}
\sx _x = V_x ( b_x + b_x ^* ) , \quad \sy _x = \i V_x  (b _x - b_x ^* ) , \quad \sz _x = 2 b_x b_x ^* - \1 \label{eqn:inversion}.
\end{align}
Since the Pauli matrices form a basis for $\C^2 $ and all polynomials in the operators $b^\#$ are in the smallest $\star$-subalgebra containing $\{ b^\#(\psi) \vert \psi \in \ell ^2 (\Lambda ) \}$, the relations (\ref{eqn:inversion}) show that this $\star$-subalgebra is indeed all of $\O_\Lambda$.

Finally, to prove (\ref{eqn:inversion}) note that $2 b^* _y b_y - \1 = 2 \su_y \sd _y - \1 = \sz _y$. The other two relations follow from this and the equations $( \sz_x )^2 = \1$ and $\sigma_x ^{(\pm)} = ( \sx _x \pm  \i \sy _x )/2$. This completes the proof of Proposition \ref{prop:rep}. \qed

As previously noted, this proves:
\begin{proposition} There exists a unitary operator $\UJW : \H_\Lambda \to \Gamma ( \fhL )$, called the Jordan-Wigner transformation satisfying
\begin{align}
\UJW \sx _x \UJW ^{-1} =  S_x (a_x + a_x^* ), \quad \UJW \sy _x \UJW^{-1} = \i S_x (a_x - a_x ^*) , \quad \UJW \sz _x \UJW ^{-1} = 2a_x ^* a_x - \1
\end{align}
where
\begin{align}
S_x =\begin{cases} \1  & x = A, \\ \prod_{y \in [A, x[ } ( 2 a_x ^* a_x - 1 ) & \mathrm{else}. \end{cases}
\end{align} \label{prop:jw}
\end{proposition}

Consider now the open XY chain confined to the interval $\Lambda = [-M, M]$ as defined in Section \ref{sec:openxy}, with left part $ \Lambda_l := [-M, -N-1]$, central part $\Lambda_c := [-N, N]$ and right part $ \Lambda_r := [N+1, M]$. The \emph{Jacobi matrix} $ h$ associated to our XY chain is a bounded operator on $\ell^2 (\Z)$ given by
\begin{align} \label{eqn:jacobidef}
(h u)(n) = J_n u(n+1) +J_{n-1} u(n-1)  + v_n u(n) ,
\end{align}
for $u \in \ell^2 (\Z)$. For any subset $\A$ of the integers we denote for $k \in \A$ the element $\delta_k \in \ell^2 (\A)$ which is $1$ at $k$ and $0$ elsewhere.

Recall that $\fh_\Lambda = \ell^2 (\Lambda) $ and we define $\fh _{l/c/r} = \ell^2 (\Lambda_{l/c/r})$. They are the single particle Hilbert spaces for the free Fermi gas on $\ell^2 ( \Lambda / \Lambda_{l/c/r})$. Let $h_\Lambda$ and $h_{l/c/r}$ be the restrictions of the Jacobi matrix $h$ to $\fh_\Lambda$ and $\fh_{l/c/r}$, respectively. The $h_\Lambda$ and $h_{l/c/r}$ are the single particle Hamiltonians. For simplicity of notation, we identify the operator $h_{l}$ acting on $\fh_{l}$ with the operator $h_{l} \oplus 0 \oplus 0$ acting on $\fh = \fh_l \oplus \fh_c \oplus \fh_r$, and we make similar identifications for $h_c$ and $h_r$. Note that
\begin{align}
h_\Lambda = h_0 + v ,
\end{align}
where $h_0 = h_l + h_c + h_r$ and $v = v_r + v_l$ with
\begin{gather}
v_l = J_{-N - 1} ( \vert \delta_{- N - 1} \rangle \langle \delta_{-N} \vert + \vert \delta_{- N} \rangle \langle \delta_{-N-1} \vert ), \\
v_r =  J_{N} ( \vert \delta_{N +1} \rangle \langle \delta_{N} \vert + \vert \delta_{N} \rangle \langle \delta_{N+1} \vert ).
\end{gather}

Up to an irrelevant additive constant, we have the following identities which are an immediate consequence of Proposition \ref{prop:jw}:
\begin{align}
\UJW H_{l/c/r} \UJW^{-1} = \d \Gamma (h_{l/c/r} ), \quad \UJW V_{l/r} \UJW ^{-1} = \d \Gamma (v_{l/r} ) , \quad \UJW H \UJW^{-1} = \d \Gamma (h).
\end{align}
For the fluxes we have
\begin{gather}
\UJW \Phi_{l} \UJW^{-1} = - \i J_{-N-1} J_{-N-2} ( a^* _{-N }a_{-N -2} - a^* _{-N-2 }a_{-N} )  -\i J_{-N-1} \lambda_{-N-1} ( a^* _{-N} a_{-N-1} - a_{-N-1}^* a_{-N} ) \label{eqn:fluxleft} \\
\UJW \Phi_{r} \UJW^{-1} = - \i J_{N} J_{N+1} ( a^* _{N }a_{N+2} - a^* _{N+2 }a_{N} )  -\i J_{N} \lambda_{N+1} ( a^* _{N} a_{N+1} - a_{N+1}^* a_{N} ) . \label{eqn:fluxright}
\end{gather}

In the fermionic representation, it is easy to see that the XY chain is time reversal invariant. If $j$ is the standard complex conjugation on $\ell^2 ( \Lambda)$, then conjugation with the anti-unitary operator
\begin{align}
\theta = U_{\JW}^{-1} \Gamma(j ) U_{\JW}
\end{align}
is a time reversal for the confined XY chain under which the initial state is invariant.

For the remainder of the paper we will work only in the Fermionic representation of the XY chain. We will slightly abuse notation and write $H_\Lambda=\d \Gamma (h_\Lambda)$, $H_{l/c/r} = \d \Gamma (h_{l/c/r} )$. Note that the proofs of Propositions \ref{prop:basic}, \ref{prop:esfails} and \ref{prop:finiteentropic} carry over in this representation without change.

\subsection{Basic Formulas}
In this section $\Lambda = [-M, M]$ is fixed and we will drop the respective subscript and write $h$ for $h_\Lambda$, $H$ for $H_\Lambda$, $\fh$ for $\fh_\Lambda$, etc. We would like to derive some basic formulas for the entropic functionals. First, let us record a few basic identities that will prove useful in this section and the sequel; for the proofs, we refer the reader to \cite{JOPP}. We have,
\begin{align}
\tr ( \Gamma (A) ) = \det ( 1 + A ), \label{eqn:trdet}
\end{align}
which holds for any linear $A : \fh \to \fh$.

Some of the basic algebraic properties of the annihilation and creation operators are summarized in (see Proposition 6.2 of \cite{JOPP}),
\begin{proposition}\label{prop:anncreat} The following holds:
\begin{enumerate}[label=(\roman{*}), ref=(\roman{*}),font=\normalfont]
\item $\norm{ a^\# (\psi )} = \norm{\psi}$ for any $\psi$,
\item For any $A \in \O_\fh$,
\begin{align}
\Gamma(A) a^* ( \psi) = a^* (A \psi) \Gamma (A) , \qquad \Gamma(A^*) a ( A \psi) = a ( \psi) \Gamma (A^*),
\end{align}
\item If $U$ is unitary,
\begin{align}
\Gamma(U) a^\# ( \psi) \Gamma ( U^*) = a ^\# ( U \psi),
\end{align}
\item For any $A \in \O_\fh$,
\begin{align}
[ \d \Gamma (A) , a^* ( \psi ) ] = a^* ( A \psi ) , \qquad [ \d \Gamma (A) , a (\psi) ] = - a ( A^* \psi),
\end{align}
\item $a^* (\psi) a ( \phi) = \d \Gamma ( \vert \psi \rangle \langle \phi \vert ).$
\end{enumerate}
\end{proposition}

 Define $k = -\beta_l h_l - \beta_r h_r $ and $k_t = e^{\i t h} k e^{-\i th}$. The initial state of the XY chain is mapped to the density matrix, which we also denote by $\omega$,
\begin{align}
\omega = \frac{ e^{- \beta_l H_l - \beta_r H_r}}{ \tr ( e^{- \beta_l H_l - \beta_r H_r } ) } = \frac{e^{ \d \Gamma (e ^k ) } }{\det ( 1 + e^k ) }. \label{eqn:finitestate}
\end{align}

The state $\omega$ is \emph{quasi-free} with density $T = ( \1+ e^{-k} )^{-1}$. That is, 
\begin{align}
\omega = \frac{1}{Z_T}\Gamma \left( \frac{T}{\1 - T} \right),
\end{align}
where
\begin{align}
Z_T = \tr \left( \Gamma \left( \frac{T}{\1 - T} \right) \right).
\end{align}
We have the following (see Proposition 6.6 in \cite{JOPP}):
\begin{proposition} \label{prop:quasifree} Since $\omega$ is quasi-free,
\begin{enumerate}[label=(\roman{*}), ref=(\roman{*}),font=\normalfont]
\item If $\phi_1, ... \phi_n , \psi_1 , ..., \psi_n \in \fh$, then,
\begin{align}
\omega ( a^* ( \phi_n ) ... a^* ( \phi_1) a (\psi_1 ) ... a ( \psi_m) ) = \delta_{nm} \det [ \langle \psi_i , T \phi_j \rangle ],
\end{align}
\item $\omega_T ( \Gamma(A)) = \det ( \1  + T( A - \1 ) ),$
\item $\omega_T ( \d \Gamma (A)) = \tr (T A ).$
\end{enumerate}
\end{proposition}

By the definition of the entropic functionals and the identity (\ref{eqn:trdet}) we have for $p < \infty$,
\begin{align}
e_{p, t} (\alpha ) = \log \frac{ \det ( 1 + ( e^{(1 - \alpha ) k/p} e^{2 \alpha k_{-t} / p } e^{ (1 - \alpha ) k/p } )^{p/2} ) } { \det (1 + e^k ) }.
\end{align}
We also have
\begin{align}
e_{\infty, t} (\alpha) = \log \frac{ \det ( 1 + e^{(1 - \alpha) k + \alpha k_{- t} } )}{\det ( 1 + e^k ) } ,
\end{align}
and, 
\begin{align}
\ES _t ( \alpha ) = \log \frac{ \det ( 1 + e^{k/2} e^{ \alpha ( k_t - k ) } e^{k/2} ) }{\det ( 1 + e^k ) }.
\end{align}
Define for $p \in ]0, \infty [$,
\begin{align}
\K _{p, t} (\alpha, u ) = \frac{1}{2} \left( e^{ -(1 - \alpha ) k_{tu} / p} \left[ \1 + \left( e^{(1 - \alpha) k_{tu} /p} e^{2 \alpha k_{- t (1 - u ) } / p} e^{ ( 1 - \alpha ) k_{tu} / p}  \right)^{-p/2} \right]^{-1} \right) e^{(1 - \alpha ) k_{tu}  / p } + \mathrm{h.c.}, \label{eqn:kp}
\end{align}
where $\mathrm{h.c.}$ stands for the hermitian conjugate of the first term. Define
\begin{align}
\K _{\infty, t } ( \alpha, u) = \left( \1 + e^{ - ( 1 - \alpha ) k_{ty} - \alpha k_{-t (1 - u ) } } \right)^{-1} , \label{eqn:kinf}
\end{align}
and
\begin{align}
\K_{\ES , t } (\alpha, u ) = - \left( \1 + e^{ - \alpha ( k_{t ( 1 - u ) } - k_{-tu} ) } e^{ - k_{-tu} } \right) ^{-1} .\label{eqn:kes}
\end{align}
We have,
\begin{lemma} \label{lem:entropyformulae}
For $p \in ]0, \infty]$,
\begin{align}
e_{p, t} (\alpha ) = t \int_0^\alpha \d \gamma \int_0 ^1 \d u \tr \left( \K_{p,t} (\gamma, u ) \i [ k, h ] \right), \label{eqn:eptform}
\end{align}
and
\begin{align}
\ES _t (\alpha) = t \int_0 ^\alpha \d \gamma \int_0^1 \d u \tr \left( \K _{\ES , t }  ( \gamma, u ) \i [ k, h ] \right). \label{eqn:esform}
\end{align}
\end{lemma}

We will need the following result of \cite{HP}, of which we provide a proof:
\begin{lemma}
Let $f(z)$ be an analytic function in a disc of radius $r$ centered at some $\lambda \in \R$. Let $F (\alpha)$ be a Hermitian matrix valued function in $\alpha \in ( -\delta, \delta)$ for some $\delta >0$. If $F(\alpha)$ is differentiable at $\alpha = 0$ and the eigenvalues of $F(0)$ are contained in $(\lambda - r, \lambda + r)$, then
\begin{align}
\frac{\d}{\d \alpha} \tr f ( F ( \alpha ) ) \bigg \vert_{\alpha = 0} = \tr \left( f ' ( F(0) ) F' (0) \right).
\end{align}
\end{lemma}
\proof By continuity of $F(\alpha)$ at $\alpha = 0$ and min-max principles, the eigenvalues of $F(\alpha)$ are all contained in $( \lambda -r, \lambda +r)$ for $| \alpha | < \delta_0 $ where $ \delta_0 > 0$ is a constant possibly smaller then $\delta$. By the analyticity of $f(z)$, the function $\alpha \to f ( F(\alpha ) )$ is defined by a norm convergent power series $f ( F (\alpha ) ) = \sum_{n = 0 } ^\infty c_n ( F (\alpha) -\lambda )^n$ for $ | \alpha | < \delta_0$. By this fact and the continuity of the trace,
\begin{align}
&\left|  \frac{\tr f ( F (\alpha ) ) - \tr f ( F (0 ) ) }{ \alpha} - \tr \left( f ' ( F(0) ) F' (0) \right) \right| \nonumber \\
&  \leq \sum_{n = 0} ^\infty | c_n |  \left| \tr \left( \frac{ (F(\alpha ) - \lambda )^n- (F(0) +\lambda )^n }{\alpha} - n (F(0)-\lambda)^{n-1} F' (0) \right)\right| \nonumber \\
&=  \sum_{n = 0} ^\infty | c_n |  \left| \tr \left( \frac{ F(\alpha) - F(0)}{\alpha} P_n (\alpha ) - n (F (0)-\lambda)^{n-1} F' (0) \right) \right|  \nonumber \\
&\leq \sum_{n=0} ^\infty |c_n|  \left| \tr P_n (\alpha)\left( \frac{ F(\alpha) - F(0)}{\alpha} - F' (0) \right) \right|  + \sum_{n=0} ^\infty |c_n |  \left| \tr F'(0) \left( P_n (\alpha) - P_n (0) \right) \right|  \label{eqn:powerseries},
\end{align}
with $P_n (\alpha) = (F(\alpha) - \lambda )^{n-1} + (F(\alpha) - \lambda )^{n-2} ( F(0) -\lambda ) + ... + (F(0) - \lambda )^{n-1}$ for $n \geq 1$ and $P_0 (\alpha ) = \1$ (here we have used cyclicity of the trace to obtain the equality in the above computation). Continuity of $F$ gives us the estimate that, by taking $\delta_0$ smaller if necessary, there is an $ \alpha_0 \in ( -\delta_0, \delta_0)$ so that $\norm{P_n (\alpha ) } \leq n\norm{F( \alpha_0) -\lambda }^{n-1}$ holds for every $| \alpha | < \delta_0/2$. By differentiation, we conclude that for any $\eps > 0$ for we have $\norm{P_n (\alpha) - P_n (0) } \leq \varepsilon n(n-1) \norm{ F(\alpha_0) - \lambda }^{n-2}$ for every $| \alpha| < \delta_0/2$ (again taking $\delta_0 $ smaller if necessary).

Since $\sum_n n(n-1) |c_n | \norm{F_n (\alpha_0 ) - \lambda}^{n-2}$ converges, these estimates together with the differentiability of $F$ at $\alpha = 0$ show that the last line of (\ref{eqn:powerseries}) goes to $0$ as $\alpha \to 0$. \qed

This lemma implies,
\begin{corollary}\label{cor:derivative}
Let $F(\alpha)$ be a differentiable function for $\alpha \in \R$ taking values in Hermitian strictly positive matrices on $\fh$. Then for any $p \in ]0, \infty[$,
\begin{align}
\frac{\d } {\d \alpha } \tr \log (\1 + F(\alpha ) ^p )= p \tr \left( \left( \1 + F(\alpha)^{-p}\right)^{-1} F(\alpha)^{-1} F' (\alpha ) \right). \label{eqn:trder1}
\end{align}

Let $F(\alpha)$ be a differentiable function for $\alpha \in \R$ taking values in Hermitian matrices on $\fh$. Then,
\begin{align}
\frac{\d}{\d \alpha} \tr \log \left( \1 + e^{F(\alpha)} \right) = \tr \left( ( \1 + e^{-F(\alpha) } )^{-1} F' (\alpha) \right). \label{eqn:trder2}
\end{align}
\end{corollary}

We may now proceed with the computation of the basic formulas for the entropic functionals.

\noindent \emph{Proof of Lemma \ref{lem:entropyformulae}. } We will complete the computation for $ 0 < p < \infty$. Recall that for matrices $\log \det A = \tr \log A$. By this fact,
\begin{align}
e_{p , t } (\alpha ) &= \tr \log \left( \1 + \left( e^{(1 - \alpha ) k/p } e^{2 \alpha k_{-t} / p } e^{(1 - \alpha ) k / p } \right)^{p/2} \right) - \tr \log (1 + e^k )  \nonumber \\
&= \int_0 ^ \alpha\d \gamma \frac{ \d } {\d \gamma } \tr \log  \left( \1 + \left( e^{(1 - \gamma ) k/p } e^{2 \gamma k_{-t} / p } e^{(1 - \gamma) k / p } \right)^{p/2} \right) . \label{eqn:form3}
\end{align}
Let $F(\gamma) = e^{ (1 - \gamma) k / p } e^{2 \gamma k_{-t} / p } e^{(1 - \gamma) k/p }$. By (\ref{eqn:trder1}), the derivative in the above integral equals
\begin{align}
&\tr \left(    (1 + F(\gamma)^{-p/2} )^{-1} F(\gamma) ^{-1} \left( e^{ (1 - \gamma ) k/p }( k_{-t} - k ) e^{ 2 \gamma k_{-t} / p } e^{(1 -\gamma ) k / p } + e^{(1 - \gamma )k/p} e^{2 \gamma k_{-t} /p } (k_{-t} - k ) e^{ (1 - \gamma ) k / p }   \right) \right) \nonumber \\
&=  \tr \left(    (1 + F(\gamma)^{-p/2} )^{-1}\left( e^{ (1 - \gamma ) k/p }( k_{-t} - k ) e^{-(1 -\gamma ) k / p } + e^{-(1 - \gamma )k/p}  (k_{-t} - k ) e^{ (1 - \gamma ) k / p }   \right) \right) , \label{eqn:form1}
\end{align}
where the last equality follows from cyclicity of the trace. We compute next,
\begin{align}
k_{-t} - k = \int_0 ^t \d s \frac{ \d }{\d s } k_{-s} = \i \int_0 ^t \d s e^{- \i s h } [k, h] e^{ i s h} = \i t \int_0 ^1 \d s e^{ -\i s t h } [k, h] e^{ \i s t h} . \label{eqn:form2}
\end{align}
Inserting (\ref{eqn:form1}) and (\ref{eqn:form2}) into (\ref{eqn:form3}) and some simple algebra yields the claimed formula.

The computation for $p = \infty $ follows the same strategy. Taking $F( \alpha) = (1 - \alpha)k +\alpha k_{-t}$ in (\ref{eqn:trder2}) yields,
\begin{align}
e_{\infty , t} ( \alpha ) &= \int_0 ^ \alpha \d \gamma \tr \left( ( \1 + e^{ -( 1 - \gamma ) k - \gamma k_{-t} } )^{-1} ( k_{-t} - k ) \right) \nonumber \\
%&= \int_0 ^ \alpha \d \gamma \tr \left( ( \1 + e^{ -( 1 - \gamma ) k - \gamma k_{-t} } )^{-1} ( k_{-t} - k ) \right) \nonumber \\
&=t  \int_0 ^ \alpha \d \gamma \int_0 ^1 \d u  \tr \left( ( \1 + e^{ -( 1 - \gamma ) k - \gamma k_{-t} } )^{-1} e^{ - \i u t h} \i [k, h ] e^{ \i u t h}\right) \nonumber \\
&= t \int_0 ^ \alpha \d \gamma \int_0 ^1 \d u  \tr \left( ( \1 + e^{ -( 1 - \gamma ) k_{tu} - \gamma k_{-t(1 - u)} } )^{-1}  \i [k, h ]\right)
\end{align}
which is the formula in question. The computation for $\ES_t (\alpha)$ is similar and can be found in \cite{JLP}. \qed

Evaluating the derivatives computed in the proof of Lemma \ref{lem:entropyformulae} at $\alpha = 0$ and using part (iii) of Proposition \ref{prop:quasifree} to evaluate $ S ( \omega_t \vert \omega)$ in terms of $k$ and $k_{-t}$ yields (v) of Proposition \ref{prop:finiteentropic}.

\section{The extended XY chain: thermodynamic limit} \label{sec:therm}

Recall that our open XY chain restricted to $\Lambda = [-M, M]$ consists of a left part $\Lambda_l = [- M , - N -1 ]$, a center part $\Lambda_c = [-N, N]$ and a right part $\Lambda_r = [N+1, M]$. We now consider the limit $ [ -M , M ] \to \Z$ in which we keep the central system $\Lambda_c = [-N, N]$ fixed and take $M \to \infty$. Recall that we are working exclusively in the Fermionic representation of the XY chain. To denote the dependence of the various objects under consideration on the size of $\Lambda$ we use the subscript $M$ and write, for example, $\O_M , \fh _M, h_M$, etc., whereas before we wrote $\O_\Lambda, \fh_\Lambda, h_\Lambda$, etc. 

The algebra of observables of the extended XY chain is denoted by $\O$ and is the norm closure of the local observables
\begin{align}
\O_{\loc} = \bigcup _{M} \O_M ,
\end{align}
where we have identified $\O_{M_1}$ with the appropriate subalgebra of $\O _{M_2}$ for $M_1 < M_2$. Recall here that $\O_M = \O_{\Gamma ( \fh_M )}$ and that
\begin{align}
\Gamma( \fh_{M_2} ) = \Gamma\left( \ell^2(  \Lambda_{M_2}  \backslash \Lambda_{M_1} ) \oplus \ell^2 ( \Lambda_{M_1} ) \right) = \Gamma \left( \ell^2 ( \Lambda_{M_2} \backslash\Lambda_{M_1} ) \right) \otimes \Gamma \left( \ell^2 ( \Lambda_{M_1}) \right) .
\end{align}
For the dynamics of the XY chain we have,
\begin{proposition} \label{prop:dynamics}
For any $A \in \O_{\loc}$ the limit
\begin{align}
\tau ^t (A) = \lim_{M \to \infty} \tau^t_M (A) =  \lim_{M\to \infty}e^{ \i t H_M} A e^{-\i t H_M}  \label{eqn:dyn}
\end{align}
exists in norm and the convergence is uniform for $t$'s in bounded sets. Furthermore, $\tau^t$ uniquely extends to a strongly continuous group of $\star$-automorphisms of $\O$. This group is called the dynamics of the extended XY chain.
\end{proposition}
In the above proposition, strong continuity means that the function $t \to \tau^t (A)$ is norm continuous for every $A \in \O$. 

For the inital state of the XY chain, we have,
\begin{proposition}\label{prop:state}
Let $\omega_M$ be the initial state in (\ref{eqn:finitestate}). Then for $A \in \O_{\loc}$ the limit
\begin{align}
\omega (A) = \lim_{M \to \infty} \omega_M (A) \label{eqn:statelimit}
\end{align}
exists and $\omega$ uniquely extends to a state on $\O$. This state is called the initial state of the extended XY chain.
\end{proposition}

The \emph{extended XY chain} is described by the $C^*$ dynamical system $ (\O, \tau^t, \omega )$. This system is time reversal invariant, as the time reversal defined in Section \ref{sec:jw} which is initially defined for every $A \in \O_{\loc}$ extends to all of $\O$ by the bounded linear transformation (BLT) theorem \cite{RS1}. We denote this time reversal by $\Theta$.

\vspace{2 pt}

\noindent\emph{Proof of Proposition \ref{prop:dynamics}.} We first prove that the norm convergence of the limit is uniform for bounded $t$. If $A \in \O _{\loc}$ then $A$ is a finite sum of polynomials in the annihilation and creation operators $\{ a^\# (\psi) \vert \psi \in \fh_{M'} \}$ for some $M'$. It therefore suffices to prove the uniform convergence when $A = a^* ( \phi_n ) ...a^* ( \phi_1 ) a ( \psi_1 ) ... a ( \psi_m)$. For $M > M'$, we have by Proposition \ref{prop:anncreat}(iii),
\begin{align}
e^{\i t \d \Gamma (h_M ) }   a^* ( \phi_n ) ...a^* ( \phi_1 ) a ( \psi_1 ) ... a ( \psi_m) e^{ - \i t \d \Gamma (h_M ) } =   a^* ( e^{ \i t h_M} \phi_n ) ...a^* ( e^{ \i t h_M}  \phi_1 ) a ( e^{ \i t h_M}  \psi_1 ) ... a ( e^{ \i t h_M}  \psi_m).
\end{align}
Since $\norm{ a^\# ( e^{\i t h_M} \psi )} = \norm{\psi}$ for any $\psi$, it suffices to prove that, for any $\psi \in \fh_{M'}$, the norms of the sequence $e^{ it h_M } \psi$ are uniformly Cauchy for bounded $t$. However this follows directly from the fact that $h_M \to h$ strongly in $\ell^2 (\Z)$ (recall $h$ is the full Jacobi matrix defined in (\ref{eqn:jacobidef})) and the estimate,
\begin{align}
\norm{ e^{ \i t h_M } \psi - e^{ \i t h} \psi } \leq \sum_{k=0}^n \frac{1}{k!} | t|^k \norm{ (h_M ^k - h^k) \psi } + \norm{\psi} \sum_{k=n+1}^\infty \frac{1}{k!} \norm{h}^k + \norm{h_M}^k. \label{eqn:wknd2}
\end{align}
Since the $\norm{h_M}$'s are uniformly bounded, the second sum can be made smaller than any $\eps >0$ for $n$ large enough, uniformly for bounded $t$, and then since $h_M^k \to h^k$ strongly for every $k$, the first sum is also under control.

Since $\norm{e^{\i t H_M} A e^{ - \i t H_M}} \leq \norm{A}$ holds for every $M$ and $A \in \O_{\loc}$, $\tau^t$ extends uniquely to all of $\O$ by the BLT theorem \cite{RS1}. Since $\tau^t$ satisfies the group property and the $\star$-automorphism properties on $\O_{\loc}$, it satisfies the same properties on $\O$. Strong continuity on $\O_{\loc}$ follows from an $\eps /3$ argument using the fact that the convergence is uniform for $t$'s in bounded sets. Strong continuity on all of $\O$ then follows immediately from a second $\eps /3$ argument. \qed

\vspace{2 pt}

\noindent\emph{Proof of Proposition \ref{prop:state}.} Let $A \in \O_{\loc}$ be given. As in the proof of Proposition \ref{prop:dynamics}, it suffices to consider $A = a^* ( \phi_n ) ...a^* ( \phi_1 ) a ( \psi_1 ) ... a ( \psi_m)$, with the $\phi_i$'s and $\psi_j$'s in $\fh_{M'}$ for some $M'$. By Proposition \ref{prop:quasifree}(i) we have for $M > M'$,
\begin{align}
\omega_T ( a^* (\phi_n ) ... a^* (\phi_ 1 ) a ( \psi_1 ) ... a ( \psi_m ) ) = \delta_{nm} \det [ \langle \psi_i, T_M \phi_j \rangle ] ,\label{eqn:quasifree}
\end{align}
with $T_M = ( \1 + e^{ - \beta_l h_{l, M} - \beta_r h_{r,M}} ) ^{-1}$. The existence of the limit (\ref{eqn:statelimit}) is then immediate from the strong convergence of $h_{l/r, M}$ to $h_{l/r}$. The estimate $ | \omega_M ( A ) | \leq \norm{A}$ which holds for every $M$ implies that $\omega$ extends uniquely to all of $\O$.
\qed 

The observables $\Phi_{l/r}$ and $\sigma$ are in $\O_{\loc}$. Let
\begin{align}
\Sigma^t =\frac{1}{t} \int_0 ^t \sigma_s \d s ,
\end{align}
and
\begin{align}
\ES _t (\alpha ) = \log \omega \left( e^{- \alpha t \Sigma^t } \right) .
\end{align}
We prove the following two basic propositions about the observables and entropic functionals in the thermodynamic limit.
\begin{proposition}We have,
\begin{enumerate}[label=(\roman{*}), ref=(\roman{*}),font=\normalfont]
\item \begin{align}
\lim_{M \to \infty} \Sigma^t_M = \Sigma^t
\end{align}
in norm and the convergence is uniform for t's in bounded sets.
\item $\lim_{M \to \infty}  S ( \omega_{t, M} \vert \omega_M ) = -t \omega \left( \Sigma^t \right)$
\item $\tau^t \left( \Sigma^{-t} \right) = \Sigma^t$
\item $\Sigma^t = - \tau ^t \left( \Theta ( \Sigma^t ) \right) $. In particular, $\mathrm{sp}{ \left( \Sigma^t \right) }$ is symmetric w.r.t. to the origin.
\item \begin{align}
\lim_{M \to \infty} \ES_{t, M} (\alpha ) = \ES _{t} (\alpha) \label{eqn:eslim}
\end{align}
and the convergence is uniform for $t$'s and $\alpha$'s in bounded sets.
\end{enumerate} \label{prop:thermobasic}
\end{proposition}
\proof  (i) The function $\tau_M^s ( \sigma)$ converges to $\tau^s (\sigma)$ uniformly for $s \in [0, t]$ due to Proposition \ref{prop:dynamics} and the claim follows. \newline
(ii) The definition of $\omega$ and (i) imply that $- t \omega (\Sigma^t ) = \lim_{N \to \infty} \lim_{M \to \infty} -t \omega_M (\Sigma ^t _N ) =  \lim_{M \to \infty}  -t \omega_M ( \Sigma^t _M)$. The second equality follows from the uniform norm continuity of the state $\omega$. By Proposition  \ref{prop:basic}, $-t \omega_M ( \Sigma^t _M ) = S ( \omega_{t, M} \vert \omega_M )$ and the claim follows. \newline
(iii) From Proposition \ref{prop:basic} and (i) we have $\Sigma^t = \lim_{M \to \infty} \Sigma^t_M = \lim_{M \to \infty} \tau^t _M  \left( \Sigma^{-t}_M \right)$. On the other hand, $\tau^t \left( \Sigma^{-t} \right) = \lim_{N \to \infty} \lim_{M \to \infty} \tau^t_M \left( \Sigma^{-t}_N \right)$ and therefore (iii) follows from the inequality $\norm{ \tau^t_M ( \Sigma^{-t}_N - \Sigma^{-t}_M )} \leq \norm{ \Sigma^{-t}_N - \Sigma^{-t}_M}$, which implies
\begin{align}
\lim_{N \to \infty} \lim_{M \to \infty} \tau^t_M ( \Sigma^{-t} _N ) = \lim_{M \to \infty} \tau^t_M ( \Sigma^{-t}_M ).
\end{align}
 \newline
(iv)  We have,
\begin{align} - \tau^t ( \Theta ( \Sigma^t ) ) = \lim_{M \to \infty} - \tau^t ( \Theta ( \Sigma^t_M ) ) = \lim_{M \to \infty} - \tau^t ( \Theta_M ( \Sigma^t_M ) ) = \lim_{M \to \infty} \lim_{N \to \infty} -\tau_N^t (\Theta_M (\Sigma^t_M ) ).
\end{align}
The RHS equals the limit $\lim_{M \to \infty} -\tau_M ^t ( \Theta_M (\Sigma^t_M ) )$ which equals $\lim_{M \to \infty} \Sigma_M^t$ by Proposition \ref{prop:basic}, which proves (iv). \newline
(v) Note that the RHS of (\ref{eqn:eslim}) equals $ \lim_{N \to \infty} \lim_{M \to \infty} \log  \omega_M ( e^{ - \alpha t \Sigma^t_N } )$ by definition. By the inequality 
\begin{align}
| \omega_M ( e^{ - \alpha t \Sigma^t_M} - e^{ - \alpha t \Sigma_N^t }) | \leq \norm{ e^{ - \alpha t \Sigma^t_M} -e^{ - \alpha t  \Sigma^t_N} },
\end{align} where the RHS goes to $0$ as $N, M \to \infty$,
\begin{align}
\lim_{N \to \infty} \lim_{M \to \infty} \log \omega_M ( e^{ -\alpha t \Sigma_N^t } ) = \lim_{M \to \infty} \log \omega_M (e^{ - \alpha t \Sigma^t_M} ). \label{eqn:wknd1}
\end{align}
This limit is the LHS of (\ref{eqn:eslim}). The desired uniform convergence follows easily by expanding the exponential on the RHS of (\ref{eqn:wknd1}) and the uniform convergence of the $\Sigma^t_M$ (recall the argument using the estimate (\ref{eqn:wknd2}) in the proof of Proposition \ref{prop:dynamics}).
\qed

\begin{proposition}For the entropic functionals, we have \label{prop:thermfunct}  
\begin{enumerate}[label=(\roman{*}), ref=(\roman{*}),font=\normalfont]
\item For all $\alpha \in \R$ and $p \in ]0, \infty ]$ the limits
\begin{align}
e_{p , t }(\alpha ) = \lim_{M \to \infty} e_{p, t, M} (\alpha) ,
\end{align}
exist and are finite.
\item $e_{p , t } (0) = e_{p, t} (1) = 0$.
\item The functions  $\alpha \to e_{p, t } (\alpha)$ are real-analytic and convex in $\alpha$ and jointly continuous in $ (p, t, \alpha)$. The function $\alpha \to \ES_{t} (\alpha)$ is real-analytic in $\alpha$.
\item $e_{p, t } (\alpha ) = e_{p, -t} (\alpha)$.
\item $e_{p, t } (\alpha ) = e_{p, t } ( 1 - \alpha )$.
\item The function $]0, \infty ] \ni p \to e_{p, t } (\alpha ) $ is continuous and decreasing.
\item $e'_{p, t } (0) = \ES _t ' (0) = - e ' _{p, t} (1) = - t \omega ( \Sigma^t )$.
\item 
\begin{align}
e_{2, t } (0) '' = \ES ''_t (0) = \int_0 ^t \int_0 ^t \omega ( (\sigma_s - \omega ( \sigma_s ) ) ( \sigma_u - \omega ( \sigma_u ) ) ) \d s \d u .\label{eqn:infinitederiv}
\end{align}
\item As $M \to \infty$, the sequence $\P_{t, M}$ converges weakly towards a Borel probability measure $\P _t$ on $\R$ and,
\begin{align}
e_{2, t} (\alpha ) = \log \int_{\R} e^{ - t \alpha \phi } \d \Phi_t (\phi ) .
\end{align}
All the moments of $\P _{t, M}$ converge to corresponding moments of $\P _t$. The measure $\P _t$ is the full counting statistics of the extended XY chain.
%\item $e_{\infty, t} (\alpha ) = \sup ( S(\rho \vert \omega ) - \alpha t \rho ( \Sigma^t ) ),$ where the supremum is taken over all states $\rho$.
\end{enumerate}
\end{proposition}
\proof   Recall the formulas (\ref{eqn:eptform}) and (\ref{eqn:esform}) and that $h_M \to h$ strongly. The $\K_{\#, t, M} (\gamma, u ) $ also have strong limits which we denote $\K_{\#, t } (\gamma, u)$, where $\#$ stands for either $p \in ]0, \infty]$ or $\ES$. Note that their strong limits are given by (\ref{eqn:kp}), (\ref{eqn:kinf}) and (\ref{eqn:kes}), but with the finite dimensional objects replaced with their infinite dimensional counterparts (i.e., the $k_M$ and $h_M$ replaced by $k$ and $h$ which are operators on $\ell^2 (\Z)$). Note that $k_t$ is given by $e^{\i t h} k e^{ -\i t h}$, and not the dynamics of Proposition \ref{prop:dynamics} (the $\tau^t$ act on $\O$ which is a different object than the bounded operators on $\ell^2(\Z)$ of which $h$ is an element).

 Since the commutators $[k_M, h_M] = [k, h]$ do not depend on $M$ and are finite rank, the traces in (\ref{eqn:eptform}) and (\ref{eqn:esform}) are the sum of only finitely many terms. Explicitly, one has (recall that we previously defined $e_{\ES, t, M} (\alpha) = \ES_{t, M} (\alpha)$),
\begin{align}
e_{\#, t, M} (\alpha) = t \sum_{k \in \A} \int_0 ^\alpha \d \gamma \int_0 ^1 \d u \langle \delta_k ,  \K _{\#, t, M} ( \gamma, u ) \i [k, h] \delta_k \rangle
\end{align}
where $\A = [ -N - 2 , - N +1 ] \cup [ N -1 , N+2 ] \subseteq \Z$. In particular, the size of $\A$ is finite and does not depend on $M$. It is easy to see that there is a constant $C$ depending only on $\alpha$ and $\#$ s.t.
\begin{align}
\norm{ \K_{\#, t, M} ( \gamma, u ) } \leq C ,
\end{align}
for all $( \gamma, u) \in [0, \alpha] \times [0, 1]$ (here, if $\alpha <0$, $[0, \alpha ] := [\alpha, 0]$). It follows by dominated convergence that,
\begin{align}
\lim_{M \to \infty} e_{\#, t, M} (\alpha ) &= t \sum_{k \in \A} \int_0 ^\alpha \d \gamma \int_0 ^1 \d u \lim_{M \to \infty} \langle \delta_k ,  \K _{\#, t, M} ( \gamma, u ) \i [k, h] \delta_k \rangle \nonumber \\
&= t \sum_{k \in \A} \int_0 ^\alpha \d \gamma \int_0 ^1 \d u \langle \delta_k ,  \K _{\#, t} ( \gamma, u ) \i [k, h] \delta_k \rangle \nonumber \\
&= t \int_0^\alpha  \d \gamma \int_0^1 \d u \tr \left( \K_{\#, t } (\gamma, u ) \i [k, h] \right), \label{eqn:thermfunct}
\end{align}
which proves (i). 

The $\K_{p, t } (\gamma, u)$ are jointly norm continuous in $(p, t, u, \gamma)$ and so the functions $\langle \delta_k, \K_{\#, t } (\gamma, u) \delta_k \rangle $ are jointly continuous in $(p, t, u, \gamma)$. Since the sum in the second last line of (\ref{eqn:thermfunct}) is finite, the functions $e_{p, t} (\alpha)$ are jointly continuous in $(p, t, \alpha)$.  Let us be careful in the proof of the analyticity of the entropic functionals. We claim that the analyticity of the functions
\begin{align}
\alpha \to  \langle \delta_k ,  \K _{\#, t} ( \gamma, u ) \i [k, h] \delta_k \rangle 
\end{align}
follows directly from Lemma \ref{lem:painful}. From this, the analyticity of the $e_{\#, t} ( \alpha)$ is immediate.  If $\# = \infty$, then
\begin{align}
\K_{\infty, t} ( \alpha, s) = f ( F_1 ( \alpha, s, t ) ) ,
\end{align}
with $f(z) = (1 + z)^{-1}$ and $F_1 ( \alpha, s, t) = e^{ ( 1 - \alpha ) k_{ts} - \alpha k_{- t ( 1 - s ) } }$. It is clear that $F_1 ( \alpha, s, t) \geq e^{ - ( 1  + 2 | \alpha | ) \norm{k}}$. The series expansion for $F_1 ( \alpha, s, t)$ converges for every $\alpha, s$ and $t$, and the coefficients of the power series expanded about some $\alpha_0$ are dominated  by the coefficients appearing infront of the $| \alpha - \alpha_0 |$ in the series expansion of $e^{ (1 + 2 | \alpha - \alpha_0 |  + 2 | \alpha_0 | )\norm{k}}$, and so Lemma \ref{lem:painful} applies.

If $\# = p$, with $p \in ]0, \infty [$, then
\begin{align}
\K_{p, t} ( \alpha, s) = F_2 ( \alpha, s, t ) f ( F_1 ( \alpha, s, t ) ) F_3 ( \alpha, s, t ) +  F_3 ( \alpha, s, t ) f ( F_1 ( \alpha, s, t ) ) F_2 ( \alpha, s, t ),
\end{align}
with $f(z) = (1 + z^{-p/2} )^{-1}$, $F_2 ( \alpha, s, t) = e^{ - (1 - \alpha ) k_{ts} /p }$, $F_3 = F_2 ^{-1}$ and,
\begin{align}
F_1 ( \alpha, s, t) = e^{(1 - \alpha) k_{ts} /p} e^{2 \alpha k_{- t (1 - s ) } / p} e^{ ( 1 - \alpha ) k_{ts} / p}  \geq e^{ 1 ( 1 + 4 | \alpha | ) \norm{k} / p }.
\end{align}
The dominating series are obtained by the expansions of $e^{ (1 + | \alpha - \alpha_0 | + | \alpha_0 | ) K / p }$ for $F_2$ and $F_3$,  and\\  $e^{ (1 + 4 | \alpha - \alpha_0 | + 4 | \alpha_0 | ) \norm{k} /p }$ for $F_1$, and so Lemma \ref{lem:painful} applies. 

For $ \# = \ES$, we have
\begin{align}
\K_{\ES, t} ( \alpha, s ) = - F_2 ( \alpha, s, t) f ( F_1 ( \alpha, s, t) ) F_3 ( \alpha, s, t)
\end{align}
with $f(z) = (1 + z )^{-1}$,  $F_2 ( \alpha, s, t) = e^{ - k_{ - ts } /2 }$, $F_3 = F_2 ^{-1}$ and,
\begin{align}
F_1 ( \alpha, s, t ) = e^{ - k_{-ts} /2} e^{ - \alpha ( k_{t (1 - s ) } - k_{-ts} )} e^{ - k_{-ts} /2}  \geq e ^{ - (1 + 2 | \alpha | ) \norm{k}}.
\end{align}
The dominating series are given by the constant $e^K$ for $F_2$ and $F_3$, and  $e ^{ (1 + 2 | \alpha - \alpha_0 | + 2 | \alpha_0 | ) \norm{k}}$ for $F_1$. This completes the proof of analyticity.

%%In similar fashion, the real-analyticity of the functions $\K_{\#, t} (\gamma, u)$ in $\gamma$ yield the real-analyticity of the %%$e_{p, t} (\alpha)$ in $\alpha$. 
The remaining parts of (ii)-(vi) follow from the corresponding properties of the $e_{\#, t, M} (\alpha)$ collected in Proposition \ref{prop:finiteentropic}.

For (vii) and (viii) we need the following (see Appendix B in \cite{JOPP}),
\begin{theorem}[Vitali's convergence theorem for analytic functions]\label{thm:vit}
Let $D (0, \eps) \subseteq \C$ be the disc of radius $\eps$ centered at $0$, and let $F_n : D (0, \eps) \to \C$ be a sequence of analytic functions such that
\begin{align} 
\sup_{\substack{z \in D (0, \eps) \\ n} } | F_n (z ) | < \infty. \label{eqn:vitbd}
\end{align}
Suppose that the limit
\begin{align}
\lim_{n \to \infty} F_n (z) = F(z) , \label{eqn:vitlim}
\end{align}
exists for every $z \in D (0, \eps) \cap \R$. Then the limit (\ref{eqn:vitlim}) exists for all $z \in D (0, \eps)$ and is an analytic function on $D (0, \eps)$. Moreover, as $n \to \infty$, the derivatives of $F_n $ converge uniformly on compact subsets of $D (0, \eps)$ to the corresponding dervatives of $F$.
\end{theorem}

The arguments that gave the analyticity of the $e_{\#, t} (\alpha)$ also apply to the finite volume $e_{\#, t, M} (\alpha)$, by replacing the infinite volume versions of the $\K_{\#, t} ( \gamma, u)$ with the finite volume $\K_{\#, t, M} ( \gamma, u)$ (the argument needs only to be modified by replacing $\norm{k}$ with $\sup_{M} \norm{k_M} < \infty$ in the bounds appearing there, and also by fixing $t$ and allowing the $F_j ( \alpha, s, \cdot)$'s to be indexed by $M$ in their third argument instead of $t$). It follows from Lemma \ref{lem:painful} that for each $\#$ there is an $r_0 > 0$, which does not depend on $M$ s.t. the functions $e_{\#, t, M} ( \alpha)$ all have analytic extensions to the disc $D (0, r_0 )$, and so by the Vitali convergence theorem,  Proposition \ref{prop:finiteentropic} and (i), we conclude (vii) and also the equality
\begin{comment}
For example, it is clear that
\begin{align}
\K _{\infty, t ,M } ( \gamma, u) = \left( \1_M + e^{ - ( 1 - \gamma ) k_{tu,M} - \gamma k_{-t (1 - u ) ,M} } \right)^{-1}
\end{align}
is analytic if $|\gamma|$ is small enough so that the matrix appearing above is invertible. Since the norms of the $h_M$ are uniformly bounded, $\gamma$ can be taken smaller than some $M$-independent $r_0$.

We define the functions, which we continue to denote by $e_{\#, t, M} (\alpha)$, for $| \alpha | < r_0$ by
\begin{align}
e_{\#, p, t} (\alpha) = t \int_\gamma \d \gamma \int_0 ^1 \d u \tr \left( \K_{\#, t, M} (\gamma, u ) \i [ k, h] \right) ,
\end{align}
where $\gamma$ is any path from $0$ to $\alpha$ lying in the disc $D_{r_0}$. These functions coincide with the previous definition of $e_{\#, t, M} (\alpha)$ for real $\alpha$ and are analytic in the disc $D_{r_0}$. 

The hypotheses (\ref{eqn:vitbd}) and (\ref{eqn:vitlim}) hold for $e_{\#, t, M} (\alpha)$, and so from Theorem \ref{thm:vit}, \end{comment} 
\begin{align}
e_{2, t} (0)'' = \ES_{t} '' (0) =\lim_{M \to \infty} \int_0 ^t \int_0 ^t \omega_M ( (\sigma_{s, M} - \omega_M ( \sigma_{s, M} ) ) ( \sigma_{u, M} - \omega_M ( \sigma_{u, M} ) ) ) \d s \d u 
\end{align}
Similar arguments as in the proof of Proposition \ref{prop:thermobasic} allow us to conclude that the integrand converges uniformly to $\omega( (\sigma_s - \omega(\sigma_s ) ) (\sigma_u - \omega( \sigma_u)) )$ for $ (s, u) \in [0, t]\times[0, t]$ which proves (viii).

To prove (ix), consider the sequence of functions
\begin{align}
\alpha \to\int e^{-\alpha t \phi} \d \P _{t, M} (\phi ) &= e^{ e_{2, t, M} (\alpha) } = \frac{ \det ( \1 + e^{(1- \alpha ) k_M }e^{ - \i t h_M } e^{ \alpha k_M } e^{ i t h_M } ) }{\det ( \1 + e^{k_M} ) } \nonumber \\
&=\det \left( \1 + ( \1 + e^{ - k_M } )^{-1} ( e^{ - \alpha k_M } e^{ -\i t h_M } e^{ \alpha k_M } e^{ \i t h_M } - \1 ) \right),
\end{align}
which are all entire analytic. The bound $| \det ( \1 + A ) | \leq e^{ \norm{A}_1 }$, where $\norm{A}_1$ denotes the trace norm of $A$, together with
\begin{align}
e^{ - \alpha k_M } e^{- \i t h_M } e^{\alpha k_M } e ^{ \i t h_M } - \1= \int_0 ^t e^{ - \alpha k_M } e^{ - \i s h_M } \i [ e^{ \alpha k_M}, v] e^{ \i s h_M } \d s
\end{align}
(with $v$ finite rank) imply that for any bounded set $B \subseteq \C$,
\begin{align}
\sup_{\substack{ \alpha \in B \\ M}} | e^{ e_{2, t, M } (\alpha ) }| < \infty .
\end{align}
Here we have again used the uniform boundedness of the $\norm{h_M}$'s. By the Vitali convergence theorem, the sequence of characteristic functions of the measures $\P _{t,M}$ converges pointwise to an entire analytic function, and the convergence is uniform on bounded sets. The existence of and the convergence to the weak limit $\P_t$ follows from the lemma following this proof, and the convergence of the moments follows from Vitali's theorem. This proves (ix). \qed 

The following lemma is Corollary 1 to Theorem 26.3 in \cite{Bi}
\begin{lemma}
Suppose that $\mu_n$ are probability measures with characteristic functions $\phi_n$ and that \\ $\lim_{n \to \infty} \phi_n (t) = g(t)$ for each $t$ where the limit function $g$ is continuous at $0$. Then $g$ is the characteristic function of a measure $\mu$ which is the weak limit of the $\mu_n$.
\end{lemma}

\begin{comment}
\begin{theorem}
Let $\mu_n$, $\mu$ be probability measures with characteristic functions $\phi_n$ and $\phi$. The measures $\mu_n$ converge weakly to $\mu$ if and only if $\phi_n (t ) \to \phi(t)$ for each $t$.
\end{theorem}
\end{comment}

\section{The large time limit} \label{sec:time}
\subsection{Operator theory preliminaries}
We first collect a few facts about spectral and scattering theory which we will use without proof. The reader is referred to \cite{Ja} (especially Section 4.9) for proofs and a more complete exposition. Let $A$ be a self-adjoint operator on a Hilbert space $\K$. The spectrum of $A$ is denoted $\sp (A)$ and its absolutely continuous part $\sp _{\ac} (A)$. The projection onto the absolutely continuous subspace of $A$ is denoted $\1_\ac (A)$. For any $\psi_1 , \psi_2 \in \K$, the boundary values
\begin{align}
\langle \psi_1 , (A - E \pm \i 0 ) ^{-1} \psi_2 \rangle := \lim_{\eps \dto 0 } \langle \psi_1 , (A - E \pm \i \eps )^{-1} \psi_2 \rangle , 
\end{align}
exist and are finite for Lebesgue a.e. $E \in \R$. Whenever we write $\langle \psi_1 , (A - E \pm \i 0 ) ^{-1} \psi_2 \rangle $ we assume that the limit exists and is finite. If $\nu_{\psi} $ is the spectral measure of $A$ for $\psi$, then the Radon-Nikodyn derivative of its absolutely continuous part is
\begin{align}
\d \nu _{\psi, \ac} (E)= \frac{1}{\pi} \langle \psi, (A - E - \i 0 )^{-1} \psi \rangle \d E.\label{eqn:boundaryac}
\end{align}

Let $\delta_l := \delta_{-N - 1 }$ and $\delta_r := \delta_{N+1}$. We denote by $\nu_{l/r}$ the spectral measure of $h_{l/r}$ for $\delta_{l/r}$. By (\ref{eqn:boundaryac}),
\begin{align}
d \nu _{l/r, \ac} = \frac{1}{\pi} F_{l/r} (E) \d E
\end{align}
where
\begin{align}
F_{l/r} (E) = \im G_{l/r} (E), \qquad G_{l/r} (E) = \langle \delta_{l/r} , (h_{l/r} - E - i0 )^{-1} \delta_{l/r} \rangle
\end{align}
We will also denote, for $z \in \C \backslash \R $, the resolvent $G_{l/r} (z) =  \langle \delta_{l/r} , (h_{l/r} -z )^{-1} \delta_{l/r} \rangle$.

By the spectral theorem, we may identify $\fh _{\ac} (h_0)$ with $L^2 (\R, \d \nu_{l, \ac} ) \oplus L^2 (\R, \d \nu_{r, \ac} )$ and $h_0\rest_{\fh_{\ac} (h_0) } $ with the operator of multiplication by the variable $E \in \R$. The set $\Sigma_{l/r, \ac} = \{ E \in \R \vert F_{l/r} (E) > 0 \}$ is an essential support of the absolutely continuous spectrum of $h_{l/r}$. Furthermore, let $J_{-N-1} = J_l$ and $J_N = J_r$. We set
\begin{align}
\E = \Sigma_{l, \ac} \cap \Sigma_{r, \ac} .
\end{align}

Let us recall basic facts from trace-class scattering theory, the proofs of which can be found in \cite{RS3}, Section XI.3 (see especially Theorem XI.8). The basic existence result is,
\begin{theorem}[Kato-Rosenblum theorem]
If $A$ and $B$ are bounded self-adjoint operators and $A-B$ is trace-class, then the wave operators
\begin{align}
w_{\pm}(A, B) = \slim_{t \to \pm \infty} e^{ \i t A} e^{ - \i t B}\1 _{\ac} (B)
\end{align}
exist and are complete. That is,
\begin{align}
\ran w_{-} (A, B) = \ran w_{+}(A, B) = \1_{\ac} (A).
\end{align}
Moreover, $\spec_{\ac} (A) = \spec_{\ac} (B)$ and the essential supports of the absolutely continuous spectrum of $A$ and $B$ coincide.
\end{theorem}
Note that
If the wave operators exist and are complete, then the scattering matrix
\begin{align}
s (A, B) = w_+ (A, B)^* w_- (A, B) ,
\end{align}
is a unitary operator $\fh_\ac (B)$. We have also, as a consequence of completeness (see Propositions 1 and 3 of Section XI.3 of \cite{RS3}),
\begin{align}
\1_{\ac}( B) =w_\pm^* (A, B) w_\pm (A, B) , \qquad \1_{\ac}(A) = w_\pm (A, B) w_\pm ^* (A, B),
\end{align}
and the adjoints are given by
\begin{align}
w_\pm^* (A, B) = w_\pm (B, A) = \slim_{t \to \pm \infty} e^{ \i t B} e^{- \i t A} \1_{\ac} (A).
\end{align}
Moreover, the wave operators satisfy the \emph{intertwining property} (see Proposition 1 of Section XI.3 of \cite{RS3}),
\begin{align}
w_\pm^* (A, B) A = B w_\pm^* (A, B).
\end{align}
In our case, $h-h_0$ is finite rank and therefore trace-class, and we denote the wave operators $w_\pm := w_\pm (h, h_0)$ and the scattering matrix $s := s(h, h_0)$.

\subsection{Formulas for wave operators and the scattering matrix}
We now derive a formula for the scattering matrix. First we compute  $w_{\pm} ^*$. We follow the methodology of \cite{JKP}. Define $\chi_l = \delta_{-N}$ and $\chi_r = \delta_N$.
\begin{proposition}\label{prop:waveoperators}
Let $g = g_l \oplus g_c \oplus g_r \in \fh$ be given. Then,
\begin{align}
w_{\pm} ^* g = g^{(\pm )}_l \oplus g^{(\pm)}_r
\end{align}
with,
\begin{gather}
g^{(\pm)}_l (E) = g_l (E)  - J_l \langle \chl, (h - E \mp \i 0)^{-1} g \rangle  \nonumber  \\
g^{(\pm)}_r (E) = g_r (E) - J_r \langle \chr , (h - E \mp \i 0)^{-1} g \rangle.  
\end{gather}
\end{proposition}
\proof Let any $f = f_l \oplus f_r \in \fh_{\ac} (h_0) = \fh_{\ac} (h_l ) \oplus \fh_{\ac} (h_r)$ be given. We will compute $\langle f , w_{+} ^* g \rangle$. The computation for $w_-^*$ is identical. We have,
\begin{align}
\lim_{ t \to \infty} \langle f, e^{ \i t h_0} e^{ - \i t h }g \rangle &= \lim_{t \to \infty} \langle e^{ \i t h} e^{ - \i t h_0} \1_{ac} (h_0 ) f , g \rangle \nonumber \\
&= \langle w_+ f, g \rangle = \langle f, w_+ ^* g \rangle .
\end{align}
Note that,
\begin{align}
\langle f , e^{ \i t h_0} e^{ - \i t h} g \rangle = \langle f , g \rangle - \i \int_0 ^t \langle f, e^{ \i s h_0} v e^{ -\i s h } g \rangle \d s.
\end{align}
We require the following lemma (see Lemma 5 in Section XI.6 of \cite{RS3})
\begin{lemma} \label{lem:abel}
Let $\phi$ be a bounded measureable function and suppose $\lim_{t \to \infty} \int_0^t \phi(s) \d s = a$ exists. Then $ a = \lim_{\eps \dto 0} \int_0^\infty e^{ - \eps s} \phi(s) \d s = \lim_{\eps \dto 0} \int_0^\infty e^{ - \eps s^2} \phi(s) \d s.$
\end{lemma}

Therefore, 
\begin{align}
\langle f, w_+ ^* g \rangle = \langle f, g\rangle - \lim_{\eps \dto 0} \i \left( L_l ( \eps)  + L_r (\eps ) \right)
\end{align}
where
\begin{align}
L_{l/r} ( \eps ) &= \int_0^\infty e^{ - \eps s} \langle f , e^{ \i s h_0 } v_{l/r} e^{- \i s h} g \rangle \d s \nonumber \\
&= \int_0 ^\infty e^{ - \eps s }  J_{l/r}  \left( \langle f , e^{ \i s h_0 } \delta_{l/r} \rangle \langle \chlr , e^{ -\i s h } g \rangle + \langle f , e^{\i s h_0 } \chlr \rangle \langle \delta_{l/r} , e^{ -\i s h } g \rangle \right)  \d s .
\end{align}
Since $f \in \fh_{\ac} (h_0)$ it follows that $ \langle f , e^{- \i s h_0 } \chlr \rangle = 0 $. 

We further compute,
\begin{align}
L_{l/r} ( \eps ) &= \int_0 ^\infty e^{ - \eps s }  J_{l/r}  \langle f , e^{\i s h_0 } \delta_{l/r} \rangle \langle \chlr , e^{ -\i s h } g \rangle  \d s\nonumber \\
&= J_{l/r} \int_\R  \widebar{f}_{l/r} (E) \left[ \int_0^\infty \langle \chlr, e^{ - \i s ( h - E - \i \eps ) } g \rangle \d s \right] \d \nu_{l/r , \ac} (E) \nonumber \\
&=  - \i  J_{l/r} \int_\R  \widebar{f}_{l/r} (E) \langle \chlr , (h - E - \i \eps )^{-1} g \rangle \d \nu_{l/r , \ac} (E)
\end{align}
The interchange of the order of integration is clearly justified. Let $H_{l/r} (E + \i \eps) = \langle \chlr , (h - E - \i \eps )^{-1} g \rangle$. Since $\lim_{\eps \dto 0} H_{l/r} (E + \i \eps) =: H_{l/r} (E)$ exists and is finite for Lebesgue a.e. $E$, we have by Egoroff's theorem that for any $n$ there are measureable sets $L_n $ and $ R_n$ with $| \R \backslash L_n | < 1/n$, $| \R \backslash R_n | < 1/n$ and $H_{l/r} (E + \i \eps ) \to H_{l/r} (E)$ uniformly on $L_n / R_n$. Clearly, the set
\begin{align}
\bigcup_{ n > 0 } \left\{ f_l \oplus f_r \in \fh_{\ac} (h_0)  \vert \supp f_l \subseteq L_n , \supp f_r \subseteq R_n \right\}\label{eqn:denseset}
\end{align}
is dense in $\fh_{\ac} (h_0)$. Suppose that $f$ belongs to this set. Then $H_{l/r} (E + \i \eps ) \to H_{l/r}(E)$ uniformly on the support of $f_{l/r}$ and therefore, the inequality
\begin{align}
 | f_{l/r} (E)| | H_{l/r}(E) - H _{l/r} (E + \i \eps ) | \leq C_{f} | f_{l/r}(E) | \in L^1 (\R, \d \nu_{l/r, \ac} ),
\end{align}
which holds for a constant depending only on $f$ and all small $\eps$ implies, by dominated convergence,
\begin{align}
\lim_{\eps \dto 0} \int_{\R}|  \widebar{f}_{l/r} | | H_{l/r} (E + i \eps ) - H_{l/r} (E) | \d \nu_{l/r, \ac} (E) = 0 .
\end{align}
$H_{l/r} (E)$ is bounded on the support of $f_{l/r}$, which implies that $H_{l/r} (E) f_{l/r} (E) \in L^1 (\R, \d \nu_{l/r, \ac} )$. Therefore, the formula
\begin{align}
\langle f, w_+ ^* g \rangle &= \int_\R \widebar{f}_l (E) \left( g_l (E) - J_l \langle \chl , (h - E - \i 0)^{-1} g \rangle \right) \d \nu_{l, \ac} (E) \nonumber \\
& + \int_\R \widebar{f}_r (E) \left(g_r (E) - J_r \langle \chr , (h - E - \i 0 )^{-1} g \rangle \right) \d \nu_{r, \ac} (E),
\end{align}
holds for the dense set of $f$ given in (\ref{eqn:denseset}). The claim follows. \qed

The following will be useful in the next subsection.
\begin{corollary} \label{cor:waveform}
For $a \in \{ l, r \}$,
\begin{align}
w_-^* h_0 \delta_a (E) &= ( E \delta_{l, a} + J_l J_a \langle \chl, (h - E + \i 0)^{-1} \cha \rangle + E J_l J_a \langle \chl , (h - E + \i 0)^{-1} \cha \rangle \langle \delta_a , (h_0 - E + \i 0)^{-1} \delta_a \rangle ) \nonumber \\
& \oplus ( E \delta_{r, a } + J_r J_a \langle \chr , (h - E + \i 0)^{-1} \cha \rangle + E J_r J_a \langle \chr , (h - E + \i 0)^{-1} \cha \rangle \langle \delta_a (h_0 - E + \i 0)^{-1} \delta_a \rangle )  \label{eqn:qq1}
\end{align}
and
\begin{align}
w_-^* \cha (E) &= ( - J_l \langle \chl , (h - E + \i 0)^{-1} \cha \rangle ) \oplus ( -J_r \langle \chr , (h - E + \i 0)^{-1} \cha \rangle ). \label{eqn:qq2}
\end{align}
\end{corollary}
\proof (\ref{eqn:qq2}) follows directly from Proposition \ref{prop:waveoperators} and (\ref{eqn:qq1}) follows from Proposition \ref{prop:waveoperators} and the identity
\begin{align}
\langle & \chb, ( h - E + \i 0)^{-1} h_0 \delta_a \rangle \nonumber \\
&= - J_a  \langle \chb, (h - E + \i 0)^{-1} \cha \rangle - E J_a \langle \chb , (h - E + \i 0)^{-1} \cha \rangle  \langle \delta_a , (h_0 - E + \i 0)^{-1} \delta_a \rangle ,
\end{align}
which holds for $a, b \in \{ l, r \}$. \qed

We can now compute the scattering matrix.
\begin{theorem} \label{thm:scat}
Let $g = g_l \oplus g_r \in \fh_{\ac} (h_0)$ be given. Then,
\begin{align}
sg = g^{(s)}_l \oplus g_r ^{(r)}
\end{align}
with 
\begin{gather}
g^{(s)}_l (E) = g_l (E) + 2 \i J_l ^2 F_l (E) \langle \chl, ( h - E - \i 0)^{-1} \chl\rangle g_l (E) + 2 \i J_l J_r F_r (E) \langle \chl, (h - E - \i 0)^{-1} \chr \rangle g_r (E) \nonumber \\
g^{(s)}_r (E) = g_r (E) + 2 \i J_r J_l F_l (E) \langle \chr, ( h - E - \i 0)^{-1} \chl \rangle g_l (E) + 2 \i J_r^2 F_r (E) \langle \chr, (h - E - \i 0)^{-1} \chr \rangle g_r (E).\label{eqn:scatform}
\end{gather}
\end{theorem}

There is a subtlety associated with the formulas in (\ref{eqn:scatform}). For example, in the formula for $g^{(s)}_l (E)$, the function on the right $g_r (E)$ is defined only for $\mu_{r, \ac}$-a.e. $E$, and may not be defined for a set of nonzero $\mu_{l, \ac}$ measure. What we mean by the formula (\ref{eqn:scatform}) is that $g_r \equiv 0$ outside of $\Sigma_{r, \ac}$. Similar statements hold for the formula for $g_r^{(s)}$.

Let $\langle \cdot , \cdot \rangle_2$ denote the standard inner product on $\C^2$. For any $f \in \fh_{\ac} (h_0)$ let $f(E)$ denote the vector $( f_l (E), f_r (E) ) \in \C^2$. Then for $f, g \in \fh_\ac (h_0)$ we have,
\begin{align}
\langle f, g \rangle = \int_\R \langle V(E) f(E) , V(E) g(E) \rangle_2 \d E ,
\end{align}
where $V(E)$ is the $2 \times 2$ diagonal matrix 
\begin{align}\label{eqn:vee}
V (E) = \left( \begin{matrix}\sqrt{\frac{ F_l (E) }{\pi } } &0\\ 0& \sqrt{\frac{F_r (E) } { \pi } } \end{matrix} \right).
\end{align}
Multiplication by the matrix $V(E)$ is a unitary operator $V : \fh_\ac (h_0) \to L^2 ( \R , \rho_l (E) \d E ) \oplus L^2 ( \R , \rho_r (E) \d E )$ with $\rho_{l/r} (E)$ the characteristic function of $\Sigma_{l/r, \ac}$.

Theorem \ref{thm:scat} implies that the scattering matrix acts by multiplication by a $2 \times 2$ matrix on $\fh_{\ac} (h_0)$. However, it is more convenient to consider the operator $V s V^{-1}$ which is a unitary operator on $V \fh_{\ac} (h_0 ) = L^2 ( \R , \rho_l (E) \d E ) \oplus L^2 ( \R , \rho_r (E) \d E )$ and acts as multiplication by the $2\times 2$ matrix
\begin{align}
s(E) =  \left( \begin{matrix}s_{ll} (E)& s_{lr} (E)\\ s_{lr} (E)  &s_{rr} (E)\end{matrix}\right)
\end{align}
where
\begin{align}
s_{ll} (E) &= 1+ 2\i J_l^2 \langle \chl , (h - E - \i 0)^{-1} \chl \rangle F_l (E) \nonumber \\
s_{lr} (E) &= 2\i J_l J_r \langle \chl , (h - E - \i 0)^{-1} \chr \rangle \sqrt{F_l (E) F_r  (E)} \nonumber \\
s_{rl} (E) &= 2\i J_r J_l \langle \chr , (h - E - \i 0)^{-1} \chl\rangle \sqrt{ F_r (E) F_l (E) }\nonumber \\
s_{rr} (E) &= 1 + 2\i J_r^2 \langle \chr , (h - E - \i 0)^{-1} \chr \rangle F_r (E). \label{eqn:unitscat}
\end{align}
This is a slight abuse of notation, and perhaps what we call $s_{ab} (E)$ should really be $[V s V^{-1}]_{ab} (E)$, but formulas appearing later are more simple and natural with this notation. With this convention, the matrix $s(E)$ is unitary for each $E$  w.r.t. the standard inner product on $\C^2$.

Since
\begin{align}
\langle \chl , (h - E - \i 0)^{-1} \chr \rangle = \langle \chr, (h - E - \i 0)^{-1} \chl \rangle ,
\end{align}
the scattering matrix is symmetric for Lebesgue a.e. $E \in \E$; that is, $s_{lr}(E) = s_{rl} (E)$. From the resolvent identity $A^{-1} - B^{-1} = A^{-1} (B - A ) B^{-1}$ we derive
\begin{align}
\langle \chi_l, (h - E - \i 0)^{-1} \chr \rangle = \frac{ J_l^2 \langle \chl , ( h -  E - \i 0)^{-1} \chl \rangle \langle \delta_l , ( h_0 - E - \i 0)^{-1} \delta_l \rangle \langle \chl , (h_0 - E - \i 0)^{-1} \chr \rangle }{ 1 - J_r ^2 \langle \delta_r, (h_0 - E - \i 0)^{-1} \delta_r \rangle \langle \chr, (h_0 - E - \i 0)^{-1} \chr \rangle}.
\end{align}
The denominator is non-zero for Lebesgue a.e. $E \in \E$, as the functions $\langle \cha , (h_0 - E - \i 0)^{-1} \chb \rangle$ are purely real for Lebesgue a.e. $E \in \R$. These functions are analytic outside of the spectrum of $h_0 \rest_{\fh_c}$ (which is necessarily a finite set of points) and are nonvanishing. It follows from Theorems 3.17 and 5.12 of \cite{Ja} that $ \langle \chl , ( h -  E - \i 0)^{-1} \chl \rangle$ is non-zero for Lebesgue a.e. $E \in \R$, and so we conclude that the scattering matrix is not diagonal for Lebesgue a.e. $E \in \E$.

\begin{comment}

From the symmetry of the scattering matrix we conclude that $\Theta( s)$ (recall $\Theta$ is the time-reversal of the extended XY chain described in Section \ref{sec:therm}) acts on $V \fh_{\ac} (h_0)$ by multiplication by the $2 \times 2$ matrix $s^* (E)$. Here we are using the fact that the conjugation on each space $\ell^2( ] - \infty , -N -1 ] )$ and $\ell^2 ( [N+1, \infty [ )$ maps to conjugation on $L^2 ( \R, \d \nu_{l, \ac} )$ and $L^2 ( \R , \d \nu_{r, \ac} )$. This itself is an immediate consequence of the fact that each $\delta_k \in \ell^2( ] - \infty , -N -1 ] )$ (resp.,  $\ell^2 ( [N+1, \infty [ )$) can be written as $\delta_k = P(h_l) \delta_l$ (resp., $P(h_r) \delta_r$) where $P$ is a polynomial with  real coefficients.
\end{comment}

%Comparing this with  (\ref{eqn:scatform}), the reader will notice that the off-diagonal elements of (\ref{eqn:unitscat}) differ from those appearing in %(\ref{eqn:scatform}) by multiplication or division by $(F_l (E) / F_r (E) )^{1/2}$.

\noindent\emph{Proof of Theorem  \ref{thm:scat}. }  Let $f = f_l \oplus f_r, g= g_l \oplus g_r \in \fh_{\ac} (h_0)$ be given. The beginning of our proof is similar to the manipulations appearing in the proof of Theorem XI.42 in \cite{RS3}.
\begin{align}
\langle f, (s - \1 ) g \rangle &= \langle f , (w_+ ^* w_- - w_-^* w_- ) g  \rangle \nonumber \\
&= \langle ( w_+ - w_- ) f , w_- g \rangle \nonumber \\
&= \lim_{t \to \infty} \langle ( e^{ \i t h } e^{ - \i t h_0} - e^{ - \i t h} e^{ \i t h_0} ) f , w_- g \rangle \nonumber \\
&= \lim_{t \to \infty} -\i  \int_{-t} ^t  \langle e^{\i s h} v e^{ - \i s h_0} f , w_- g \rangle  \d s \nonumber \\
&= \lim_{\eps \dto 0}- \i \int_\R e^{ - \eps s^2 } \langle e^{\i s h } v e^{ - \i s h_0} f, w_- g \rangle \d s
\end{align}
The last equality is Lemma \ref{lem:abel}. Let us compute the integrand. We have,
\begin{align}
\langle e^{\i s h } v e^{ - \i s h_0 } f, w_- g \rangle% &= J_l \left( \langle e^{ - \i s h_0 }f , \delta_{- N} \rangle \langle e^{ \i s h} \delta_l , w_- g \rangle +  %\langle e^{ - \i s h_0 }f , \delta_{l} \rangle \langle e^{ \i s h} \delta_{-N} , w_- g \rangle        \right) \nonumber \\
%&+ J_r \left(  \langle e^{ - \i s h_0 }f , \delta_{N} \rangle \langle e^{ \i s h} \delta_r , w_- g \rangle     +  \langle e^{ - \i s h_0 }f , \delta_{ r} \rangle \langle %e^{ \i s h} \delta_N , w_- g \rangle    \right) \nonumber \\
&= J_l \langle e^{ - \i s h_0} f , \delta_l \rangle \langle e^{ \i s h } \chl , w_- g \rangle + J_r \langle e^{ - \i s h_0} f , \delta_r \rangle \langle e^{ \i s h } \chr , w_- g \rangle . \label{eqn:dif1}
\end{align}
By the intertwining property of the wave operators, $ \langle e^{ \i s h } \chlr, w_- g \rangle =  \langle w_-^*  e^{ \i s h } \chlr , g \rangle = \langle w_-^* \chlr, e^{ - \i s h_0} g \rangle $. Therefore, the formula in Proposition \ref{prop:waveoperators} gives
\begin{align}
\langle f, (s - \1 ) g \rangle &= \lim_{\eps \dto 0}  \i \left( H_{ll} (\eps) +  H_{rl} (\eps) +  H_{lr} (\eps) +  H_{rr} (\eps) \right)
\end{align}
with,
\begin{align}
H_{ab} (\eps ) =  J_a J_b \int_\R e^{ - \eps s^2 }   \left[ \int_\R e^{ \i s E} \widebar{f}_a (E) \d \nu_{a, \ac}(E)   \right] \left[ \int_\R e^{ - \i s E'}\langle \cha , (h - E' - \i 0) \chb \rangle g_b (E') \d \nu_{b, \ac} (E')\right] \d s ,
\end{align}
 Let $G_{ab} (E') =\langle \cha , (h - E' - \i 0) \chb \rangle $. Our next manipulations are similar to the proof of Plancherel's theorem in \cite{LL} (Thm 5.3).  Fubini's theorem allows us to interchange the order of integration and compute,
\begin{align}
H_{ab} (\eps) &= J_a J_b \int_\R \widebar{f}_a (E) \left[ \int_\R G_{ab} (E') g_b (E') \left[ \int_\R e^{ - \eps s^2 } e^{ \i s (E - E' ) } \d s \right] \d \nu_{b, \ac} (E') \right] \d \nu_{a, \ac} (E) \nonumber \\
&= \sqrt{\pi} J_a J_b \int_\R \widebar{f}_a (E) \left[ \int_\R  G_{ab} (E' ) g_b (E') \eps^{-1/2} \exp \left( - \frac{(E - E')^2}{4 \eps} \right) \d \nu_{b, \ac} (E') \right] \d \nu_{a, \ac} (E) .
\end{align}
Since $G_{ab} (E') g_b(E') \frac{ \d \nu_{b, \ac}}{\d E' } (E')$ is an $L^1 (\R, \d E' )$ function, 
\begin{align}
\int_\R  G_{ab} (E' ) g_b (E') \eps^{-1/2} \exp \left( - \frac{(E - E')^2}{4 \eps} \right) \d \nu_{b, \ac} (E') \to 2 \sqrt{\pi} G_{ab} (E) g_b (E) \frac{\d \nu_{b, \ac} }{\d E} (E)
\end{align}
strongly in $L^1 (\R, \d E)$ as $\eps \dto 0$ (see, e.g., Thm 2.16 and Thm. 5.3 in \cite{LL}). Let $L_n / R_n$ be the set $ \{ E \vert \frac{ \d \nu_{l/r, \ac}}{\d E} (E) > n \}$. As $ n \to \infty $ both $|L_n| \to 0$ and $ | R_n | \to 0$. The set
\begin{align}
\bigcup_{ n > 0 } \left\{ f_l \oplus f_r \in \fh_{\ac} (h_0 ) \vert \supp f_l \subseteq L_n^{\mathsf{c}} , \supp f_r \subseteq R_n^{\mathsf{c}} , \norm{f_l}_{\infty} < \infty , \norm{f_r}_{\infty} < \infty \right\} \label{eqn:densesetscat}
\end{align}
is dense in $\fh_{\ac} (h_0)$. Fix such an $f$. Then the functions
\begin{align}
f_{l/r} (E) \frac{ \d\nu_{ l/r , \ac} } {\d E} (E)
\end{align}
are bounded and so by H\"older's inequality,
\begin{align}
\lim_{\eps \dto 0} H_{ab} (E) = 2  J_a J_b \int_\R \widebar{f}_a (E) \langle \cha, (h - E - \i 0)^{-1} \chb \rangle F_{b} (E) g_b (E) \d \nu_{a, \ac} (E).
\end{align}
We have therefore shown that the formula
\begin{align}
\langle f, (s - \1 ) g \rangle = \langle f , g^{(s)}_l \oplus g^{(s)}_r \rangle .
\end{align}
holds for the dense set of $f$ in (\ref{eqn:densesetscat}). The claim follows. \qed

We record here the following corollary of the unitarity of the scattering matrix; while we will not use it later, it is worth noting for possible future reference.
\begin{corollary} \label{cor:funnyform}
For Lebesgue a.e. $E$,
\begin{align}
J_{r/l}^2 &| \langle \chlr , ( h - E - \i 0)^{-1} \chrl \rangle |^2 F_{l/r} (E) F_{r/l} (E) \nonumber \\
&= F_{l/r}(E) \im \left[ ( J_{l/r} \langle \delta_{l/r} , (h - E - \i 0)^{-1} \chlr \rangle - 1) \langle \chlr, (h - E + \i 0)^{-1} \chlr \rangle  \right].
\end{align}
In particular, for $\nu_{l/r , \ac}$-a.e. $E$,
\begin{align}
J_{r/l}^2 &| \langle \chlr , ( h - E - \i 0)^{-1} \chrl \rangle |^2 F_{r/l} (E) \nonumber \\
&= \im \left[ ( J_{l/r} \langle \delta_{l/r} , (h - E - \i 0)^{-1} \chlr \rangle - 1) \langle \chlr, (h - E + \i 0)^{-1} \chlr \rangle  \right].
\end{align}
\end{corollary}
\proof Unitarity of the scattering matrix implies that $ |s_{ll} (E) |^2 + | s_{lr} (E) |^2 = 1$. With the formulas established in (\ref{eqn:unitscat}), this implies
\begin{align}
J_r^2 & | \langle \chl , ( h - E - \i 0)^{-1} \chr \rangle |^2 F_l (E) F_r (E) \nonumber \\&= F_l (E) \big(\im [ \langle \chl , ( h - E - \i 0)^{-1} \chl \rangle ] 
- J_l ^2 F_l (E) | \langle \chl , (h - E - \i 0)^{-1} \chl \rangle |^2 \big). \label{eqn:dif3}
\end{align}
On the other hand the identity 
\begin{align}
\langle \delta_l , ( h - E - \i 0)^{-1} \chl \rangle = - J_l \langle \delta_l, (h_0 - E - \i 0)^{-1} \delta_l \rangle \langle \chl, (h - E - \i 0)^{-1} \chl \rangle
\end{align}
 gives
\begin{align}
\im \big[ &(J_l  \langle \delta_l,   ( h - E - \i 0)^{-1} \chl \rangle - 1 ) \langle \chl, (h - E + \i 0)^{-1} \chl \rangle \big] \nonumber \\
&= \im [ \chl, (h - E - \i 0)^{-1} \chl \rangle ] - J_l ^2 F_l (E) | \langle \chl, (h - E - \i 0)^{-1} \chl \rangle |^2 
\end{align}
which, when combined with (\ref{eqn:dif3}), yields the claim. The proof of the other equality is identical. \qed

\subsubsection{The scattering matrix for two directly coupled chains} 

The formula for the scattering matrix given in \cite{JLP} is slightly different from that derived here. We explain here where the difference arises. In \cite{JLP} the XY chain consists only of a left part and right part and does not have a central part. To be more precise, the $h_0$ we are considering is replaced by $\widetilde{h}_0 = \widetilde{h}_r + \widetilde{h}_l$ where $\widetilde{h}_l = h \rest_{\ell^2 ( ] - \infty, 0] )}$ and $\widetilde{h}_r = h \rest_{\ell^2 ( [1, \infty [ )}$, and the formula given in \cite{JLP} is for $s (h, \widetilde{h}_0 )$.

The formula obtained is that $s ( h, \widetilde{h}_0)$ acts as multiplication by the $2\times 2$ matrix
\begin{align}
\widetilde{s} (E) =  \left( \begin{matrix}\widetilde{s}_{ll} (E)& \widetilde{s}_{lr} (E)\\ \widetilde{s}_{lr} (E)  & \widetilde{s}_{rr} (E)\end{matrix}\right)
\end{align}
where,
\begin{align}
\widetilde{s}_{ll} (E) &= 1 + 2 \i J_0 ^2 \langle \delta_1 , ( h - E - \i 0)^{-1} \delta_1 \rangle \widetilde{F}_l (E)\nonumber \\
\widetilde{s}_{lr} (E) &=  2\i J_0 ( J_0 \langle \delta_1 , ( h - E - \i 0)^{-1} \delta_0 \rangle -1 ) \sqrt{\widetilde{F}_l (E) \widetilde{F}_r  (E)} \nonumber \\
\widetilde{s}_{rl} (E) &= 2\i J_0 ( J_0 \langle \delta_0, ( h - E - \i 0)^{-1} \delta_1 \rangle - 1) \sqrt{ \widetilde{F}_r (E) \widetilde{F}_l (E) }\nonumber \\
\widetilde{s}_{rr} (E) &= 1 + 2 \i J_0^2 \langle \delta_0 , (h - E - \i 0)^{-1} \delta_0 \rangle \widetilde{F}_r (E)
\end{align}
with $\widetilde{F}_{l/r} (E) = \im \langle \delta_{0/1} , ( \widetilde{h}_{l/r} - E - \i 0)^{-1} \delta_{0/1} \rangle$. First note that the $\chlr$ have been replaced by $\delta_{1/0}$ - this is expected as these vectors play the same role in linking the left/right chain to the rest of the system through the interaction potential (i.e., these vectors appear in `equivalent' places in $v = h - h_0$ and $\widetilde{v} = h - \widetilde{h}_0$).  There is also the presence of an extra term in the off-diagonal elements. This difference arises from the fact that when there is a central part of the chain, the $\chrl$ are not in the cyclic subspace for $h_{l/r}$ and $\delta_{l/r}$ whereas when there is no central part, the corresponding vectors $\delta_{1, 0}$ are precisely the cyclic vectors for the other part of the chain.

 The exact same methodology outlined in the proofs of Proposition \ref{prop:waveoperators} and Theorem \ref{thm:scat} will yield the above formulas. Let us describe where the difference arises in the derivation.

Suppose that one repeats the proof of Theorem \ref{thm:scat} for $h$ and $\widetilde{h}_0$. When one gets to (\ref{eqn:dif1}) and substitutes for the terms $\langle w_-^* ( h, \widetilde{h}_0 ) \delta_{0/1}, e^{- \i s \widetilde{h}_0 } g \rangle$  (these are the terms that take the place of the $\langle w_-^* \chlr , e^{ - \i s h_0} g \rangle$ which appear in the original derivation) one will first see a difference between the two formulas. The proof of Proposition \ref{prop:waveoperators} applied to $h$ and $\widetilde{h}_0$ yields
\begin{align}
w_-^* ( h, \widetilde{h}_0 ) \delta_{0/1} = [\delta_{0,0/1} - J_0 \langle \delta_0, (h - E + \i 0)^{-1} \delta_{0/1} \rangle ] \oplus [ \delta_{1,0/1} - J_0 \langle \delta_1, (h - E + \i 0)^{-1} \delta_{0/1} \rangle ] \label{eqn:dif2}
\end{align}
where $\delta_{a, b}$ is 1 if $a= b$ and $0$ otherwise. Compare this to
\begin{align}
w^* \chlr = [ - J_l \langle \chl , ( h - E + \i 0)^{-1} \chlr \rangle ] \oplus [ - J_r \langle \chr , ( h - E + \i 0)^{-1} \rangle ].
\end{align}
The extra $\delta_{0, 0/1}$ and $\delta_{1, 0/1}$ appearing in (\ref{eqn:dif2}) lead to the slightly different formulas.

We also note that the main results of this paper (i.e., Theorems \ref{thm:ness} and \ref{thm:main}) remain valid in this case, and the proofs carry over with little change.
\begin{comment}
\begin{align}
\langle w_-^* ( h, \widetilde{h}_0 ) \delta_{0/1} , e^{ - \i s h_0 } g \rangle &= \int_\R \d \mu_{0, 1, \ac} (E') g_{l/r} (E') e^{ - \i s E'} ( 1 - J_0 \langle \delta_{0/1} , (h - E' - \i 0)^{-1} \delta_{1/0} \rangle \nonumber \\
&- \int_\R \d \mu_{1/0, \ac} (E') g_{r/l} (E') e^{ - \i s E'} J_0 \langle \delta_{0/1} , (h - E' - \i 0)^{-1} \delta_{0/1} \rangle,
\end{align}
where $\mu_{0/1, \ac}$ is the spectral measure for $\widetilde{h}_{l/r}$ and $\delta_{0/1}$. Compare this to
\begin{align}
\langle w_- ^*  \chlr , e^{-\i s h_0 } g \rangle = 
\end{align}
\end{comment}

\subsection{Reflectionless Jacobi matrices} \label{sec:refl}

We will call a Jacobi matrix $h$ \emph{reflectionless} if the scattering matrix $s(E)$ is off-diagonal for Lebesgue a.e. $E \in \E$. A priori, this definition depends on the choice of $N$ in the formula for $s(E)$, but we will see in this section that $s(E)$ is off-diagonal for Lebesgue a.e. $E \in \E$ for some $N$ $\iff$ $s(E)$ is off-diagonal for Lebesgue a.e. $E \in \E$ for all $N$.

The word reflectionless comes from the fact that the reflection coefficients $ | s_{ll} (E) |^2$ and $ |s_{rr} (E) |^2$ vanish if $s(E)$ is off-diagonal. The reflection coefficients have the interpretation of describing the probability that a wave packet coming in from the left/right in the distant past is reflected and exits via the left/right in the distant future. There is a huge literature devoted to reflectionless Jacobi matrices and we mention here only \cite{BRS}, \cite{R} and \cite{T} and the references therein (the paper \cite{BRS} contains a substantial list). 

There are several equivalent definitions of reflectionless appearing in the literature, and in the remainder of this section we will discuss their relation to ours. Parts of our discussion will follow \cite{BRS}. First we will require some notation. We denote the elements of the Green's function for $z \in \C \backslash \R$,
\begin{align}
g_{nm} (z) = \langle \delta_n, ( h - z ) \delta_m \rangle.
\end{align}
The limits,
\begin{align}
g_{nm} (E \pm \i 0) = \lim_{ \eps \dto 0 } \langle \delta_n, ( h - E \mp \i \eps)^{-1} \delta_m \rangle \label{eqn:refl1}
\end{align}
exist for Lebesgue a.e. $E \in \R$. We denote by $h_n ^+$ the operator $h \rest_{ \ell^2 ( [ n+1 , \infty [ )}$ and by $h_n ^- $ the operator $h \rest_{\ell^2 (  ] \infty, n -1 ] )}$. The Weyl $m$-functions are
\begin{align}
m_n ^\pm (z) = \langle \delta_{n \pm 1} , (h_n ^\pm - z )^{-1} \delta_{n \pm 1} \rangle
\end{align}
for $z \in \C \backslash \R$, and the functions $m_n ^\pm (E + \i 0)$ and $m_n ^\pm (E - \i 0)$ are defined as in (\ref{eqn:refl1}). In what follows we will sometimes adopt the shorthand $g_{nm} (E) = g_{nm} (E + \i 0)$, $\widebar{g}_{nm} (E) = g_{nm} (E - \i 0)$ to denote the limits whenever they exist, and the same for the $m_n ^\pm$. We have the formula \cite{T} (see also \cite{BRS}, although their notation for the $m$ functons are different than ours),
\begin{align}
g_{nn} (z) = \frac{ -1 } { J_n ^2 m_n ^+ (z) - m_{n+1} ^{-} (z)^{-1}} = \frac{-1}{J_{n-1}^2 m_{n} ^- (z) - m_{n-1} ^+ (z) ^{-1}}. \label{eqn:refl4}
\end{align}
If $\A \subset \R$ is Borel, then $h$ is called \emph{measure theoretically reflectionless} (see \cite{BRS}) on $\A$ if for Lebesgue a.e. $E \in \A$ and every $n$,
\begin{align}
\re \left[ g_{nn} (E + \i 0) \right] = 0. \label{eqn:refl2}
\end{align} 
Similarly, for $\A$ Borel, a Jacobi matrix $h$ is called \emph{spectrally reflectionless} on $\A$ if for Lebesgue a.e. $E \in \A$ and every $n$,
\begin{align}
J_n^2 m_n^+ ( E + \i 0) \widebar{m_{n+1}^- } (E + \i 0) =1. \label{eqn:refl3}
\end{align}
Direct computation using (\ref{eqn:refl4}) shows that if (\ref{eqn:refl3}) holds for $n$ and $E$, then  (\ref{eqn:refl2}) holds for  $n$ and $E$, and so,
\begin{align}
h\mbox{ is spectrally reflectionless on } \A \implies h \mbox{ is measure theoretically reflectionless on } \A .
\end{align}
It is well known that (see, e.g., \cite{GKT, SY, T} and Theorem 7.4.1 of \cite{Si} for what is probably the most readable proof) that
\begin{align}
(\ref{eqn:refl2}) \mbox{ for } E \mbox{ and three consecutive } n \implies (\ref{eqn:refl3}) \mbox{ for } E \mbox{ and one } n.
\end{align}
It is also true that \cite{BRS}
\begin{align}
(\ref{eqn:refl3}) \mbox{ for } E \mbox{ and one } n  \implies (\ref{eqn:refl3}) \mbox{ for } E \mbox{ and all } n ,
\end{align}
and so $h$ is spectrally reflectionless on $\A$ iff it is measure theoretically reflectionless on $\A$.

Returning to our scattering matrix,  using (\ref{eqn:refl4}) we find that
\begin{gather}
s_{rr} (E) = 1 + 2 \i J_N^2 g_{NN} ( E  ) \im m^+_N (E ) = \frac{ J_N ^2 \widebar{m_N^+} (E) m_{N+1} ^- (E) - 1 }{ J_N ^2 m_{N+1} ^+ (E) m_N ^- (E) - 1 }, \nonumber \\
s_{ll} (E) = 1 + 2 \i J_{- N -1 }^2 g_{-N -N} (E) \im m^-_{-N} (E) = \frac{J_{-N -1 } ^2 \widebar{ m_{-N} ^- } (E) m_{-N -1}^+ -1}{J_{-N -1}^2 m_{-N} ^- m_{-N -1 } ^+ -1 }. \label{eqn:refl5}
\end{gather}
It follows that for Lebesgue a.e. $E \in \Sigma_{r, \ac}$,
\begin{align}
s_{rr} (E) = 0 \iff J_N ^2 \widebar{m_N^+} (E) m_{N+1} ^- (E) = 1
\end{align}
and for Lebegue a.e. $E \in \Sigma_{l, \ac}$.
\begin{align}
s_{ll} (E) = 0 \iff J_{-N -1 } ^2 \widebar{ m_{-N} ^- } (E) m_{-N -1}^+  = 1
\end{align}
It therefore follows by the above discussion that,
\begin{align}
h \mbox{ is reflectionless } \iff h \mbox{ is measure theoretically reflectionless on } \E \iff h \mbox{ is spectrally reflectionless on } \E.
\end{align}
Moreover, since, by the Kato-Rosenblum theorem, $\Sigma_{l/r, \ac}$ does not depend on the choice of $N$ it follows that whether or not $h$ is reflectionless does not depend on the choice of $N$. Additionally, analoguous computations to those in (\ref{eqn:refl5}) hold in the case of two directly coupled chains, and so the discussion extends to this case (in particular, our definition of reflectionless does not even depend on whether or not there is a central system).

\subsection{Non-equilibrium steady state}

From here on we assume that $h$ has purely absolutely continuous spectrum. Concerning the existence of a non-equilibrium steady state, we have \cite{AP},
\begin{theorem} \label{thm:ness}
If $h$ has purely absolutely continuous spectrum, then for any $A \in \O$ the limit
\begin{align}
\langle A \rangle_+ = \lim_{t \to \infty} \omega ( \tau^t (A) )  \label{eqn:ness}
\end{align}
exists. The state $\omega_+ ( \cdot ) = \langle \cdot \rangle_+$ is called the non-equilibrium steady state (NESS) of the quantum dynamical system $ ( \O , \tau^t , \omega )$. The steady state heat fluxes are
\begin{align}
\langle \Phi_l \rangle_+ = - \langle \Phi_r \rangle _+ = \frac{1}{4 \pi } \int_\E E | s_{lr} (E) |^2 \frac{ \sinh ( \Delta \beta E /2 ) } { \cosh ( \beta_r E /2 ) \cosh ( \beta_l E /2 ) } \d E , \label{eqn:flux}
\end{align}
and the steady state entropy production is
\begin{align}
\langle \sigma \rangle_+ = - \beta_l \langle \Phi_l \rangle_+ - \beta_r \langle \Phi_r \rangle_+ = \Delta \beta \langle \Phi_l \rangle_+ ,
\end{align}
where $ \Delta \beta = \beta_r - \beta_l$.
\end{theorem}

\proof First suppose $A \in \O_{\loc}$. The existence of the limit (\ref{eqn:ness}) is the assertion that the limit
\begin{align}
\lim_{t \to \infty} \lim_{M' \to\infty} \lim_{M \to \infty} \omega_M (\tau_{M'} ^t (A ) ) 
\end{align}
exists. By linearity, we may assume that $A = a^* (\phi_n ) ... a^* (\phi_1) a (\psi_1) ... a (\psi_n).$ It follows from Proposition \ref{prop:quasifree} that
\begin{align}
\lim_{M' \to\infty} \lim_{M \to \infty} \omega_M (\tau_{M'} ^t (A ) )  = \det \left[ \langle e^{  \i t h} \psi_i , T e^{ \i t h} \phi_j \rangle  \right]
\end{align}
with $T = (1 + e^{ \beta_l h_l + \beta_r h_r })^{-1}$. Since $h_0$ commutes with $T$ and $\fh_{\ac} (h) = \fh$,
\begin{align}
 \lim_{t \to \infty} \langle e^{ \i t h} \psi, T e^{ \i t h} \phi \rangle &= \lim_{t \to \infty}  \langle e^{ - \i t h_0} e^{ \i t h} \psi, T e^{ - \i t h_0} e^{ \i t h} \phi \rangle \nonumber \\
&= \lim_{t \to \infty}  \langle e^{ - \i t h_0} e^{ \i t h} \1_{\ac} (h) \psi, T e^{ - \i t h_0} e^{ \i t h} \1_{\ac} (h) \phi \rangle   \nonumber \\
&= \langle w_-^* \psi, T w_-^* \phi \rangle
\end{align}
since $h_0$ commutes with $T$. This proves the existence of (\ref{eqn:ness}) for $\O_{\loc}$. An $\eps /3$ argument extends the result to $\O$.

We compute $\langle \Phi_l \rangle_+$. Our computations above have proven the formula
\begin{align}
\omega_+ ( a^*(\psi) a( \phi) ) = \langle w_-^* \psi, T w_-^* \phi \rangle.
\end{align}
Therefore, (\ref{eqn:fluxleft}) implies
\begin{align}
\langle \Phi_l \rangle_+ = 2  J_{l} \im \langle w_-^* h_0 \delta_l , T w_-^* \chl \rangle .
\end{align}
\begin{comment}
Applying Proposition \ref{prop:waveoperators},
\begin{align}
w_-^* h_0 \delta_l &= ( E + J_l ^2 \langle \chl, (h - E + \i 0)^{-1} \chl \rangle + E J_l ^2 \langle \chl , (h - E + \i 0)^{-1} \chl \rangle \langle \delta_l , (h_0 - E + \i 0)^{-1} \delta_l \rangle ) \nonumber \\
& \oplus ( J_r J_l \langle \chr , (h - E + \i 0)^{-1} \chl \rangle + E J_r J_l \langle \chr , (h - E + \i 0)^{-1} \chl \rangle \langle \delta_l (h_0 - E + \i 0)^{-1} \delta_l \rangle )    \nonumber \\
%w_-^* h_0 \delta_l &= ( E - J_l \langle \chl, (h - E + \i 0)^ {-1} h_0 \delta_l \rangle ) \oplus ( -J_r  \langle \chr , ( h - E + \i 0 )^{-1} %h_0 \delta_l \rangle ), \nonumber \\
w_-^* \chl &= ( - J_l \langle \chl , (h - E + \i 0)^{-1} \chl \rangle ) \oplus ( -J_r \langle \chr , (h - E + \i 0)^{-1} \chl \rangle ).
\end{align}
where we have used the identity
\begin{align}
\langle & \chlr, ( h - E + \i 0)^{-1} h_0 \delta_l \rangle \nonumber \\
&= - J_l  \langle \chlr, (h - E + \i 0)^{-1} \chl \rangle - E J_l \langle \chlr , (h - E + \i 0)^{-1} \chl \rangle  \langle \delta_l , (h_0 - E + \i 0)^{-1} \delta_l \rangle .
\end{align}
\end{comment}
Let
\begin{align}
H_{ab} (E) = \langle \cha , (h - E - \i 0)^{-1} \chb \rangle .\label{eqn:hab}
\end{align}
Direct computation using the formulas in Corollary \ref{cor:waveform} yields,
\begin{align}
2  J_{l} \im \langle w_-^* h_0 \delta_l , T w_-^* \chl \rangle &= \int_\R \d \nu_{l, \ac}  (E) \frac{2 E }{1 + e^{\beta_l E} } \left( J_l ^2 \im \left[ H_{ll} (E) \right] - J_l ^4 | H_{ll} (E) |^2 F_l (E) \right) \nonumber \\
&-  \int_\R \d \nu _{r, \ac} (E) \frac{2 E } {1 + e^{ \beta_r E} } J_r^2 J_l ^2 | H_{lr} (E) |^2 F_l (E) \nonumber \\
&= \int_\R \frac{\d E}{2 \pi } E \left( ( 1 - | s_{ll} (E) |^2 ) \frac{1}{1 + e^{\beta_l E} } -  | s_{lr} (E) |^2 \frac{1}{1 + e^{ \beta_r E } } \right).
\end{align}
The formula in question then follows from the identity $1 = |s_{ll} (E) |^2 + | s_{lr} (E) |^2$ which is a consequence of the unitarity of the scattering matrix. The computation of $\langle \Phi_r \rangle_+$ is similar. \qed

If the Lebesgue measure of $\E$, denoted $| \E |$, is $0$ then obviously there is no energy transfer between the left and right parts of the chain. If $| \E | > 0$, then since the scattering matrix is not diagonal for Lebesgue a.e. $E \in \E$, it follows from (\ref{eqn:flux}) that $\langle \sigma  \rangle_+> 0$ iff $ \beta_l \neq \beta_r$. That is, the steady state entropy production is strictly positive if initially the left and right parts of the chain are at different temperature.

In order for results similar to those of  (\ref{thm:ness}) to hold (i.e., existence of NESS, strict positivity of entropy production), it was necessary to take the thermodynamic limit before taking the large time limit (see Section 5.1 in \cite{JOPP}, and \cite{L} for the solution to Exercise 5.1 appearing there; we repeat some of the discussion in \cite{JOPP} here). More precisely, if $( \O ' , {\tau '}^t  , \omega ')$ is a finite dimensional quantum system, then the limit
\begin{align}
\lim_{t \to \infty} \omega' \left( {\tau ' }^t (A) \right)
\end{align}
does not exist except in trivial cases. However, the Ces\`aro limit
\begin{align}
\omega ' _+ (A) = \lim_{T \to \infty} \frac{1}{T} \int_0 ^T \omega' \left( { \tau ' } ^t (A) \right) \d t \label{eqn:ces}
\end{align}
exists for every observable $A \in \O$,  and $\omega_+$ is a steady state of the system. It is easy to see that
\begin{align}
\omega_+ \left( \i [H', A ] \right) = 0 
\end{align}
for every observable $A \in \O$, where $H'$ is the Hamiltonian of the system. From the above we conclude that it was necessary to take the thermodynamic limit of the XY chain before the large time limit. We also comment that the above discussion applies to infinite dimensional systems that are  `confined' - that is, $H'$ has pure point spectrum.

From this discussion it is reasonable to expect that we can extend Theorem \ref{thm:ness} in the case that $h$ has, in addition to its absolutey continuous spectrum, some pure point spectrum. Of course, the limit in (\ref{eqn:ness}) must be replaced with the Ces\`aro limit (\ref{eqn:ces}) and then the formula (\ref{eqn:flux}) holds \cite{AJPP}. The extension of our proof to this case is easy and requires only the Riemann-Lebesgue lemma.

We now complete a computation which will be useful later. Let
\begin{align}
k_0 (E) = \left( \begin{matrix} - \beta_l E& 0\\ 0  & - \beta_r E\end{matrix}\right) \label{eqn:knot}.
\end{align}

\begin{proposition}\label{prop:main}
Recall the definition of $V$ in (\ref{eqn:vee}). Let $\T$ be a bounded operator on $V \fh_{\ac} (h)$ that acts by multiplication by a $2 \times 2 $ matrix
\begin{align}
\T (E) =  \left( \begin{matrix}\T_{ll} (E)& \T_{lr} (E)\\ \T_{rl} (E)  &\T_{rr} (E)\end{matrix}\right) .
\end{align}
Then,
\begin{align}
\tr \left( V^{-1} \T V w_- ^* \i [k, h] w_- \right) = - \int_\E \tr_{\C ^2} \left( \T (E) ( s^* (E) k_0 (E) s (E) - k_0 (E) ) \right) \frac{ \d E}{2 \pi } .\label{eqn:entropicscattering}
\end{align}
\end{proposition}

%%%\remark The RHS of (\ref{eqn:entropicscattering}) must be interpreted appropriately. Specifically, the trace is with respect to %%the standard inner product on $\C^2$. Therefore, the same transformation that was applied to the scattering matrix to obtain the %%formula (\ref{eqn:unitscat}) (i.e., multiplying or dividing the off-diagonal elements by an appropriate factor) must also be applied %%%to $\T (E)$. This will be clear in the proof.

\proof Recalling that $[k , h] = -\beta_l [h_l , v_l] - \beta_r [ h_r, v_r ]$, and using cyclicity of the trace we have ,
\begin{align}
\tr ( V^{-1} \T V w_- ^* \i [k , h] w_- ) &= \i\beta_l J_l ( \langle \delta_l , h_l w_- V^{-1} \T V w_-^* \chl \rangle - \langle \chl , w_- V^{-1} \T V w_-^* h_l \delta_l \rangle ) \nonumber \\
&+ \i \beta_r J_r ( \langle \delta_r , h_r w_- V^{-1} \T V w_-^* \chr \rangle - \langle \chr , w_- V^{-1} \T V w_- ^* h_r \delta_r \rangle ) \nonumber \\
&= \i\beta_l J_l ( \langle w_- ^* h_l \delta_l , V^{-1} \T V w_-^* \chl \rangle - \langle w_- ^* \chl ,V^{-1} \T V w_-^* h_l \delta_l \rangle ) \nonumber \\
&+ \i \beta_r J_r ( \langle w_-^* h_r \delta_r ,  V^{-1} \T V w_-^* \chr \rangle - \langle w_-^* \chr ,  V^{-1} \T V w_- ^* h_r \delta_r \rangle ) .\label{eqn:ent1}
\end{align}
We would like to compute these four terms. Let $H_{ab} (E)$ be as in (\ref{eqn:hab}). Note that $V^{-1} \T V$ acts on $\fh_{\ac} (h_0)$ as multiplication by the $2 \times 2$ matrix $V^{-1} (E) \T (E) V (E)$. Using this and the formulas in Corollary \ref{cor:waveform} to directly compute the inner products appearing above yields,
\begin{align}
&\tr ( V^{-1} \T V w_- ^* \i [k , h] w_- ) \nonumber \\
&= \int_\R \T _{ll} (E)  E \bigg\{ \beta_l  \left( 2 J_l^4 | H_{ll} (E) |^2 F_l (E)  - 2 J_l^2 \im \left[ H_{ll} (E) \right] \right) +  \beta_r 2 J_r ^2 J_l ^2 | H_{rl} (E) |^2 F_r (E) \bigg\} \d \nu_{l, \ac} (E)  \nonumber \\
&+ \int_\R  \T_{lr} (E) \sqrt{\frac{F_r (E) }{F_l (E) } } E \bigg\{ \beta_l  \left( 2 J_r J_l ^3 \widebar{H}_{lr} (E) H_{ll} (E) F_l (E) - J_r J_l \i \widebar{H}_{lr} (E) \right) \nonumber \\
& + \beta_r   \left(   2 J_l J_r^3 H_{rl} (E) \widebar{H}_{rr} (E) F_r (E)  + \i J_l J_r H_{rl} (E) \right) \bigg\} \d \nu_{l, \ac} (E) \nonumber \\
&+ \int_\R \T _{rr} (E)  E \bigg\{ \beta_r  \left( 2 J_r^4 | H_{rr} (E) |^2 F_r (E)  - 2 J_r^2 \im \left[ H_{rr} (E) \right] \right) +  \beta_l 2 J_l ^2 J_r ^2 | H_{lr} (E) |^2 F_l (E) \bigg\} \d \nu_{r, \ac} (E)  \nonumber \\
&+ \int_\R  \T_{rl} (E) \sqrt{\frac{F_l (E) }{F_r (E) } } E \bigg\{ \beta_r  \left( 2 J_l J_r ^3 \widebar{H}_{rl} (E) H_{rr} (E) F_r (E) - J_l J_r \i \widebar{H}_{rl} (E) \right) \nonumber \\
& + \beta_l   \left(   2 J_r J_l^3 H_{lr} (E) \widebar{H}_{ll} (E) F_l (E)  + \i J_r J_l H_{lr} (E) \right) \bigg\} \d \nu_{r, \ac} (E) \nonumber \\
&= \int_\R  E \bigg\{  \T_{ll} (E)  \left( \beta_l ( | s_{ll} (E) |^2 -1 ) + \beta_r  | s_{lr} (E) | ^2 \right) + \T_{lr} (E)  \left( \beta_l s_{ll} (E) \widebar{s}_{lr} (E) + \beta_r \widebar{s}_{rr}(E) s_{rl} (E) \right) \nonumber \\ 
&+ \T_{rr} (E) \left( \beta_r ( | s_{rr} (E) |^2 -1 ) + \beta_l | s_{lr} (E) |^2   \right) + \T_{rl} (E) \left(  \beta_r s_{rr} (E) \widebar{s}_{rl} (E) + \beta_l s_{lr} (E) \widebar{s}_{ll} (E)  \right) \bigg\} \frac{\d E}{2 \pi } ,
\end{align}
%%%\endgroup
which is easily seen to be the formula in question. \qed

An identical computation gives, using the symmetry of the scattering matrix,
\begin{corollary} For $\T$ as above, \label{cor:timereversed}
\begin{align}
\tr \left( V^{-1} \T V w_+ ^*  \i [k, h] w_+ \right) = - \int_\E \tr_{\C^2} \left( \T (E) ( s (E) k_0 (E) s^* (E) - k_0 (E) ) \right) \frac{\d E}{2 \pi}.
\end{align}
\end{corollary}

\subsection{The Gallavotti-Cohen functional}
The Gallavotti-Cohen functional is defined by
\begin{align}
\GC _t (\alpha) = \log \omega_+ ( e^{ - \alpha t \Sigma^t } )
\end{align}
and describes fluctuations of the mean entropy production rate $\Sigma^t$ with respect to the NESS $\omega_+$.  It is the direct quantization of the Gallavotti-Cohen functional of classical nonequilibrium statistical mechanics \cite{JPR}. Note that
\begin{align}
\GC_t (\alpha ) = \lim_{s \to \infty} \log \omega_s ( e^{ -  \alpha t \Sigma^t } ) = \lim_{s \to \infty} \lim_{M' \to \infty} \lim_{M \to \infty}\log  \omega_{s, M} ( e^{ - \alpha t \Sigma^t_{M'}} )= \lim_{s \to \infty} \lim_{M \to \infty} \log \omega_{s, M} ( e^{ - \alpha t \Sigma^t_M } ).
\end{align}
Note also that the formula
\begin{align}
\omega_{s, M} ( e^{ - \alpha t \Sigma^t_M } ) = \frac{ \det ( \1 + e^{ k_{-s, M} /2 } e^{ \alpha ( k_{t, M} - k_M ) }e^{ k_{-s, M}/2} ) }{\det ( \1 + e^{k_M} ) }
\end{align}
holds. This allows us to apply the arguments in the proofs of Lemma \ref{lem:entropyformulae} and Proposition \ref{prop:thermfunct} to obtain,
\begin{align}
\lim_{M \to \infty} \log \omega_{s, M} ( e^{- \alpha t \Sigma^t_M } ) = -t \int_0 ^\alpha \d \gamma \int_0 ^1 \d u \tr \left( \left( \1 + e^{ - \gamma ( k_{t (1 - u)} - k_{- t u } )} e^{ - k_{- (s + tu) } } \right)^{-1} \i [ k , h] \right) .
\end{align}
Since $k$ is bounded, commutes with $h_0$ and $ \ran ( w^*_\pm) \subseteq \fh_{\ac} (h_0)$ the strong limit
\begin{align}
\slim_{s \to \pm \infty} k_s = \slim_{s \to \pm \infty} e^{ \i s h} e^{ - \i s h_0} k e^{ \i s h_0} e^{ - \i s h} = w_{\pm} k w_{\pm}^* = k_\pm
\end{align}
exists. An application of dominated convergence yields,
\begin{align}
\GC _t (\alpha ) = t \int_0 ^\alpha \d \gamma \int_0 ^1 \d u \tr \left( \K _{\GC , t } ( \gamma , u ) \i [k, h] \right), \label{eqn:gcf}
\end{align}
with
\begin{align}
\K_{\GC, t} ( \alpha, u) = - \left( \1 + e^{ - \alpha ( k_{t (1 - u) } - k_{-tu } ) } e^{ - k_- } \right)^{-1} .
\end{align}

\subsection{The entropic functionals in the large time limit}

Recall the definition of $k_0 (E)$ in (\ref{eqn:knot}). Let
\begin{align}
K_\alpha (E) = e^{ k_0 (E) /2} e^{ \alpha ( s^* (E) k_0 (E) s (E) - k_0 (E) ) } e^{ k_0 (E) /2 } , \label{eqn:wknd4}
\end{align}
and 
\begin{align}
K_{\alpha, p} (E) = \left( e^{ k_0 (E) (1 - \alpha ) /p } s(E) e^{ k_0 (E) 2 \alpha / p } s^* (E) e^{ k_0 (E) (1 - \alpha ) /p } \right) ^{p/2} , \label{eqn:wkdn5}
\end{align}
for $ p \in ]0, \infty ]$ and let
\begin{align}
K_{\alpha, \infty} (E) = \lim_{p\to \infty} K_{\alpha, p} (E) = e^{ (1 - \alpha ) k_0 (E) + \alpha s(E) k_0 (E) s^* (E) } ,\label{eqn:wknd6}
\end{align}
where the formula for the limit follows from Corollary 2.3 in \cite{JOPP}. 

\begin{comment}
For $E \in \R$ and $\alpha \in \R \backslash \{ 0 \}$,
\begin{align}
K_\alpha (E) = K_0 (E) \iff e^{ \alpha ( s^* (E) k_0 (E) s (E) - k_0 (E) ) }  = \1 \iff [s(E), k_0 (E) ],
\end{align}
and for $p \in ]0, \infty [$,
\begin{align}
K_{\alpha, p} = K_0 (E) \iff e^{ s(E) k_0 (E) s^* (E) 2 \alpha / p } = e^{ k_0 (E) 2 \alpha /p } \iff [ s(E), k_0 (E) ].
\end{align}
Similarly, for $E \in \R$ and $\alpha \in \R \backslash \{ 0 \}$
\begin{align}
K_{\alpha, \infty} (E) = K_0 (E) \iff e^{ (1 - \alpha ) k_0 (E) + \alpha s (E) k_0 (E) s^* (E) } = e^{ k_0 (E) } \iff [ s(E), k_0 (E) ].
\end{align}
\end{comment}
Since,
\begin{align}
[s (E), k_0 (E) ] =  \left( \begin{matrix}s_{ll} (E) \beta_l E & s_{lr} (E) \beta_l E \\ s_{rl} (E) \beta_r E  &s_{rr} (E) \beta_r E \end{matrix}\right) -  \left( \begin{matrix}s_{ll} (E) \beta_l E & s_{lr} (E) \beta_r E \\ s_{rl} (E) \beta_l E  &s_{rr} (E) \beta_r E \end{matrix}\right) ,
\end{align}
$[s (E) , k_0 (E) ] = 0$ iff $\beta_l = \beta_r$ or $E = 0$ or $s(E)$ is diagonal. We recall that $s(E)$ is not diagonal for Lebesgue a.e. $E \in \E$.
\begin{comment}We have therefore proven,
\begin{lemma} \label{lem:iff}
For $E, \alpha \in \R \backslash \{ 0 \}$, and $p \in ]0, \infty ]$,
\begin{align}
K_{\alpha} (E) = K_0 (E) \iff K_{\alpha, p} (E) = K_0 (E) \iff [s(E), k_0 (E) ] = 0 \iff  \beta_l = \beta_r \mbox{ or } s(E) \mbox{ is diagonal.}
\end{align}
\end{lemma}
\end{comment}

The main result concerning the large time limit of the entropic functionals is,
\begin{theorem}\label{thm:main}
Suppose that $h$ has purely absolutely continuous spectrum. Then the following holds:
 \begin{enumerate}[label=(\roman{*}), ref=(\roman{*}),font=\normalfont]
\item For $\alpha\in\R$ and $p\in ]0, \infty]$, 
\[
e_{p,+}(\alpha)=\lim_{t\rightarrow\infty}\frac{1}{t}e_{p,t}(\alpha)
=\int_{\cal E}\log\left(\frac{\det(1+ K_{\alpha,p}(E))}{\det(1+K_{0,p}(E))}\right)\frac{\d E}{2\pi}, 
\]
\[
\lim_{t\rightarrow\infty}\frac{1}{t}\ES_t(\alpha)
=\lim_{t\rightarrow\infty}\frac{1}{t}\GC_t(\alpha)
=e_{+}(\alpha)
=\int_{\cal E}\log\left(\frac{\det (1+K_{\alpha}(E))}{\det (1+ K_0(E))}\right)\frac{\d E}{2\pi}.
\]
These functionals are identically zero iff $|{\cal E}|=0$ or $\beta_l=\beta_r$. In what follows we
assume that $|{\cal E}|>0$ and $\beta_l\not=\beta_r$. 
\item The function $\R\ni\alpha\mapsto e_{p,+}(\alpha)$ is real analytic and strictly convex.
Moreover, $e_{p,+}(0)=0$, $e_{p,+}^\prime(0)=-\langle\sigma\rangle_+$, and
\begin{equation*}
e_{p,+}(\alpha)=e_{p,+}(1-\alpha).
\end{equation*}
\item The function $\R\ni\alpha\mapsto e_+(\alpha)$ is real-analytic and strictly convex.
Moreover, it satisfies $e_+(0)=0$, $e_{+}^\prime(0)=-\langle \sigma \rangle_+$, and 
\begin{equation}
e_{+}^{\prime\prime}(0)=e_{2,+}^{\prime\prime}(0)
=\lim_{T\rightarrow\infty}\frac{1}{T}\int_0^T\left\{
\frac12\int_{-t}^t\left<(\sigma_s-\langle\sigma\rangle_+)(\sigma-\langle\sigma\rangle_+)
\right>_+\d s\right\}\d t.
\label{sat-ti}
\end{equation}
\item 
$e_+(1)>0$ unless $h$ is reflectionless.  If $h$ is reflectionless then 
\[
e_{+}(\alpha)=\int_{\cal E}\log\left(\frac{\cosh((\beta_l(1-\alpha)+\beta_r\alpha)E/2)
\cosh((\beta_r(1-\alpha)+\beta_l\alpha)E/2)}{\cosh(\beta_l E/2)\cosh (\beta_r E/2)}\right)\frac{\d E}{2\pi},
\]
and  $e_+(\alpha)=e_+(1-\alpha)$.
\item The function $]0,\infty]\ni p\mapsto e_{p,+}(\alpha)$ is continuous and decreasing. It is
strictly decreasing for $\alpha\not\in\{0,1\}$ unless $h$ is reflectionless. If $h$ is reflectionless, 
then $e_{p,+}(\alpha)$ does not depend on $p$ and is equal to  $e_+(\alpha)$. 
\end{enumerate}
\end{theorem}
%Remarks??

\remarkk The time-reversal invariance of our system together with Corollary \ref{cor:timereversed} implies that we can switch $s(E)$ and $s^*(E)$ whenever they appear in the equations (\ref{eqn:wknd4})-(\ref{eqn:wknd6}) when inserted into the formulas in part (i). This is can also be seen directly by conjugation because $\widebar{s} (E) = s^* (E)$.
%%%\end{remark}

\proof We begin with part (i). Note the existence of the strong limits
\begin{align}
\K_+ (\alpha) = \slim_{t \to \infty} \K_{\ES , t} (\alpha, u) = \slim_{t \to \infty} \K_{\GC , t } (\alpha, u ) = - \left( \1 + e^{ - \alpha ( k_+ - k_- ) }e^{- k_-} \right)^{-1},
\end{align}
and, for $p \in ]0, \infty [$,
\begin{align}
\K_{p, + } (\alpha) &= \slim_{t \to \infty} \K_{p, t} (\alpha, u) \nonumber \\
&= \frac{1}{2}e^{- (1 - \alpha) k_+ /p } \left( \1 + ( e ^{(1 - \alpha ) k_+ / p } e^{ 2 \alpha k_- / p } e^{ ( 1 - \alpha ) k_+ / p } )^{-p/2} \right)^{-1} e^{(1 - \alpha) k_+ /p} + \mbox{h.c.} ,
\end{align}
and
\begin{align}
\K_{\infty, +} (\alpha) = \slim_{t \to \infty}\K_{\infty , t} (\alpha, u) = \left( \1 + e^{ - ( 1 - \alpha) k_+ - \alpha k_- } \right)^{-1}.
\end{align}
This follows immediately from the definitions of the $\K_{\#, t}$ and $k_\pm$, where $\#$ stands for $\ES, \GC$ or $p \in ]0, \infty]$. By (\ref{eqn:thermfunct}) and (\ref{eqn:gcf}),
\begin{align} \label{eqn:mainfinite}
\frac{1}{t} e_{\#, t} (\alpha) = \int_0 ^\alpha \d \gamma \int_0^1 \tr \left( \K_{\#, t} ( \gamma, u ) \i [k, h] \right) =  \sum_{k \in \A} \int_0 ^\alpha \d \gamma \int_0 ^1 \d u \langle \delta_k ,  \K _{\#, t} ( \gamma, u ) \i [k, h] \delta_k \rangle ,
\end{align}
where $e_{\GC, t} (\alpha ) := \GC_t (\alpha)$ and $\A = [ - N - 2 , - N +1 ] \cup [ N-1 , N+2 ] \subseteq \Z$. The aforementioned existence of the strong limits and dominated convergence implies,
\begin{align}
\lim_{t \to \infty} \frac{1}{t} e_{\#, t} (\alpha) &=  \sum_{k \in \A} \int_0^\alpha  \langle \delta_k, \K_{\#, +} ( \gamma ) \i [k, h] \delta_k \rangle \nonumber \\
&= \int_0^\alpha \d \gamma \tr ( \K_{\#, +} ( \gamma ) \i [k, h] ).
\end{align}
In particular,
%follows immediately from the definitions of the $\K_{\ES / \GC , t}$ and $k_\pm$. We then have
\begin{align}
e_+ (\alpha ) = \lim_{t \to \infty} \frac{1}{t} \ES _t (\alpha) = \lim_{t \to \infty} \frac{1}{t} \GC_t (\alpha) = \int_0^\alpha \tr ( \K_+ (\gamma ) \i [k, h] ) \d \gamma .
\end{align}

By the definition of the scattering matrix and Proposition \ref{prop:main},
\begin{align}
\tr &( \K _+ (\gamma) \i [ k, h ] ) = - \tr ( ( \1 + e^{ - \gamma (s^* k s - k ) } e^{ - k} )^{-1} w_- ^* \i [k, h] w_- ) \nonumber \\
&=\int_\E \tr \left( ( \1 + e^{ - \gamma (s^* (E) k_0 (E) s(E) - k_0 (E) )} e^{ - k_0 (E) } )^{-1} ( s^* (E) k_0 (E) s(E ) - k_0 (E) ) \right) \frac{\d E}{2 \pi}.
\end{align}
At this point, we can essentially reverse the computation completed in Lemma \ref{lem:entropyformulae}. By Corollary \ref{cor:derivative},
\begin{align}
\tr&\left(\left(\1 +e^{-\gamma(s^\ast(E)k_0(E)s(E)-k_0(E))}
e^{-k_0(E)}\right)^{-1}\left(s^\ast(E)k_0(E)s(E)-k_0(E)\right)\right)\\[3mm]
&=\frac{\d\ }{\d\gamma}\tr\log \left(\1+e^{k_0(E)/2}
e^{\gamma (s^* (E)k_0(E)s(E)-k_0(E))}e^{k_0(E)/2}\right).
\end{align}
For matrices $A$ and $B$ the equality $ 1 + e^{A} e^{B} = e^{-B/2} ( 1 + e^{ B/2} e^A e^{B/2} ) e^{B/2}$ implies
\begin{align}
\norm{( \1 + e^{ - \gamma (s^* (E) k_0 (E) s(E) - k_0 (E) )} e^{ - k_0 (E) } )^{-1}} \leq \exp \left[ \max_{E \in \E, \mbox{ }a \in \{l, r\} } |\beta_a E| \right] ,
\end{align}
and so, Fubini's theorem yields (recall that $\E$ is bounded),
\begin{align}
e_+ (\alpha) &= \int_0^\alpha \tr ( \K_+ (\gamma) \i [k, h] ) \d \gamma \nonumber \\
&= \int_\E  \left[ \int_0^\alpha \frac{\d}{\d \gamma} \tr\log \left( \1 + e^{ k_0 (E) /2} e^{ \gamma (s^* (E) k_0 (E) s(E) - k_0 (E) ) } e^{ k_0 (E) /2} \right) \d \gamma \right] \frac{\d E}{2\pi} \nonumber \\
&= \int_\E \log \frac{ \det ( \1 + e^{ k_0 (E) /2} e^{ \alpha (s^* (E) k_0 (E) s(E) - k_0 (E) ) } e^{ k_0 (E) /2 } ) } {\det ( 1 + e^{ k_0 (E) } ) } \frac{\d E } {2 \pi} .\label{eqn:main1} ,
\end{align}
which is the formula in question.
For $p \in ]0, \infty ]$ \begin{comment} note that, Proposition \ref{prop:main} yields,
\begin{align}
\K_{p, + } (\alpha) &= \slim_{t \to \infty} \K_{p, t} (\alpha, u) \nonumber \\
&= \frac{1}{2}e^{- (1 - \alpha) k_+ /p } \left( \1 + ( e ^{(1 - \alpha ) k_+ / p } e^{ 2 \alpha k_- / p } e^{ ( 1 - \alpha ) k_+ / p } )^{-p/2} \right)^{-1} e^{(1 - \alpha) k_+ /p} + \mbox{h.c.}
\end{align}
and 
\begin{align}
\K_{\infty, +} = \slim_{t \to \infty}\K_{\infty , t} (\alpha, u) = \left( \1 + e^{ - ( 1 - \alpha) k_+ - \alpha k_- } \right)^{-1}.
\end{align}
The derivation of the claimed formulae then follows the strategy of the derivation of the formula for $e_+ ( \alpha )$. Dominated convergence yields
\begin{align}
e_{p, +} (\alpha ) = \int_0 ^ \alpha \tr ( \K_{p, +}( \gamma)  \i [k, h] \d \gamma ,
\end{align}
and Proposition \ref{prop:main} yields
\end{comment}
\begin{align}
\tr ( \K_{p, +} ( \gamma) \i [k, h] ) = - \int_\E \tr \left( \T_{\gamma ,  p} (E) ( s^* (E) k_0 (E) s(E) - k_0 (E) ) \right) \frac{\d E}{2 \pi}
\end{align}
where, for $p < \infty$,
\begin{align}
\T_{\gamma , p} (E) &= \frac{1}{2}e^{- (1 - \gamma) s^*k_0 s (E)/p } \left( \1 + ( e ^{(1 -\gamma )s^*k_0 s(E)/ p } e^{ 2 \gamma  k_0 (E) / p } e^{ ( 1 - \gamma) s^*k_0 s (E)/ p } )^{-p/2} \right)^{-1} e^{(1 - \gamma) s^*k_0 s (E) /p} \notag \\
& + \mbox{h.c.},
\end{align}
and
\begin{align}
\T_{\gamma ,\infty }(E) = \left( \1 + e^{ - (1 - \gamma ) s^* (E) k_0 (E) s(E) -\gamma k_0 (E) } \right)^{-1},
\end{align}
and $s^*k_0 s (E) = s^*(E) k_0(E) s(E)$. By Corollary \ref{cor:derivative} for $p \in ]0, \infty[$,
\begin{align}
\tr \left( \T_{\gamma, p} (E )  (k_0 (E) - s^* (E) k_0 (E) s (E)  ) \right) &= \frac{ \d}{ \d \gamma} \tr \log \left( 1 + \left(  e^{ (1 - \gamma) s^*k_0 s (E) /p } e^{ 2 \gamma k_0 (E) } e^{ ( 1- \gamma) s^* k_0 s (E) /p } \right)^{p/2}  \right) \notag \\
&=  \frac{ \d}{ \d \gamma} \tr \log \left( 1 + \left(  e^{ (1 - \gamma) k_0  (E) /p } s(E) e^{ 2 \gamma k_0 (E) } s^* (E) e^{ ( 1- \gamma) k_0 (E) /p } \right)^{p/2}  \right) \notag \\
&= \frac{\d}{\d \gamma}  \tr \log \left( \1 +  K_{\gamma, p} (E) \right) ,
\end{align}
where the second equality is just the fact that $s(E)$ is unitary. We have also,
\begin{align}
-  \tr \left( \T_{\gamma, \infty} (E )  (s^* (E) k_0 (E) s (E) - k_0 (E) ) \right) &= \frac{ \d}{ \d \gamma} \log \tr \left( 1 + e^{ - ( 1 - \gamma ) s^^ (E) k_0 (E) s (E) - \gamma k_0 (E) } \right) \nonumber \\
&=  \frac{ \d}{ \d \gamma} \log \tr \left( 1 +  e^{  (1 - \gamma ) k_0 (E) + \gamma s(E) k_0 (E) s^* (E)  } \right) \notag \\
&=  \frac{ \d}{ \d \gamma} \log \tr \left( 1 + K_{\gamma, \infty} ( E) \right).
\end{align}
It is easy to see that, for $p < \infty$,
\begin{align}
\norm{ \T_{\gamma, p} } \leq \exp \left[ | 1- \gamma | p^{-1} \max_{E \in \E, \mbox{ } a \in \{l, r\} } | \beta_a E | \right] ,
\end{align}
and
\begin{align}
\norm{ \T_{\gamma, \infty} } \leq 1 ,
\end{align}
and so Fubini's theorem yields for every $p$,
\begin{align}
e_{p, +} (\alpha) &=  \int_0^\alpha \tr ( \K_{p, +} (\gamma ) \i [ k, h] ) \d \gamma = \int_\E \left[ \int_0 ^\alpha \frac{\d} {\d \gamma }  \tr \log K_{\gamma, p} (E)  \d \gamma \right] \frac{\d E}{2 \pi } = \int_\E \log \frac{ \det ( \1 + K_{\alpha, p} (E) ) }{\det (\1 + K_0 (E) ) } \frac{ \d E}{2 \pi }. \label{eqn:mainept}
\end{align}
Note that if $G(\alpha)$ is a differentiable matrix-valued function taking values in invertible matrices, then,
\begin{align}
\frac{\d}{\d \alpha} G^{-1} ( \alpha ) = - G^{-1} ( \alpha ) G' (\alpha ) G^{-1} ( \alpha).
\end{align}
This and dominated convergence yield, after some simple algebra,
\begin{align}
e''_+ ( \alpha ) &= \int_\E \frac{\d}{\d \alpha} \tr  \left( e^{ k_0 (E) } \left( e^{ k_0 (E) } + e^{ - \alpha (s^* (E) k_0 (E) s (E) - k_0 (E) ) } \right)^{-1} ( s^* (E) k_0 (E) s (E) - k_0 (E))   \right) \frac{ \d E}{2 \pi } \nonumber \\
&= \int_\E \tr \left( (e^{ \alpha A(E)  } + e^{ - k_0 (E) } )^{-1} A(E) (e^{ - \alpha A(E) }+ e^{  k_0 (E) } )^{-1} A(E) \right) \frac{ \d E}{2 \pi} \label{eqn:e+2der}
\end{align}
where $A(E) = s^* (E) k_0 (E) s (E) - k_0 (E)$. The integrand is non-negative for every $E$ and vanishes iff $A(E) = 0 \iff [s (E), k_0 (E) ] = 0$. Since $e_+ ( 0) = 0$, it follows that $e_+ (\alpha)$ vanishes identically iff $[s(E), k_0 (E) ] = 0$ for Lebesgue a.e. $E \in \E$ or $ | \E | = 0$.

An explicit computation shows that $[ s (E) , k_0 (E) ] = 0$ iff $\beta_l = \beta_r$ or $s(E)$ is diagonal.  Since $s(E)$ is not diagonal for Lebesgue a.e. $E \in \E$ it follows that $e_+ (\alpha)$ vanishes identically iff $ | \E | = 0$ or $\beta_l = \beta_r$.

Similarly, with $B(E) = s^* (E) k_0 (E) s (E)$,
\begin{align}
e''_{2, +} (\alpha) &= - \int_\E \frac{\d}{\d \alpha} \tr \bigg(  \left[ (1 + e^{ - (1 - \alpha) B(E)} e^{ - \alpha } k_0(E) )^{-1} + (1 + e^{ - \alpha k_0 (E)}e^{ (1 - \alpha ) B(E) } )^{-1} \right]\notag \\
& \times  (B(E) - k_0 (E) ) \bigg) \frac{\d E}{4 \pi} \nonumber \\
&= \int_\E \tr \bigg( ( e^{ ( 1 - \alpha ) B(E) } + e^{ - \alpha k_0(E)} )^{-1} (B(E) - k_0(E) ) ( e^{ \alpha k_0 (E) } + e^{ -( 1 - \alpha ) B(E) } )^{-1} \notag \\
&\times (B(E) - k_0 (E) ) + \mbox{h.c.}   \bigg) \frac{ \d E}{4 \pi}.
\end{align}
The integrand is non-negative and vanishes iff $[s (E), k_0 (E) ] = 0$ and so as in the case of $e_+( \alpha)$, we conclude that $e_{2, +} ( \alpha)$ vanishes iff $ | \E | = 0$ or $\beta_l = \beta_r$.

If $e_{2, +} (\alpha)$ vanishes identically, then $[s (E), k_0 (E) ] = 0$ and it follows that $e_{p, +} ( \alpha)$ vanishes identically. On the other hand, if $e_{2, +} ( \alpha )$ doesn't vanish identically then the above computation shows it is a strictly convex function. In (ii) we will see that $e_{2, + } ( 1) = e_{2, +} (0) = 0$ and so $e_{2, + } ( \alpha ) < 0$ for $ \alpha \in ]0, 1[$ and $e_{2, +} ( \alpha) > 0$ for $ \alpha \notin [0, 1]$. In (v) we will see that the function $ ]0, \infty] \ni p \to e_{p, +} (\alpha)$ is decreasing, and so it follows that $e_{p, +} (\alpha)$ does not vanish identically if $e_{2, +} (\alpha)$ does not vanish identically. This completes the proof of part (i).

We now prove parts (ii) and (iii). The analyticity of the entropic pressure functionals follows directly from the analyticity of the functions
\begin{align}
\langle \delta_k , \K_{\#, + } ( \alpha) \i [k, h] \delta_k \rangle.
\end{align}
The same arguments which led to the analyticity of the finite time entropic functionals in Proposition \ref{prop:thermfunct} yields that the above functions are analytic (note that here the argument is easier as there is no integral and no $t$, so the $F_j$'s will be taken to be functions only of $\alpha$).

Convexity and the symmetry $e_{p, + } ( \alpha) = e_{p, +} (1 - \alpha )$ follow from the fact that these properties are satisfied by the finite time functionals, of which these are pointwise limits. The fact that $e_{p, +} (0) = 0 $ follows from the formulas in (i), or that the finite time functionals satisfy the same equality. The formula (\ref{eqn:e+2der}) shows that $e_+ ( \alpha )$  is strictly convex unless either $ | \E | = 0$ or $[s (E), k_0 (E) ] = 0$ for Lebesgue a.e. $E \in \E$ iff $\beta_l = \beta_r$. 

As limits of convex functions, the $e_{p, +} (\alpha)$ are convex. Their second derivatives are analytic and therefore are either identically $0$ or have a set of isolated $0$'s. If they vanish identically then the $e_{p, +} (\alpha)$ are linear; however, since $e_{p, +} (1) = e_{p, +} (0) = 0$, they then must vanish identically if they are linear. Hence, the $e_{p, + } (\alpha)$ are strictly convex if they do not vanish identically.

Since $\K_{p, + } ( 0) = ( 1 + e^{-k_+ } )^{-1}$ for $p \in ]0, \infty]$ and $\K_+ ( 0 ) = - ( 1 + e^{- k_-} )^{-1}$, we have by the first line of (\ref{eqn:main1})
\begin{align}
e'_+  (0) = - \tr \left( ( 1 + e^{ -k_-} )^{-1} \i [k, h] \right) = - \tr \left( ( 1 + e^{-k} )^{-1} w_- ^* \i [k, h] w_-  \right) = - \langle \sigma \rangle_+ ,
\end{align}
and by (\ref{eqn:mainept}),
\begin{align}
e'_{p, +} = \tr \left( ( 1 + e^{- k_- })^{-1} \i [k, h] \right) = \tr \left( (1 + e^{-k} )^{-1} w_+ ^* \i [k, h] w_+ \right) = - \tr \left( (1 + e^{-k} )^{-1} w_- ^* \i [k, h] w_- \right) = - \langle \sigma \rangle_+ ,
\end{align}
where the second to last equality follows from the time reversal.

The application of Lemma \ref{lem:painful} in the proof of Proposition \ref{prop:thermfunct} yields that there is an $\eps >0$ so that the functions $\ES_t ( \alpha) /t$ and $e_{2, t} ( \alpha ) /t$  have analytic extensions to the disc $D ( 0, \eps)$ and are uniformly bounded in $t >0$ on this disc. The same arguments apply to the functions $\GC_t ( \alpha)/t$. The Vitali convergence theorem implies,
\begin{align}
e''_{2, +} ( 0 ) = \lim_{t \to \infty} \frac{1}{t} e''_{2, t} ( 0 ) , \qquad e''_{+} (0) = \lim_{t \to \infty} \frac{1}{t} \ES ''_{t} (0) = \lim_{t \to \infty} \frac{1}{t} \GC ''_t (0).
\end{align}
By Proposition \ref{prop:thermfunct}, $e_{2, t}'' (0) = \ES '' _t (0)$ for each $t$ and so $e '' _{2, +} (0) = e '' _+ (0)$.  By the Vitali convergence theorem,
\begin{align}
\GC '' _t  ( 0 ) = \lim_{s \to \infty} \lim_{M \to \infty} \frac{ \d^2 }{ \d \alpha ^2 } \log \omega_{s, M} \left( e ^{ - \alpha t \Sigma_M ^t } \right) \bigg\vert_{\alpha = 0}.
\end{align}
\begin{comment}
For twice differentiable matrix-valued functions $T ( \alpha )$ with $T (0) > 0$, 
\begin{align}
\frac{ \d ^2 } {\d \alpha^2 } \log \tr \left( T ( \alpha ) \right) \bigg\vert_{\alpha = 0 } = \tr \left( T'' (0)  \right) / \tr \left( T ( 0 )\right) - \left( \tr \left( T' (0 ) \right) / \tr \left( T (0) \right) \right)^2 ,
\end{align}
\end{comment}
Direct computation yields,
\begin{align}
 \frac{ \d^2 }{ \d \alpha ^2 }  \log \omega_{s, M} \left( e ^{ - \alpha t \Sigma_M ^t } \right) \bigg\vert_{\alpha = 0} &= \omega_{s, M} \left(  \left( t \Sigma_M ^t \right)^2 \right) - \left( \omega_{s, M} \left( t \Sigma_M ^t \right) \right)^2 \nonumber \\
&= \int_0 ^t \int_0 ^t \omega_{s, M} ( \sigma_{u, M} \sigma_{v, M} ) - \omega_{s, M} ( \sigma_{u, M} ) \omega_{s, M} ( \sigma_{v, M} ) \d u \d v ,
\end{align}
and so by dominated convergence,
\begin{align}
\frac{1}{t} \GC''_t (0) &= \frac{1}{t} \int_0 ^t \int_0 ^t \lim_{s \to \infty} \lim_{M \to \infty} \left( \omega_{s, M} ( \sigma_{u, M} \sigma_{v, M} ) - \omega_{s, M} ( \sigma_{u, M} ) \omega_{s, M} ( \sigma_{v, M} ) \right) \d u \d v \nonumber \\
&=  \frac{1}{t}\int_0 ^t \int_0 ^t \langle \sigma_u  \sigma_v \rangle _+ - \langle \sigma_u \rangle_+ \langle \sigma_v \rangle _+ \d u \d v \nonumber \\
&=  \frac{1}{t} \int_0 ^t \int_0 ^t \langle \sigma_{u - v } \sigma \rangle_+ - \langle \sigma \rangle_+ ^2 \d u \d v \nonumber \\
&= \frac{1}{2} \int_{-t} ^t \left(  \langle \sigma_s \sigma \rangle_+ - \langle \sigma \rangle_+ ^2 \right) \left( 1 - \frac{ |s| }{t} \right) \d s \nonumber \\
&= \frac{1}{2} \int_{-t} ^t \langle ( \sigma_u - \langle \sigma \rangle_+ \rangle ) ( \sigma - \langle \sigma \rangle _+ ) \rangle_+ \d s .
\end{align}
In the third and fifth lines we have used $\omega_+ \circ \tau ^t = \omega_+$, and the fourth line follows from a change of variable. Integration by parts yields,
\begin{align}
\frac{1}{t} \GC'' _t (0) = \frac{1}{t} \int_0 ^t \left[  \frac{1}{2}   \int_{-s} ^s \langle ( \sigma_u - \langle \sigma \rangle_+ ) ( \sigma - \langle \sigma \rangle _+  ) \rangle_+ \d u \right] \d s ,
\end{align}
which completes the proofs of (ii) and (iii).

We turn to the proof of (iv). We will compare $e_+(1)$ and $e_\infty (1)$. For any $2 \times 2$ matrix, $\det (\1 + A ) = 1 + \tr (A) + \det (A)$. We have also,
\begin{align}
\det \left( K_{\alpha, \infty} (E) \right) &= \exp \left[ \tr \left( ( 1 - \alpha ) k_0 (E) + \alpha s(E) k_0 (E) s^* (E) \right) \right] \nonumber \\
&= \exp \left[ \tr \left( k_0 (E) \right) \right]\nonumber \\
&= \det \left( K_{\alpha } (E) \right), 
\end{align}
The equality
\begin{align}
\tr \left( e^{k_0 (E) + \alpha ( s(E) k_0 (E) s^* (E) - k_0 (E) ) } \right) = \tr \left( e^{k_0 (E) + \alpha ( s^* (E) k_0 (E) s(E) k_0 (E)  - k_0 (E) ) } \right)
\end{align}
follows by conjugation and so,
\begin{align}
\det \left( \1 + K_\alpha (E) \right)  - \det \left( \1 + K_{\alpha, \infty} (E) \right) = \tr \left(  e^{ k_0 (E) } e^{ \alpha A(E) }  \right) - \tr \left( e^{ k_0 (E) + \alpha A(E)} \right)   ,
\end{align}
with $A(E)$ as before. By the Golden-Thompson inequality (see Corollary 2.3 and Exercise 2.8 in \cite{JOPP}; see also \cite{L}), the RHS is strictly greater than $0$ unless $ s^* (E) k_0 (E) s (E)$ and $k_0 (E)$ commute. It is easy to see that, if $\beta_l \neq \beta_r$, this can happen only if $s(E)$ is diagonal or off-diagonal or $E = 0$. For example, one can compute,
\begin{align}
[ k_0 (E) s^* (E) k_0 (E) s (E) ]_{1,2} &= E^2 ( \beta_l ^2 \widebar{s}_{ll} (E) s_{lr} (E) + \beta_l  \beta_r s_{rr} (E) \widebar{s}_{rl} (E) ) %\nonumber \\
%[ k_0 (E) s^* (E) k_0 (E) s (E) ]_{2,1} &= E^2 ( \beta_r ^2 \widebar{s}_{rr} (E) s_{rl} (E) + \beta_l  \beta_r s_{ll} (E) %\widebar{s}_{lr} (E) )
\end{align}
and
\begin{align}
[  s^* (E) k_0 (E) s (E) k_0 (E) ]_{1,2} &= E^2 ( \beta_l \beta_r \widebar{s}_{ll} (E) s_{lr} (E) + \beta_r ^2 s_{rr} (E) \widebar{s}_{rl} (E)).% \nonumber \\
%[  s^* (E) k_0 (E) s (E) k_0 (E) ]_{2,1} &= E^2 ( \beta_r \beta_l \widebar{s}_{rr} (E) s_{rl} (E) +   \beta_l ^2 s_{ll} (E) ) %\widebar{s}_{lr} (E)
\end{align}
By unitarity of the scattering matrix, $ \widebar{s}_{ll} (E) s_{lr} (E) + s_{rr} (E) \widebar{s}_{rl} (E) = 0$. It follows that if $k_0 (E)$ and $s^* (E) k_0 (E) s (E)$ commute, then either $E = 0$ or $\beta_l = \beta_r$ or $\widebar{s}_{ll} (E) s_{lr} (E) = 0$. Since the scattering matrix is unitary, this can happen only if $s(E)$ is diagonal or off-diagonal. Since $s(E)$ is not diagonal for Lebesgue a.e. $E \in \E$, it follows that $s^* (E) k_0 (E) s(E)$ and $k_0 (E)$ commute for Lebesgue a.e. $E \in \E$ iff $s(E)$ is off-diagonal for Lebesgue a.e. $E \in \E$, i.e., iff $h$ is reflectionless. In summary, we have proven that
\begin{align}
e_+ (1) \geq e_{\infty , + } (1) = 0
\end{align}
and equality holds iff $h$ is reflectionless.

If $h$ is reflectionless, then,
\begin{align}
s^* (E) k_0 (E) s (E) = \left(  \begin{matrix} - \beta_r E & 0 \\ 0 & -\beta_l E \end{matrix} \right) ,
\end{align}
and so, 
\begin{align}
\det \left( 1 + K_{\alpha} (E) \right) &= ( 1 + e^{ - \beta_r E \alpha - (1 - \alpha ) \beta_l E } ) ( 1 + e^{ - \beta_l E \alpha  - (1 - \alpha ) \beta_r E } ) \nonumber \\
&= \frac{ 4  \cosh ( ( \beta_l ( 1 - \alpha ) + \beta_r \alpha ) E /2 ) \cosh ( ( \beta_r ( 1 - \alpha ) + \beta_l \alpha )E /2 ) } { e^{ ( \beta_l + \beta_r ) E /2 }} ,
\end{align}
from which the formula in (iv) follows.

Only (v) remains. We have,
\begin{align}
\det \left( \1 + K_{\alpha , p } (E) \right) - \det ( \1 + K_{\alpha, q} (E) ) = \tr \left( K_{\alpha, p } (E) \right) - \tr \left( K_{\alpha, q } (E) \right).
\end{align}
The Araki-Lieb-Thirring inequality (see Theorem 2.2 and Exercise 2.8 in \cite{JOPP}; see also \cite{L}) implies that the RHS is strictly positive for $p < q$ unless $s^* (E) k_0 (E) s (E)$ and $k_0 (E)$ commute iff $h$ is reflectionless. If $h$ is not reflectionless, it follows that $ ] 0, \infty] \ni p \to e_{p, + } (\alpha)$ is strictly decreasing. If $h$ is reflectionless then it is trivial to check that all these functionals are identical and equal $e_+ (\alpha)$. \qed

\subsection{Large deviations}

We discuss here some consequences of Theorem \ref{thm:main}. In this section we assume that $ | \E | >0$ and $\beta_l \neq \beta_r$. Recall that,
\begin{align}
e_{2 , t} ( \alpha ) = \FCS_t ( \alpha ) = \log \int_\R e^{ - \alpha t \phi } \d \P _t ( \phi) ,
\end{align}
where $\P _t$ is the FCS measure of the extended XY chain of Proposition \ref{prop:thermfunct}. In the GNS representations for $\O$ and the states $\omega$ and $\omega_+$ (see, e.g.,  \cite{BR1}) denote the spectral measure for $\Sigma^t$ and $\omega$ and $\omega_+$ by $\P_{\ES, t}$ and $\P_{\GC, t}$, respectively. Then,
\begin{align}
\ES_t ( \alpha ) = \log \int_\R e^{ - \alpha t \phi } \d \P_{\ES, t} ( \phi ), \qquad \GC_t ( \alpha ) = \log \int _\R e^{ - \alpha t \phi } \d \P _{\GC, t} ( \phi ).
\end{align}
The large deviation rate functions are given by,
\begin{align}
I_{\FCS + } ( \theta ) &= - \inf _{\alpha \in \R} ( \alpha \theta + e_{2, + } ( \alpha ) ) \nonumber \\
I_{+ } ( \theta ) &= - \inf _{\alpha \in \R } ( \alpha \theta + e_+ ( \alpha ) ) .
\end{align}
We have,
\begin{lemma}
 \begin{enumerate}[label=(\roman{*}), ref=(\roman{*}),font=\normalfont]
\item The large deviation rate functions are non-negative, real-analytic, strictly convex and vanish at the single point $\theta = \langle \phi \rangle _+ $. 
\item The two rate functions are different unless $h$ is reflectionless.
\end{enumerate}
\end{lemma}
\proof The results follow from basic properties of Fenchel-Legendre transforms and the facts that $e '_+ ( 0) = e ' _{2, +} (0) = - \langle \phi \rangle _+$ and $e_+ (1) > 0 = e_{2, + } (1)$ if $h$ is not reflectionless. \qed

The symmetry $e_{2, +} ( \alpha ) = e_{2, + } ( 1 - \alpha )$ implies ,
\begin{align}
I_{\FCS +} ( \theta ) = I_{\FCS +} ( - \theta) + \theta .
\end{align}
If $I_+ ( \theta )$ satisfies this relation, then the symmetry $e_+ ( \alpha ) = e_+ ( 1 - \alpha )$ must hold and so $h$ is reflectionless. The consequences of Theorem \ref{thm:main} are,

\begin{corollary}\label{cor:main}
Suppose that $h$ has {\color{red} purely absolutely continuous} spectrum. 
\begin{enumerate}[label=(\roman{*}), ref=(\roman{*}),font=\normalfont]
\item The Large Deviation Principle holds:  for any open set $O\subseteq \R$,
\[
\lim_{t\rightarrow\infty}\frac{1}{t}\log{\mathbb P}_{{\rm ES},t}(O)
=\lim_{t\rightarrow\infty}\frac{1}{t}\log{\mathbb P}_{{\rm GC},t}(O)
=-\inf_{\theta\in O}I_+(\theta),
\] 
\[
\lim_{t\rightarrow \infty}\frac{1}{t}\log {\mathbb P}_{{\rm FCS},t}(O)
=-\inf_{\theta \in O}I_{{\rm FCS}+}(\theta).
\]
\item The Central Limit Theorem holds: for any Borel set $B\subseteq \R$, let
$B_t=\{\phi\,|\,\sqrt{t}(\phi-\langle\sigma\rangle_+)\in B\}$. Then
\[
\lim_{t\rightarrow\infty}{\mathbb P}_{{\rm ES},t}(B_t)
=\lim_{t\rightarrow\infty}{\mathbb P}_{{\rm GC},t}(B_t)
=\lim_{t\rightarrow\infty}{\mathbb P}_{{\rm FCS},t}(B_t)
=\frac{1}{\sqrt{2\pi D_+}}\int_B e^{-\phi^2/2D_+}\d\phi ,
\]
where the variance is $D_+=e_+^{\prime\prime}(0)$.
\end{enumerate}
\end{corollary}
\proof Since $e_+ ( \alpha )$  and $e_{2, +} ( \alpha ) $ are, in particular, differentiable, part (i) follows from the G\"artner-Ellis theorem (see, e.g., Appendix A.2 in \cite{JOPP}). In the proof of Theorem \ref{thm:main} we showed that there is an $\eps > 0$ so that the functions $e_{2, t} ( \alpha) / t $, $\ES_t ( \alpha ) /t $ and $\GC_t ( \alpha ) / t $ have uniformly bounded extensions to the disc $D(0, \eps )$. This, with the fact that $e' _+ (0) = e'_{2, +} (0) = - \langle \sigma \rangle_+$, implies that  part (ii) follows from Bryc's lemma (see Appendix A.4 in \cite{JOPP}). \qed

\section{Conclusions}

The differences between quantum and classical mechanics is due to the non-commutative structure of quantum mechanics. In quantum statistical mechanics this results in the emergence of novel entropic functionals. In our notation, the functionals $e_{p, t} ( \alpha)$ and $\ES _t ( \alpha)$ all result via different quantizations of the classical entropic functional. Note that the entropic pressure functionals all have the $\alpha \doublearrow 1 - \alpha$ symmetry which has played a key role in the recent developments in classical non-equilibrium statistical mechanics, while the direct quantizations of the Evans-Searles and Gallavotti-Cohen functionals do not. 

Theorem \ref{thm:main} shows that the Gallavotti-Cohen and Evans-Searles functionals are identical. This is not surprising, as the same phenomenon will holds in classical statistical mechanics - this is known as the \emph{principle of regular entropic fluctuations} - see \cite{JPR}. The fact that the entropic pressure functionals all equal the Evans-Searles functional if the Jacobi matrix is reflectionless is remarkable. Recall that $e_{2, t} (\alpha)$ was linked with the full counting statistics of a repeated quantum measurement protocol in Section \ref{sec:fcs}. This measurement protocol is purely of quantum origin, and has no classical analogue. That this functional should equal the one arising from classical mechanics is surprising.

Recall that our hypotheses required that the Jacobi matrix $h$ have purely absolutely continuous spectrum. The case that is most important for quantum mechanics is the Schr\"odinger case, in which all the $J_n$'s are all equal (recall that the $J_n$'s are the nearnest neighbour coupling of the spins). What is remarkable is that all known examples of this type of Jacobi matrix with purely absolutely continuous spectrum are reflectionless. For example, if $J_n = 1$ and $v_n = 0$ for every $n$, then the resulting operator is called the Laplacian, denoted $\Delta$. The discrete Fourier transform identifies $\Delta$ with the operator of multiplication by $2 \cos \theta$ on $L^2 ( [-\pi, \pi ], \d \theta )$, and so $\Delta$ has purely absolutely continuous spectrum.  Explicit computation shows that $\Delta$ is reflectonless (see, e.g., \cite{L}).

To summarize, it is currently an open problem if there exists a Jacobi matrix with $J_n = J \neq  0$ with purely absolutely continuous spectrum that is not reflectionless, and it would seem that this problem is relevent to the theory of entropic fluctuations. It is believed that such an example exists \cite{JLP}. In the non-Schr\"odinger case it is easy to construct operators that have both purely absolutely continuous spectrum that are not reflectionless - see \cite{VY}.

In the remarks following the proof of Theorem \ref{thm:ness},  we noted that the existence of a NESS could be extended to the case where $h$ has some pure point spectrum if the steady state is replaced by a Ces\`aro sum. It is currently open as to whether a result of this sort can be extended to Theorem \ref{thm:main}. That is, do the large time limits of the entropic functionals exist if $h$ has some point spectrum?

There are few models in quantum mechanics for which the existence of the large time limit entropic functionals can be proven.  The XY chain is a useful model because the functionals can be computed in closed form in terms of the scattering data and their properties can be examined. It remains to be seen that if the identification of XY chains with identical entropic functionals with reflectionless Jacobi matrices can be extended beyond exactly solvable models like the XY chain in the future.

\appendix
\section{Analyticity lemma}

We record here and prove a lemma that was used to obtain the analyticity of the entropic functionals.

\begin{lemma} \label{lem:painful}
Let $f(z)$ be a function analytic on the half-plane $\re z > 0$. Let $F_j ( \alpha, s, t ) : \R \times [0, 1] \times \R \to \O$, $j=1, 2, 3$ take values in bounded operators on a Hilbert space. Suppose that there is a strictly positive continuous function $a (\alpha)$ s.t.
\begin{align}
F_1 ( \alpha, s, t ) \geq a ( \alpha) \1
\end{align} 
for every $s$ and $t$.
% \begin{enumerate}[label=(A\arabic{*}), ref=(\roman{*}),font=\normalfont]
%\item 
Suppose that for fixed $s, t$ the function $\alpha \to F_j ( \alpha, s, t)$ is analytic in the following uniform sense: for each $\alpha_0 \in \R$, there is an $\eps > 0$ so that the series
\begin{align}
F_j ( \alpha, s, t) = \sum_{n = 0} ^\infty A_{n, j} (s, t) ( \alpha - \alpha_0 )^n
\end{align}
is absolutely convergent for $ | \alpha - \alpha_0 | < \eps$. Suppose furthermore that each function $s \to A_{n, j} (s, t)$ is continuous, and that there are numbers $a_{n, j}$ s.t. $\norm{ A _{n, j} (s, t) } \leq a_{n, j}$ for every $s, t$ and $j$ and the function
\begin{align}
H_j(\alpha) = \sum_{n=0}^\infty a_{n, j} | \alpha - \alpha_0 |^n
\end{align}
is finite for $ | \alpha - \alpha_0 | < \eps$. 

Then for any $\psi$ and $\phi$, the functions
\begin{align}
g_t (\alpha) = \int_0 ^1 \langle \psi , F_2 ( \alpha, s, t ) f ( F_1 ( \alpha, s, t) ) F_r ( \alpha, s, t ) \phi \rangle \d s
\end{align}
are real-analytic in $\alpha$. Furthermore, for each $\alpha_0 \in \R$, there is an $\eps > 0$ so that each $g_t (\alpha )$ is analytic in the disc $D ( \alpha_0, \eps )$ and
\begin{align}
\sup_{\substack{ \alpha \in D ( \alpha_0, \eps )  \\ t } } | g_t ( \alpha ) | < \infty.
\end{align}
%\end{enumerate}
\end{lemma}

\proof Fix $ \alpha_0 \in \R$ and let $\eps > 0$ be as in the hypotheses on the $F_j$ and $H$. WLOG we can suppose that $a_{n, 1} = a_{n, 2} = a_{n, 3}$ for every $n$ and we drop the subscripts and write $a_n$ and $H$ for $a_{n, j}$ and $H_j$. Let,
\begin{align}
M = \sup_{ \alpha \in D ( \alpha_0, \eps /2 ) } H ( \alpha ) < \infty, \qquad a = \inf_{\alpha \in [ \alpha_ 0 - \eps, \alpha_0 + \eps ] } a( \alpha) > 0 .
\end{align}
Choose $\lambda > M + a $. Define,
\begin{align}
\widetilde{H} ( \alpha ) =  \lambda - a  + ( H (\alpha) - a_0 ) \geq 0.
\end{align}
It follows that,
\begin{align}
\norm{ F_1 ( \alpha , s, t ) - \lambda } \leq \widetilde{H} ( \alpha ) ,
\end{align}
for $ \alpha \in D ( \alpha_0, \eps )$.
Since $\widetilde{H} ( \alpha_0 ) =  \lambda$, there is an $0 < \eps' < \eps /2 $ and a $\delta > 0$ s.t.
\begin{align}
\widetilde{H} ( \alpha ) \leq \lambda - \delta ,
\end{align}
for $\alpha \in D ( \alpha_0, \eps ')$. Since $f(z)$ is analytic in the half-plane $\re z > 0$, it follows that the series
\begin{align}
f(z) = \sum_{n = 0} ^\infty b_n ( z - \lambda )^n 
\end{align}
converges absolutely for $ | z - \lambda | < \lambda $. Note that $\eps'$ does not depend on $s$ or $t$. It follows that $f ( F_1 ( \alpha, s, t ) )$ is analytic for $ \alpha \in D ( \alpha_0, \eps ' )$.

The formal power series obtained by expanding each of $F_2 ( \alpha, s, t), f (F_1 ( \alpha, s, t ), F_3 ( \alpha, s, t )$, sandwiching them between then $\psi$ and the $\phi$ and integrating over $s$ in fact converges absolutely in the disc $D ( \alpha_0, \eps ')$ because,
\begin{align}
& \int_0^1 \norm{\psi} \norm{\phi} \left[ \sum_{n = 0}^\infty \norm{ A_{n, 2} (s, t )} | \alpha - \alpha_0 |^n  \right] \left[ \sum_{n = 0}^\infty \norm{ A_{n, 3} (s, t )} | \alpha - \alpha_0 |^n  \right] \nonumber \\
&\times  \sum_{m = 0} ^\infty |b_m | \left[ \norm{A_{0, 1} (s, t) - \lambda }+  \sum_{n = 1}^\infty \norm{ A_{n, 1} (s, t )} | \alpha - \alpha_0 |^n  \right]^m \d s \nonumber \\
&\leq \norm{\psi} \norm{\phi} H ( \alpha ) H ( \alpha ) \sum_{m = 0 } ^ \infty | b_m |  \widetilde{H} ( \alpha ) ^ m \leq \norm{\psi} \norm{\phi} M^2 \sum_{m = 0}^ \infty |b_m | ( \lambda - \delta )^m < \infty .
\end{align}
This yields the claim. \qed

\end{document}